\documentclass[%
reprint,
superscriptaddress,
longbibliography,
nofootinbib,
amsmath,amssymb,
floatfix,
aip,
author-year,
]{revtex4-2}

\usepackage[utf8]{inputenc}
\usepackage{amsthm,amsmath,amssymb}
\usepackage{graphicx}
\usepackage[table]{xcolor}

\usepackage{tikz}
\usetikzlibrary{quotes}
\usetikzlibrary{shapes.misc}
\usetikzlibrary{calc}

\usepackage{pgfplots}
\pgfplotsset{compat=1.17}
\usepgfplotslibrary{colorbrewer}
\pgfplotsset{colormap/Set1-8}

\tikzstyle{e}=[thick,>=latex,rounded corners]
\tikzset{every label/.style={rectangle,fill=none,draw=none, label distance=0pt}}

\tikzstyle{n}=[rounded rectangle,draw=white,line width=2pt,outer sep=0pt,inner sep=3pt,draw opacity=0,minimum size=5pt]
\tikzstyle{inline}=[anchor=base]
\tikzstyle{open}=[fill=Dark2-A!50]
\tikzstyle{closed}=[fill=Dark2-B!50]
\tikzstyle{unobserved}=[opacity=0.5]
\tikzstyle{conditioned}=[draw=black,line width=2pt,draw opacity=1]
\tikzstyle{unfair}=[text=Dark2-B]
\tikzstyle{bias}=[draw=Dark2-B]

\tikzstyle{faded}=[opacity=0.3]
\tikzstyle{n2}=[n,fill=Dark2-C!30]

\makeatletter
\DeclareRobustCommand{\rvdots}{%
  \vbox{
    \baselineskip4\p@\lineskiplimit\z@
    \kern-\p@
    \hbox{.}\hbox{.}\hbox{.}
  }}
\makeatother

\usepackage[colorlinks=true,%
 linkcolor = {Set1-B},%
 citecolor = {Set1-B},%
 urlcolor = {Set1-B},
 unicode]{hyperref}
\usepackage{xurl}
\usepackage{orcidlink}

\usepackage[british]{babel}

\makeatletter
\let\old@fixname\bbl@fixname
\def\bbl@fixname#1{%
  \@ifundefined{babelalias#1}%
    {\old@fixname{#1}}%
    {\edef\languagename{\csname babelalias#1\endcsname}}}
\makeatother

\DeclareMathOperator{\lognormal}{LogNormal}

\DeclareMathOperator{\E}{E}

\usepackage{pgfplots}
\pgfplotsset{compat=1.17}
\usepgfplotslibrary{colorbrewer}
\pgfplotsset{colormap/Set1-8}

\usepackage[colorlinks=true,%
 linkcolor = {Set1-B},%
 citecolor = {Set1-B},%
 urlcolor = {Set1-B},
 unicode]{hyperref}
\usepackage{xurl}

\graphicspath{ {figures/} }

\usepackage{footmisc}

\makeatletter
\def\frontmatter@thefootnote{%
 \altaffilletter@sw{\@fnsymbol}{\@fnsymbol}{\csname c@\@mpfn\endcsname}%
}%
\makeatother

\begin{document}

\title{Science of science---Citation models and research evaluation}

\thanks{%
  This is a draft. %
  The final version will be available in \emph{Handbook of Computational Social Science} edited by Taha Yasseri, forthcoming 2025, Edward Elgar Publishing Ltd. %
  The material cannot be used for any other purpose without further permission of the publisher and is for private use only. %
  Please cite as: %
  Traag, V. A. (2025). Citation models and research evaluation. In: T. Yasseri (Ed.), \emph{Handbook of Computational Social Science}. Edward Elgar Publishing Ltd.%
  }

\author{Vincent Traag\,\orcidlinkc{0000-0003-3170-3879}}
\email{v.a.traag@cwts.leidenuniv.nl}
\affiliation{Centre for Science and Technology Studies (CWTS), Leiden University, the Netherlands}

\date{\today}

\begin{abstract}

Citations in science are being studied from several perspectives, among which approaches such as scientometrics and science of science.
In this chapter I briefly review some of the literature on citations, citation distributions and models of citations.
These citations feature prominently in another part of the literature which is dealing with research evaluation and the role of metrics and indicators in that process.
Here I briefly review part of the discussion in research evaluation.
This also touches on the subject of how citations relate to peer review.
Finally, I conclude by trying to integrate the two literatures.
The fundamental problem in research evaluation is that research quality is unobservable.
This has consequences for conclusions that we can draw from quantitative studies of citations and citation models.
The term ``indicators'' is a relevant concept in this context, which I try to clarify.
Causality is important for properly understanding indicators, especially when indicators are used in practice: when we \emph{act} on indicators, we enter causal territory.
Even when an indicator might have been valid, through its very use, the consequences of its use may invalidate it.
By combining citation models with proper causal reasoning and acknowledging the fundamental problem about unobservable research quality, we may hope to make progress.

\end{abstract}


\maketitle

The study of science itself has a venerable history, and is studied from several points of view.
The field of scientometrics studies science from a quantitative perspective.
Relatedly, the field of science of science is similarly taking a quantitative perspective, but often with a somewhat different approach.
The two have much in common and share a more quantitative formal perspective on studying science, especially based on large-scale data sets of publications, their authors and their citations.
Scientometrics has been traditionally more focused on ``measuring science'', while much of science of science is more focused on ``modelling science''.
This distinction is not absolute though: some publications in what most would consider scientometrics build models, while publications in science of science sometimes also address issues of measuring.
I will review part of this literature, with a focus on citations.

Studies are often motivated by the fact that citations are considered to be relevant for how science operates: they may reflect advances in science and clarify intellectual contributions.
Moreover, citations and related aspects seem to play a role in scientists' own careers, a development that seems to have become increasingly stronger over the years.
The use of metrics in research evaluation is increasingly criticised.
The role of metrics in research evaluation and the effects of metrics are less often discussed explicitly by the scientometric and science of science literature.
In this chapter I aim to connect these two literatures, with a focus on citations.

First I discuss various observations of citation distributions, how they change over time, and how they can be modelled.
This literature is largely based on a mix of scientometrics and science of science.
Then, I review a small part of the literature on research evaluation.
This includes some aspects relevant to national research evaluations.
This also touches upon issues of comparing peer review and metrics, which I will also briefly discuss.
After reviewing this literature, I will conclude by bringing the two literatures in conversation with each other.
I deconstruct some aspects of citation dynamics and clarify that other factors play a role in citation dynamics, the implications of which are, although sometimes acknowledged, not often appreciated in this literature.
Additionally, I will consider how we can think of ``indicators'' in this context, how they can be biased, and how they can be made more accurate.
A causal understanding is key to understanding indicators, I believe.

The fundamental problem in research evaluation is that scientific quality is unobservable.
Any study on the subject therefore must acknowledge this, and this has consequences for the type of conclusions that we can draw, especially from quantitative studies.
Being more clear about our causal reasoning and being careful about what we can and cannot conclude helps to clarify that.
Having a better understanding of the overall dynamics in citations, and relying on models that capture these dynamics, we can improve what we can infer from observations.

\section{Citations}

\label{sec:citations}

\subsection{Citation distributions}
\label{sec:distribution}

One of the earliest and most commonly studied aspects in scientometrics and science of science is the distribution of the number of citations.
Various authors have tried to find theoretical distributions that could fit the empirical distribution well.
Part of the literature has tried to come up with theoretical models that could explain the observed type of distributions, and I will cover some such studies later.
One important consideration here is that there does not exist such a thing as \emph{the} distribution of citations.
Distribution of citations always refer to a particular set of papers, and the results will vary across fields, years, journals or institutions.

One of the earliest studies of citations was covered by \citet{Price1965-gg}.
He studied the number of citations to all papers covered in one of the earliest editions of the Science Citation Index, the precursor of what is currently known as the Web of Science.
\citet{Price1965-gg} finds that citations are distributed approximately as a power law:
\begin{equation}
  \Pr(C \geq c) \propto c^{-\alpha + 1},
\end{equation}
with $\alpha$ estimated to be somewhere between $2.5$--$3.0$.
This points to a highly skewed citation distribution.
Indeed, he finds that ``only 1 percent of the cited papers are cited as many as six or more times each in a year''.

Physicists became increasingly interested in citations and citation networks in the late 1990s.
\citet{Redner1998-wy} is an early study of citations in that literature and reports a power law distribution with an exponent of about $3$, similar to \citet{Price1965-gg}.
He studies a few different citation distributions: a distribution from a single year (1981) coming from a precursor of the Web of Science, and a few different volumes of \emph{Physical Review D}.
These different datasets show quite a different number of average citations and older years generally have accumulated more citations, simply as the result of having had more time to accumulate citations.
He finds that various datasets collapse onto a universal curve when dividing the number of citations by the average number of citations in that dataset.

\citet{Laherrere1998-vo} study a slightly different citation distribution, namely the citation distribution of all citations to authors, instead of citations to individual papers.
They find that a stretched exponential is the best fit for their distribution:
\begin{equation}
  \Pr(C \geq c) = \exp - \left(\frac{c}{c_0}\right)^\alpha
\end{equation}
However, they also find that a power law is a reasonable fit, with an exponent of about $3$ again.

\citet{Radicchi2008-ju} study citation distributions of a few different fields and a few different years.
They study these distributions separately and try to determine whether the distributions are universal, in the sense that after some transformation, all the distributions look alike.
They find that a simple scaling of citations with the average number of citations in the same field and the same year collapses all the distributions onto a single curve, hence finding evidence for universal scaling.
That is, they define the normalised citations $\tilde{C}_i = \frac{C_i}{\E(C_i)}$ where $C_i$ is the total number of citations received for publication $i$ and $\E(C_i)$ is the average number of citations received over all publications from the same field and the same publication year.
They find that the normalised citations $\tilde{C}_i$ are well-fitted by a lognormal distribution $\lognormal(-\frac{\sigma^2}{2}, \sigma^2)$, with $\sigma^2 \approx 1.3$, which by definition has an average of $1$.
In this study they used data from Web of Science and relied on the field definitions given by journal subject categories.
They repeat this study later with data from the American Physics Society (APS) and PACS codes to define fields \citep{Radicchi2011-wx}, again finding a similar collapse of distributions onto a universal curve.
\citet{Chatterjee2016-ih} perform a similar study of the universality of citation distributions, but then of institutions and journals.
They also find evidence for universality and find that the normalised citations are well-fitted by a lognormal distribution.
For academic institutions they find that $\sigma^2 \approx 1.7$; somewhat more skewed than the paper level citation distribution identified by \citet{Radicchi2008-ju}, while for journals the collapsed citation distributions is slightly less broad with $\sigma^2 \approx 1.4$.
For both institutions and journals, the lognormal is less able to fit the tail of the distributions, which seems to be better approximated by a power law.
Possibly, this could be a result of differences in sizes, which plays a role for institutions and journals, but not for individual paper distributions.

The universality claim by \citet{Radicchi2008-ju} was revisited by \citet{Waltman2011-jw}, who argued that citation distributions are not truly universal, and that differences can still be observed between some fields.
They study this by comparing the top 10\% of all publications based on the normalised citations to the top 10\% within each field.
If citations were perfectly universal, the overall top 10\% would overlap with the top 10\% of each field, but this is not the case.
Ignoring uncited articles does make the case for universal distribution stronger.
The probability of having zero citations may be slightly distinct from the overall citation distribution.

In a follow-up analysis \citet{Radicchi2012-vo} devise a clever way of empirically deriving a slightly different normalisation such that citation distributions across different fields collapse.
They base this on comparing the overall citation distribution to all distributions of citations per field, and find that the transformation $\left(\frac{C}{a}\right)^\alpha$ produces highly similar distributions across nearly all fields of science, where $a$ and $\alpha$ are estimated empirically.
For this universal distribution, they find it is well fit by a lognormal distribution.

As suggested by the analysis of \citet{Waltman2011-jw}, the case of zero citations may function slightly differently.
\citet{Wallace2009-lc} also consider these uncited articles when studying a century of citation distributions.
In particular, they find that $e^{\beta\frac{N_r}{N_a}}$ is a good fit for predicting the number of uncited papers in a distribution, where $N_a$ is the total number of articles published in a year and $N_r$ the total number of references to those $N_a$ articles.
This is based on a simple idea that the $N_r$ citations are randomly distributed across the $N_a$ articles, and the uncitedness is the probability of having drawn $0$ references, at least within a short time window (2 years).
They fit a stretched exponential to the citation distribution, where the probability to be cited $c$ times is
\begin{equation}
  \Pr(C) \sim P(0) \exp -\left(\frac{C}{\tau}\right)^{\alpha}
\end{equation}
where $\tau$ and $\alpha$ are estimated parameters, with $P(0)$ the separately modelled uncited publications.
This is based on the idea that different papers accumulate citations at different  rates, and that the overall distribution is a mixture of all those different rates.
It is not clarified how the stretched exponential arises as a mixture of individual Poisson processes with different rates.
One possibility is to model the distribution as a mixture of Poisson distributions with the rate of each Poisson distribution following a Gamma distribution.
That would result in a Negative Binomial distribution, which is studied by~\citet{Mingers2006-my}.
\citet{Thelwall2014-gk} find that Negative Binomial regression is a bad fit, and advise against using it, and suggest using an OLS logarithmic fit.
To cover the entire range, \citet{Wallace2009-lc} suggest that a distribution first suggested by \citet{Tsallis2000-qu} fits best:
\begin{equation}
  \Pr(c) = \frac{P(0)}{\left[ 1 + (q - 1) \lambda c \right]^{\frac{q}{q-1}}},
\end{equation}
with parameters $\lambda$ and $q$, but a clear theoretical underpinning for this distribution is lacking.

The decline of concentration in citations is described by \citet{Lariviere2009-ap}.
They find that over time, the number of uncited papers continues to decrease (except for the humanities).
Whereas in the 1920s about 70\% of the articles remain uncited within 5 years, in the 2000s this has decreased to about 10-30\%.
The citation distribution also seems to become less skewed over time.
Before the Second World War, the percentage of papers that attracted 80\% of the citations increased from a few percent to 25--30\%, and it continues to hover around that percentage, with most recent times seeing an even larger increase.

\citet{Redner2005-qk} also takes a long-term perspective, studying citation statistics from 110 years of \emph{Physical Review} journals.
He finds that the overall number of citations of all papers is well fit by a lognormal distribution.
This is in a sense surprising, since he studies the distribution of papers from multiple years (1893--2003), in which case you might expect a mixture of yearly lognormal distributions, which could result in a stronger power law tail.

\citet{Moreira2015-ww} find that a (discretised) lognormal distribution captures well citation distribution over sets of papers from authors and departments, and that the distributions are relatively stable over time.
\citet{Sinatra2016-up} find that a lognormal distribution also fits well the citation distribution over a set of papers.
\citet{Stringer2005-em} find that a lognormal distribution also fits well the citation distribution over journals, and that the distribution becomes stationary after about 10 years.
Similar to \citet{Milojevic2016-yb}, they use this to rank journals by focusing on the probability that a paper from one journal is cited more highly than a paper from another journal.
The ranking results are consistent with journal distributions being approximately lognormal.

Overall then, the most reasonable assumption seems to be that citations are distributed approximately as a lognormal.
Other observed distributions most likely arise as mixtures of a lognormal, resulting in stronger power law tails.

\subsection{Temporal decay of citation rate}

Citations generally decay over time.
Most papers tend to cite recent work more frequently than older work.
We can study this from two perspectives.
We can take a retrospective, backward looking approach \citep{Burrel2001-de}, sometimes called a synchronous approach \citep{Line1974-tt}, and study the age of references in papers.
Alternatively, we can take a prospective, forward-looking approach \citep{Burrel2001-de}, sometimes called a diachronous approach \citep{Line1974-tt}, and study at how frequently a paper is cited in the years after it is published.

The decay of citations over time is sometimes referred to as obsolescence, referring to the decline of the use of certain publications over time.
Publications need not become fully obsolete, but their usage may decline nonetheless.
As \citet{Line1974-tt} explain, there are various reasons why certain publications may become obsolete.
The work may become common knowledge in the field, sometimes referred to as obliteration by incorporation \citep{Garfield1957-qo}.
This happens for example when a theory has become eponymised, such as the Nash equilibrium \citep{McCain2011-fr}.
Alternatively, the work may have become outdated or belong to an abandoned paradigm \citep{Kuhn2012-in}.
Work may also later be found to be incorrect or inaccurate \citep{Furman2012-uq}, although some citations continue after retractions for example, seemingly unaware of the retracted status \citep{Bornemann-Cimenti2016-nf}.

Many studies take a retrospective approach.
This approach is easier to use, especially historically.
Taking a retrospective approach involves taking a paper, and checking the years of the cited references in that paper, which is relatively straightforward, especially when working with actual printed papers, which used to be the case historically.
In contrast, as \citet{Egghe2000-td} point out, a prospective approach requires one to go through all papers to check whether it has cited the paper of interest,
This is only possible if there is a proper citation database available.

\citet{Gross1927-yr} are one of the first to study how publications reference literature in earlier years.
\citet{Burton1960-xm} introduced the half-life of attention/usage in this context, although half-life was already used earlier, according to \citet{Line1970-gl}, while the concepts of growth, utility and obsolescence were introduced by \citet{Brookes1970-lk}.
\citet{Price1965-gg} introduced a measure of immediacy, later sometimes called the Price Index, which is defined as the percentage of references younger than $t$ years.

\citet{Line1970-gl}\footnote{%
Interestingly, this seems to be one of the earliest examples of a paper that appends the report by one of the referees, at least that I am aware of. %
An early example of transparent peer review, even signed by the referee!
} %
distinguishes between real and apparent obsolescence, arguing that we should control for the number of papers being published, which increases exponentially each year.
In his words ``if every item had an equal probability of being used or cited, more use would be made of more recent literature simply because there is more of it.''
He takes a retrospective approach and studies the (median) reference age.
He introduces a very simple correction to the observed obsolescence factor.
Suppose that the observed obsolescence is $a(t) = \frac{c(t - 1)}{c(t)}$, where $c(t)$ is the total citations given to articles in the year $t$, from some reference year $t' > t$.
Now suppose that the number of citations $c(t)$ has grown from year $t - 1$ to $t$ with a factor $g(t)$ such that $c(t) = g(t)c(t-1)$.
In order to correct the obsolescence $a(t)$ for this growth $g(t)$, we should then divide $c(t - 1)$ by the expected number of citations $\frac{c(t)}{g(t)}$ that were obtained had there been no growth.
Hence, the growth corrected obsolescence should then be defined as $a(t) = \frac{c(t - 1)}{\frac{c(t)}{g(t)}}$.
Assuming constant growth rates, $a(t) = a$ and $g(t) = g$, we then obtain constant corrected obsolescence $d(t) = d$.
The growth-corrected number of citations in year $t$ then simply is $c(t) = c(t - 1) d$, such that $c(t) = c(0) d^t$, and the infinite series $\sum_t c(t)$ equals $\frac{c_0}{1 - d}$.
The corrected half-life $h$ is then $\frac{\log \frac{1}{2}}{\log a + \log g}$ while the uncorrected (observed) half-life would be $\frac{\log \frac{1}{2}}{\log a}$.
Although a gross oversimplification, it nicely captures how the growth in the number of publications affects the apparent obsolescence of the literature.
With a yearly growth percentage of 5\%, a median citation age of 7 years would suggest that items might be considered for removal from the library after 7 years, while in reality they would continue to be used for almost 14 years.

\citet{Brookes1970-lk} discusses some problems with estimating the obsolescence, and relates this to geometric decay of utility such that $c(t) \propto (1 - a) a^{t - 1}$ with an annual ageing factor $a$.
\citet{Egghe1992-uo} argue against the ageing perspective from \citet{Brookes1970-lk} that assumes a constant ageing factor.
Instead, they find that ageing has a certain minimum, suggesting there is a natural peak in reference age.
They find that the most sensible distribution is then a lognormal distribution, based on finding a unique minimum in ageing, and find it fits the data well.

\citet{Avramescu1979-sr} studies retrospective reference distributions.
He suggested the following model to fit to the retrospective distribution:
\begin{equation}
  c(t) = C_0 \left[ \exp(-\alpha t) - \exp(-m \alpha t) \right]
\end{equation}
where $c(t)$ is the number of citations received $t$ years after publication, and $\alpha$ and $m > 1$ are some parameters.

Instead of working with obsolescence rates $a(t) = \frac{f(t - 1)}{f(t)}$ \citet{Egghe1994-rm} proposes a continuous counterpart for a continuous function $c(t)$, namely $a(t) = \exp (\log f(t))'$, where the prime $'$ indicates taking the derivative with respect to $t$.
This of course equals $\exp\left(\frac{f'(t)}{f(t)}\right)$ so that this is the exponent of the relative growth of $t$.
This is a ``true'' rate, as \citet{Egghe1994-rm} states, and makes intuitive sense and has some sensible properties.
However, this formulation does not seem to have been used frequently.

\citet{Stinson1987-wk} studies citation ageing from both a synchronous and diachronous perspective.
They take into account a correction for the growth of the literature, but they do not report any particular distribution.

\citet{Redner2005-qk} finds an exponential decrease in the age distribution.
As suggested by \citet{Nakamoto1988-nh}, the growth in the number of publications is also relevant in this context.
Combining the exponential decrease in age of references with an exponential growth of publication leads to a power law decrease in age overall \citep{Redner2005-qk,Egghe2005-pa}.

\citet{Vinkler1996-eb} formulates a relatively simple model for the possibility to be cited and finds that the possibility to be cited increases with the growth of the field.
Faster growing fields are hence more likely to show higher chances of citations.
This is also noticed by \citet{Hargens1984-gl} who argue that in growing fields, older work tends to gather more citations from recent work than in stable fields.
Scientists might therefore be eager to jump on the bandwagon of a newly emerging field, because it pays off to be one of the first movers in a new field.

\citet{Lariviere2008-zg} take a long-term perspective, and find that the average reference age increased over the last decades.
Similarly, in physics the average reference age was found to have increased over the last decades \citep{Sinatra2015-ag}.
\citet{Verstak2014-hc} also find that the average reference age increased over the last decades in various fields.
\citet{Lariviere2008-zg} observe some interesting peaks during both world wars.
Relatively few publications were published during those two periods, showing a dip in the number of publications.
Most papers that appeared during, and shortly after, the war therefore reference papers from before the war, resulting in a quite high reference age.

\citet{Egghe2010-cx} proposes a simple model for some observations of increasing reference age, as observed by \citet{Lariviere2008-zg}, while it still has a decreasing Price Index (i.e. the proportion of references in the last $t$ years).
The model is quite straightforward: it assumes that the literature grows exponentially, and that publications are cited completely at random.
Even in that simplest case, one already sees an increasing reference age, but a decreasing Price Index.
Hence, qualitatively, such observations do not require an explanation beyond a simple exponential growth of the literature.

\citet{Parolo2015-iq} study the prospective citation distribution and state that the nature of the decay is not well established, varying between an exponential decay and a slower power law decay.
They find that attention decays faster more recently than in earlier years.
If they renormalise time in terms of the number of papers, they find that the decay rate is stable.
Hence, the faster attention decay is simply a result of the increasing number of publications.
They only study the decay after the initial peak of citations.
Over time, the peak in citations has come increasingly faster, consistent with the increasing reference age found by \citet{Lariviere2008-zg}, according to \citet{Parolo2015-iq}.
The decay after the initial peak is best fit by an exponential function.
The half-life decreases over time, and citations taper off increasingly faster in more recent years.
Again, when rescaling time in terms of number of publications, this decrease is no longer visible.

\citet{Subelj2017-jr} find that the peak year of reference distributions (i.e. retrospective) has stayed stable in computer science and physics.
The peak year of the citation distribution (i.e. prospective) has shifted however, and is more volatile, especially for computer science.
Again, when normalising the citations based on the number of publications, all distributions seem to collapse onto a universal curve.
As \citet{Egghe2000-td} explain, growth influences ageing, but it does not cause ageing per se.
They find that increasing growth rates lead to higher obsolescence, i.e. papers tend to become obsolete more quickly.

\citet{Pan2018-yc} also study the ageing of reference distributions, and finds evidence of ``citation inflation'': papers need increasingly more citations to be part of the top 5\%.
They find that citations to recent literature and very old literature decreased, while citations to the ``middle'' part increased.

\citet{Gingras2008-zr} find that the average age of the references depends on the age of researchers.
Younger researchers initially tend to cite more recent work, but when researchers become older, their references age with them, with a turning point when researchers become about 40 years old.

\citet{Herman2004-sm,Herman2004-bq} studies scholars' literature search behaviour qualitatively.
She finds that most people only go back a few years to look for references to keep up-to-date on the most recent development.
Most scholars mentioned that they would not search the literature further back than just a couple of years, but may follow up by chasing down references from that literature.

\citet{Poncela-Casasnovas2019-gf} find that papers that reference a highly cited paper and are relatively highly cited themselves as well are published relatively shortly after each other.
This suggests something like the start of a field, where an initial publication is cited by another paper shortly afterwards, both of which play a role in the ensuing citation dynamics and the influx of authors to such a field.
Higher impact papers tend to cite younger papers and very young papers ($< 1$~year).
They find that method references are typically older.
\citet{Bertin2016-mp} find that references in the introduction of a paper are typically older.
This most likely sets the stage and background of a field for a paper.

In principle, there is a certain connection between a retrospective and a prospective approach.
The exact connection depends on the dynamics of the number of publications and the number of references throughout time.
But, in general, if the retrospective distribution remains stable throughout time, it can be used to infer the prospective distribution, while using the empirically observed publication and referencing dynamics.
\citet{Yin2017-la} provide an exact relationship between the two approaches and find that
\begin{equation}
  \Pr^\leftarrow(t_2 \mid t_1) M(t_1) = \Pr^\rightarrow(t_1 \mid t_2)L(t_2)
\end{equation}
with $\displaystyle \Pr^\leftarrow$ the prospective distribution and $\displaystyle \Pr^\rightarrow$ the retrospective distribution, where $M(t)=m(t)N(t)$ is the total number of references given at time $t$, with $m(t)$ the average number of references in year $t$ and $N(t)$ the total number of publications in year $t$, which approximately grows exponentially over time $N(t) \sim e^{\beta t}$, while $L(t) = \int_{t}^\infty {\displaystyle \Pr^\leftarrow}(t \mid \tau) M(\tau) d \tau$ is the total number of citations received by papers at time $t$.
Hence, one can derive the one distribution from the other.
Arguably, the retrospective distribution is primary and the prospective distribution is derivative.
After all, the retrospective distribution describes how researchers behave and choose to cite previous literature, while the prospective distribution is the result of that process.

\citet{Yin2017-la} find that after normalising the citations for the number of publications it is well fit by a lognormal distribution (both prospective and retrospective).
This entails that the unnormalised, crude, age distributions are a mixture of the publication and referencing dynamics and the actual lognormal decay.

\subsection{Citation models}

The models that I cover in this section attempt to capture various observations.
Some models try to explain the overall citation distribution, others target the ageing distribution of references, while others aim to model individual paper citation dynamics, sometimes with an eye on predicting future citations.

One of the first models that was introduced in this context was developed by \citet{Price1976-rb}.
It introduced the notion of cumulative advantage, based on the ideas of the Matthew effect introduced earlier by \citet{Merton1968-qu}, sometimes called a rich-get-richer effect.
The model of \citet{Price1976-rb} aims to explain the broad distribution of citations observed earlier (see also section~\ref{sec:distribution}).
The model is relatively straightforward and works as follows.
For each time step, an additional paper is added to the population, citing $m$ earlier papers.
These references are not added randomly, but are assumed to be distributed proportional to the current number of citations of each publication.
That is, the probability of paper $i$ to be cited is proportional to $C_i(t)$, the total number of citations at time step $t$.
Often some constant is added, in order to make sure that papers for which $C_i(t) = 0$ also have some non-zero probability to be cited.
The overall citation distribution is then affected by the influx of new papers at each time step, which initially have no citations, and the earlier publications which accumulate increasingly more citations.
These forces give rise to a distribution, which \citet{Price1976-rb} calls the Cumulative Advantage Distribution $c \sim (m + 1)B(c, m + 2)$ where $B(a, b)$ is the Beta function.
In the limit of large citations $c$ this approaches a power law with exponent $m + 2$, which for $m = 1$ is close to the earlier observations of citation distributions reviewed in section~\ref{sec:distribution}.
This idea of cumulative advantage was again suggested in the late 1990s in the context of complex networks by \citet{Barabasi1999-dz}, who termed this preferential attachment.

\citet{Redner2005-qk} finds some evidence for a linear preferential attachment, and suggests that a redirection mechanism could be reasonable.
That is, instead of directly connecting to a paper with probability proportional to $C_i(t)$, the idea is to pick a reference from a randomly selected paper (with probability $1 - r$) or simply reference the randomly selected paper itself (with probability $r$), leading to a linear preferential attachment.

Demonstrating that there is a cumulative advantage effect in empirical observations is not easy.
Often scholars study the relationship between the cumulative number of citations $C(t)$ after time $t$ and the additional citations in some time period $\Delta t$ after $t$.
If the cumulative number of citations $C(t)$ is correlated with this increase $\Delta C(t + \Delta t) = C(t + \Delta t) - C(t)$, this is often taken as evidence for the existence of a cumulative advantage.
However, this does not need to be the case.
The inherent problem with this approach is that some \emph{latent citation rate} of the article may affect both $C(t)$ and $\Delta C(t + \Delta t)$.
Hence, the additional citations $\Delta C(t + \Delta t)$ need not be the result of the earlier citations $C(t)$: publications that achieve a higher $C(t)$ just have a higher latent citation rate, and therefore also show a higher $\Delta C(t + \Delta t)$.
As a straightforward example, if citations accrue at a rate of $\lambda$ then $E(C(t)) = \lambda t$ and so $\E(\Delta C(t + \Delta t)) = \lambda \Delta t$, so that $\E(C(t)) = \E(\Delta C(t + \Delta t)) \frac{t}{\Delta t}$.
Hence, observing a linear growth of the additional number of citations with the initial number of citations need not indicate much more than simply the result of a constant growth rate.
In other words, the correlation between $C_i(t)$ and $\Delta C(t + \Delta t)$ may not arise due to $C_i(t)$ \emph{causing} more citations $\Delta C(t + \Delta t)$, but because they are confounded by the underlying citation rate $\lambda$.
Citation distributions might be skewed just because the underlying latent citation rates are skewed, not because of a cumulative advantage effect.
In some other contexts there is some clear experimental evidence of a Matthew effect \citep{Van_de_Rijt2014-yx}.
Additionally, there is some qualitative evidence that people browse the literature by following references \citep{Herman2004-sm,Herman2004-bq}, also leading to a type of cumulative advantage effect.
So, there might be other reasons to believe cumulative advantage is reasonable, but the empirical data analysis is more challenging.

Some models start on the basis of such a simple assumption of a constant rate of accumulation of citations, often taking a prospective perspective.
\citet{Mingers2006-my} for example proposes a Gamma mixture of Poisson distributions.
That is, each paper has a latent citation rate $\lambda_i$ and the number of citations is then simply Poisson distributed with rate parameter $\lambda_i$.
Assuming $\lambda_i$ is distributed according to a Gamma distribution, the overall number of citations is distributed as a Negative Binomial.
They fit this distribution to empirical data, and find it to be a good fit, except for some extremely highly cited papers.
Next, they suggest that the rate of attracting citations is a time dependent variable $\lambda_i(t)$, with some constant latent citation rate, modulated by some time factor, i.e. $\lambda_i(t) = \lambda_i f(t)$, with $f(t)$ the \emph{obsolescence function}.
They estimate their model for journals, estimating the overall citation distribution and the decay, which allows them to predict citations for articles in that journal, but do not predict citations for individual articles.
\citet{Burrel2001-de} considers a similar starting model, where each paper $i$ accumulates citations at a latent citation rate $\lambda_i$ modulated by some time dependence $f(t)$, so that the effective citation rate is $\lambda_i(t) = \lambda_i f(t)$.
In this model, the shape of the distribution of the time to the first-citation is independent of the mixing distribution of the latent citation rates, and only depends on the shape of the obsolescence function, suggesting that the obsolescence function follows some $S$-shaped pattern.
Moreover, after a sufficiently long time, the number of citations depends only on $\lambda_i$ and not on the obsolescence function $f(t)$.
\citet{Burrell2002-cu} follows up on this work and investigates the $n$-th citation distribution of this model.
He finds that a Gamma distribution of latent citation rates, which leads to a Negative Binomial distribution of citations, fits well the data, while relying on an obsolescence function that follows a specific Gamma distribution $\Gamma(2, 1)$, corresponding to $f(t) = 1 - e^{-t} (1 + t)$.

\citet{Higham2017-ho} propose that the rate of attracting additional citations is a separable function of preferential attachment and some obsolescence function, while taking a forward-looking prospective view.
The rate of attracting citations in year $t$ is then
\begin{equation}
  \lambda(C(t), t) = a(C(t)) f(t),
  \label{equ:separable}
\end{equation}
where $f(t)$ is some obsolescence function that depends on time $t$ only and $a(c)$ is some cumulative advantage function that depends on citations $C(t)$ only.
In particular, they use functional forms $a(c) = c^\alpha + c_0$, and $f(t) = d_0 \exp \left( - \frac{t}{\tau} \right)$.
They test if these can indeed be separated by checking various years and bins of citations $c$ against each other, and find support for the idea of separability.
They find that $\alpha$ is about $1.0$--$1.2$ while the exponential fit performs well only for $t \geq 3$.
Based on this prospective model, they also derive an expression for the retrospective distribution of references.
The separability of citation dynamics is an important observation that simplifies the modelling.
Some earlier authors also proposed separable models.
\citet{Dorogovtsev2000-jd} seems to have introduced the earliest ageing with preferential attachment model, including a separable formulation, and used $f(t) = t^{-\alpha}$ and $a(c) = c$, in terms of Eq.~\ref{equ:separable}.
\citet{Wang2009-cr} also proposed a separable model with $f(t) = \exp(- \lambda t)$ and $a(c) = c$, and find it fits well some empirical data.
Neither study explicitly addresses the separability though.
\citet{Hajra2006-po} propose changing earlier models and publish multiple papers simultaneously, instead of sequentially introducing single papers, as was usually done, and find that this improves the fit.

Although the previously discussed models consider obsolescence, they offer no theoretical explanations.
\citet{Simkin2007-jy} suggest a mathematical theory of citations that provides an explanation for ageing.
They propose that every year $t$ there are $N$ papers published that contain $N_r$ references on average.
A fraction $\alpha$ of these references goes to randomly selected papers in the preceding year $t - 1$ (with $\alpha \approx 0.1\text{--}0.15$).
This leads naturally to the first-year citations for papers published in $t - 1$ being distributed Poisson, in line with earlier discussed results from~\citet{Wallace2009-lc}, with $\alpha N_r$ expected citations.
With probability $1 - \alpha$ then, a random reference from a random publication in year $t - 1$ is followed and is cited.
Such a publication from year $t - 1$ might have cited a random publication from year $t - 2$ (with probability $\alpha$) or might have cited a random reference from that publication (with probability $1 - \alpha$), and so on.
This leads to a branching process which \citet{Simkin2007-jy} solve analytically.
They find the prospective and retrospective distribution of citations to be a power law with an exponential cut-off.
To account for large exponential cut-offs and obtain a power law scaling, \citet{Simkin2007-jy} propose to add a latent citation rate parameter for each paper.
Instead of choosing a random paper and a random reference from a random paper, scientists then choose a paper proportional to the latent citation rate.
They consider a uniform distribution of latent citation rates, and marginalise the decay over this to obtain a citation distribution across all articles.
Various latent citation rate distributions yield similar observed citation distributions.

\citet{Peterson2010-lx} created a similar model to \citet{Simkin2007-jy}, but propose as the first step to find random papers from all years instead of the preceding year only.
Their model focuses on the citation distribution, not the ageing distribution.
\citet{Goldberg2015-fl} also consider a similar copying model, and find it to be the best fitting model.

\citet{Pan2018-yc} propose a model that combines various elements from earlier models.
It takes redirection from the models by \citet{Simkin2007-jy} and \citet{Peterson2010-lx}, but also uses an initial preferential attachment.
In each time step, $n(t)$ new publications are added, which grows exponentially.
Each new publication cites directly an existing publication $j$ with probability $(a + C_j(t))f(t_j)^\alpha$ where $C_j(t)$ is the number of citations to publication $j$, which is published at time $t_j \leq t$, and with an additional $k$ random references from publication $j$, with $k$ binomially distributed.
They find their model to reproduce several stylistic features of citation networks.

\citet{Eom2011-jy} find that a shifted power-law best fits citation distributions.
They propose to model the citation network as follows.
At each step, a new paper, i.e. node, is added to the citation network.
The new paper $i$ cites a previous paper $j$ proportional to their current cumulative number of citations $C_j(t)$ and a certain decay as
\begin{equation}
  c_{ij} \propto C_j(t) + \lambda_j f(t)
\end{equation}
where $\lambda_i$ is the latent citation rate of article $j$ and $f(t)$ is some decay factor, assumed to be exponential by \citet{Eom2011-jy}.
They find this model to reproduce various distributions reasonably well.

\citet{Wang2013-tj} introduced a model that similarly combines a temporal decay with a rich-get-richer effect while also allowing for an individual article level parameter to account for variability across papers.
In a sense, this approach is similar to what was proposed by \citet{Eom2011-jy}, but they only considered aggregate properties, such as citation distributions, whereas \citet{Wang2013-tj} try to predict citation dynamics of individual papers.
This hence combines most previous elements, and is relatively similar in spirit to the model by \citet{Pan2018-yc}.
More specifically, \citet{Wang2013-tj} model the rate of attracting additional citations $c_i(t)$ at time $t$ as
\begin{equation}
  c_i(t) \propto \lambda_i C_i(t) f(t_i)
\end{equation}
with $C_i(t)$ the total number of citations up until time $t$ and $\lambda_i$ the latent citation rate of article $i$.
It might be interesting to empirically compare this model, using a multiplicative formulation, to the earlier model by \citet{Eom2011-jy}, which uses an additive formulation.
Solving the model by \citet{Wang2013-tj} leads to the result that
\begin{equation}
  C_i(t) \propto e^{\lambda_i F(t_i)} - 1
\end{equation}
where $F(t_i)$ is the cumulative distribution of $f(t_i)$.
For $t_i \to \infty$ then, we have that $F(t_i) = 1$ so that after a sufficiently long time we arrive at
\begin{equation}
  c_i(t) \propto e^{\lambda_i} - 1.
\end{equation}
This means that ultimately, after waiting long enough, the total number of citations is expected to depend only on the latent citation rate $\lambda_i$, similar to what was observed by \citet{Burrell2002-cu}.
\cite{Wang2013-tj} found their model to fit well the citation dynamics of many papers.

In response, \citet{Wang2014-zu} wrote that the predictions of the model of \citet{Wang2013-tj} were not so good and that naive predictions were more accurate.
In a rebuttal \citet{Wang2014-sh} argued that overfitting of their model should be prevented by using informative priors (in a Bayesian analysis) or by otherwise regularising the fitting procedure.
One element of dispute seems to be the purpose of models.
From the perspective of \citet{Wang2014-zu} the complexity of the model by \citet{Wang2013-tj} is simply not necessary, since a simple prediction performs equally well, while \citet{Wang2014-sh} argue that they model the dynamics that are seen in citations.
One difference between the two seems to be that, once the model of \citet{Wang2013-tj} is in place, one could in principle predict forward citations across multiple years over time.
The naive prediction that \citet{Wang2014-zu} considered was to actually assume citations after $5$ and after $30$ years simply have not changed, which is of course not informative.
In addition, \citet{Penner2013-dt} point to a problem when predicting citations, namely that many studies focus on the cumulative number of citations, which is also relevant in this particular disagreement.
\citet{Penner2013-dt} argue that comparing cumulative citations is misleading, because one can easily predict cumulative citations from earlier cumulative citations, even if the process is completely random.
That is, suppose that $c_i(t)$ is a completely random variable, with the cumulative number of citations up until time $t$ being $C_i(t) = \sum_{\tau = 0}^t c_i(\tau)$.
The correlation of $C_i(t)$ between time $t$ and $t + \Delta t$ can then be quite high and equals
\begin{equation}
  \sqrt{\frac{t}{t + \Delta t}}.
\end{equation}
Hence, if $\Delta t$ is small compared to $t$, the correlation will be high.
For small $t$, the correlation is lower.
These correlations are purely mechanical and result directly from the cumulative citations.
If, instead of the cumulative citations $C_i(t)$, we try to predict the yearly citations $c_i(t)$, we would quickly learn that the expected correlation between any $C_i(t)$ and $c_i(t)$ is zero, because the process is completely random.
Hence, when comparing the predictive capabilities of different models, the focus should be on predicting $\Delta C(t + \Delta t)$, not on predicting $C(t + \Delta t)$, which, as \citet{Wang2014-zu} also observe, can nearly trivially be predicted based on $C(t)$.

\section{Evaluation of research}

Research is regularly being evaluated, for various reasons, such as funding decisions, hiring decisions or quality assurance.
The use and misuse of citation-based metrics regularly feature in the literature on this topic.
Here I briefly review some of that literature.

The role of journals in research evaluation has been contested for quite some time.
The Journal Impact Factor (JIF)---the average number of citations to a journal in the preceding two years---was originally developed for decisions about journal collection management in libraries \citep{Lariviere2019-ib}.
From the 1990s onwards, JIFs were increasingly used in research evaluation \citep{Hicks2015-nj}.
Journals show a high heterogeneity of what they publish, and \citet{Seglen1997-sk} argued that you should not evaluate an individual article based on where it is published, similar to the adage that you should not judge a book by its cover.
The JIF became increasingly contested, resulting in a call to abandon them for research evaluation in the Declaration on Research Assessment~\citep{Dora2013-kt}.
The subject was also discussed in a workshop on Rethinking JIFs \citep{Wouters2019-cr}.
Some even talked about ``Impact Factor mania'' \citep{Casadevall2014-nx}.
Following a call to publish the full citation distribution instead of the JIF~\citep{Lariviere2016-ws}, the Journal Citation Reports now provide more detailed information.
Still, in recent times, the JIF has continued to be used in promotion and tenure decisions \citep{McKiernan2019-nu}.

The JIF was reported to feature not only when evaluating research that has already been done, but also to shape decisions of what research questions to focus on \citep{Rushforth2015-vg}.
The JIF structures discussions about what is novel and sufficiently high-quality to target high-impact journals.
The JIF was not used per se to say something about the potential novelty and quality of the science itself, but was also seen as a ``ticket'' to advance one's career.
Importantly, this shows that the JIF is not just about targeting specific journals once the research itself is already done; the research is done and shaped with impact factors in mind.
Indeed, this phenomenon has been called ``thinking with indicators'', shaping not only post-research where a manuscript should be submitted, or how something is evaluated, but also actively shaping what research is done \citep{Muller2017-bq}.
These effects of indicator usage have been reviewed more broadly by \citet{De_Rijcke2016-kf}.

One important recurrent theme in this context is that of goal displacement.
This phenomenon is sometimes known as Goodhart's law, or Campbell's law: scoring high on assessment indicators becomes more important than doing well on whatever those indicators were meant to measure.
This is closely related to the so-called constitutive effects of performance indicators \citep{Dahler-Larsen2014-kl}.
When indicators are used in practice, they may affect how people respond.
This should not be thought of as ``unintended consequences''; rather, the usage of the indicator itself defines what is evaluated.
Hence, if citations are used to evaluate research quality, they might not necessarily be misused, but rather, an indicator, such as citations, comes to represent the very object that they purport to measure.
For instance, when publishing university rankings, their very usage alone may result in such rankings becoming thought of as measures of university ``performance''.
More highly ranked universities may attract more students and more high-qualified personnel.
Such effects may not necessarily result from university ranking itself being ``correct'' indicators of performance, but because the ranking itself produces such effects.
Something similar may happen with journal impact.
Journal impact rankings and publicly visible indicators, such as the JIF, may reify through constitutive effects any initial ranking of ``journal impact''.
That is, if scholars start to judge journals by such a ranking, they might start to submit their best work to the highest ranked journal, which thereby may solidify, or even improve their ranking, while lower ranked journals may start to receive increasingly worse manuscripts, thereby potentially lowering their ranking.
In this sense, constitutive effects may function similarly to self-fulfilling prophecies.
Whether constitutive effects ameliorate or deteriorate outcomes is not clear \emph{a priori}.

\citet{Molas-Gallart2018-qw} provide a broad critique of indicators.
Citation-based indicators may not align well with research objectives, leading to an ``evaluation gap''.
They argue that scientists respond to evaluation by aiming to improve their performance as measured by indicators, similar to constitutive effects.
If such an evaluation has the desired properties, this effect might be positive, but this need not be the case.
Even without responding strategically to such incentives, evaluations may act as a selective pressure \citep{Smaldino2016-io}.
That is, it does not require constitutive effects in order to exert an influence.

\citet{Bhattacharya2020-ya} also critique metrics based on the argument that attention (i.e. citations) to novel ideas has decreased, and that evaluating people based on citations effectively selects against novelty.
They argue that more scientists are working on only incremental advances that will be more likely to be cited, instead of working on foundational groundwork.

A particular context in which metrics are sometimes used for research evaluation is in performance-based university research funding systems (PBRFS), which were reviewed by \citet{Hicks2012-tb}.
Although the distribution of funding is an important component of PBRFS, they also seem to feed into a prestige competition.
The first and perhaps most well-known PBRFS is the UK's Research Assessment Exercise (RAE), currently known as the Research Excellence Framework (REF).
In general, PBRFS aim to stimulate excellence, or fund more selectively, to allocate scarce resources more effectively.
The resource concentration has also been linked to the ``new public management'' that has become more dominant in research policy circles.
The most common unit of evaluation is the department of universities or research organisations, although some countries also evaluate individual scientists, for example for appointing professors.

There is an extensive literature discussing the potential effects of PBRFS.
\citet{Butler2003-em} performed a seminal study on the increase of publications in lower impact journals following the introduction of a PBRFS in Australia.
Another effect of introducing PBRFS is that researchers may cite each other's work more heavily \citep{Baccini2019-is}.
Some authors found evidence that self-citations increased after the introduction of an evaluation system for promotion in Italy \citep{Seeber2017-wo}.
\citet{Moed2008-ni} showed that the UK RAE exercises seemed to affect UK scholars' publishing practices.
The classical work by \citet{Butler2003-em} was revisited by \citet{Van_den_Besselaar2017-yr}, reaching different conclusions: productivity and impact both increased in the Australian case.
However, generally, causes and effects in PBRFS are rather challenging to disentangle \citep{Aagaard2017-fh}, as argued earlier by \citet{Osuna2011-se}.
\citet{Glaser2016-uk} describe the overall problem of inferring how macro level science policies affect macro level outcomes.
Their central question is: How does research governance change knowledge production?
This not only needs to be studied at the macro level, which is bound to be affected by problematic confounding effects (e.g. other changes happening simultaneously); this also needs to be studied at a micro level, providing evidence for a macro-micro-macro link.
That is, it should be made reasonable that the macro policy affects researchers' behaviour, which in turn becomes visible at the system level again.
One additional potential problem is that an increase in national productivity may also increase national citations.
Such higher within-country citations are regularly observed \citep{Schubert2006-ng,Bakare2017-xe}, similar to citations in the same language \citep{Bookstein1999-rb}.
This raises the question of how to disentangle an increase in citations due to a higher productivity from an increase in citations due to actual differences in research quality.
Whether such observations are really driven by national citation biases, or whether they are a result of more general geographical patterns, as observed by~\citet{Pan2012-da} is not clear.

\citet{Sandstrom2018-mp} study the performance of several national science systems.
They conclude that having ex post evaluation, combined with high institutional funding may be most efficient.
Ex ante evaluation, either through grant funding, or through lower professional autonomy and more university management, may result in lower efficiency, and may reinforce the existing academic elite.

\citet{Schneider2016-fs} compared the effects of PBRFS in Australia and Norway.
They find that, unlike in Australia \citep{Butler2003-em}, the introduction of a PBRFS in Norway that awarded publications did not show a decreasing impact or an increasing output in lower impact journals.
The important difference here is whether the evaluation differentiates the awards based on some impact indicator.
In the Norwegian case they differentiated between lower and higher impact tier outlets.
\citet{Bloch2016-dy} study the effects of the Norwegian model further, and conclude that due to the fractionalisation, the system may not properly reward collaboration.

In principle, evaluation at the institutional level is to be stimulated \citep{Tiokhin2021-ch}.
Institutional evaluation may alleviate some problems that might appear at the individual level, where contributions other than scholarly publications might be disregarded.
Institutions can take a broader perspective, and can for example hire someone who does not directly produce scholarly output, but who has a large indirect effect on scholarly output, for instance by maintaining critical infrastructure.
Unfortunately, one recurrent problem of evaluation at the institutional level seems to be that institutions pass down the institutional requirements directly to lower levels \citep{Glaser2007-aq}.
For example, in the UK REF system, which is an institutional evaluation, the institutions organise so-called mock-REFs to identify areas where individual scientists could improve their performance, with sometimes dire consequences for their future careers \citep{Owens2013-ku}.

When studying the causal effects of a PBRFS we should differentiate between system level effects and individual level effects.
For example, consider that we fund institutions differentially, based on some performance indicator.
After a few years, the overall performance may have increased.
At the same time, the differences between institutions may have become smaller: all institutions have increased their performance.
Differentiating between institutions then becomes more difficult, and institutions that receive more funding may not necessarily perform much better than institutions that received less funding in one year.
It may then appear the differential funding may not be predictive of individual performance, while at the same time, the differential funding did increase the overall quality.

\subsection{Peer review}
\label{sec:peer_review}

Most scientists argue that the scientific ``quality'' of a paper is a multidimensional concept \citep{Aksnes2019-kl}.
For example, in most journals peer review is based on multiple criteria, such as novelty, potential impact and methodological rigour.
In recent years, peer review has been heavily discussed, with multiple possible interventions on several fronts, such as open peer review, post-publication peer review or collaborative peer review \citep{Woods2022-xp}.
In almost any evaluative setting, the focus is on trying to evaluate research ``quality''.
The question is how either peer review or citations can reflect such ``quality''.
Let me briefly review some of the literature on peer review.

\citet{Bornmann2011-eg} provides a general overview of peer review and identifies a number of problems of peer review.
One particular problem is poor reliability: the inter-rater reliability between peer reviewers is generally low.
This was already observed earlier by \citet{Cole1981-dt}, but was also confirmed in later research again by \citet{Ernst1993-jo}, \citet{Rothwell2000-oe} and \citet{Pier2018-rj}.
The low reliability of peer review opens up the possibility of bias.
When a decision needs to be made in a difficult case, the possibility for bias becomes larger to ``tip the scale''.
On the other hand, the uncertainty in peer review can be one of its strengths.
It is difficult to know in advance how something will be evaluated by peers, so using peer review for evaluation decreases the chances of people targeting a specific indicator.
Low agreement on evaluation may also reflect different positions and considerations that reviewers may have on a manuscript.
Peer review can indeed improve the reporting of findings \citep{Goodman1994-yw}, although the textual changes are often relatively minor \citep{Klein2016-sx}.
In a sense, poor agreement demonstrates that multiple reviewers provide more comprehensive feedback than a single reviewer.
If reviewers would simply reiterate the same point, there is little added value of the additional reviewer.
Initiatives, such as the consultative peer review from eLife \citep{King2017-gr}, try to benefit from this diversity and suggest an innovative approach to consolidate the various points raised by multiple reviewers.

As said, one problem of peer review is the potential bias: factors unrelated to ``quality'' may affect peer review \citep{Lee2013-rb}.
It can be challenging to establish whether something is a bias \citep{traag_causal_2022}.
For example, simply showing that authors from a particular institution have higher peer review scores is insufficient: it is possible that such authors simply more often produce higher quality work.
Comparisons of double-blind to single-blind peer review reveal some interesting effects, where author and affiliation reputation seem to affect the acceptance of manuscripts \citep{Tomkins2017-lb,Okike2016-cv}.

Another problem that \citet{Bornmann2011-eg} identifies is that of validity: peer review might be unable to predict scientific impact or relevance.
However, the problem is that scientific impact and peer review itself may be noisy: how will we measure scientific impact?
For example, if you compare the ``best'' unfunded scholars (i.e. those with the highest scientific impact) to the scientific impact of funded scholars, as done by \citet{Van_den_Besselaar2015-qp}, it might very well be that the ``best'' unfunded scholars outperform the funded ones, not because the best unfunded are ``better'' than the funded, but simply because citations are such a noisy proxy \citep{Lai2020-bi}.
\citet{Bornmann2008-kk} analyse the citation outcomes of both accepted and rejected publications at the prestigious \emph{Angewandte Chemie International Edition}, and find that peer review outcomes predict subsequent citations.
However, this conclusion is problematic if the publication venue causally affects how frequently it is cited \citep{Traag2021-rq}.
In that case, citations do not necessarily reflect whether peer review is predictive, they may just reflect the causal effect of being published in a certain venue.
A similar problem plays in an analysis of the predictive validity of peer review when highlighting publications in a journal~\citep{Antonoyiannakis2021-od}.

\subsection{Metrics}
\label{sec:metrics}

Much research in scientometrics does not necessarily engage with the citation models that I briefly covered in section~\ref{sec:citations}.
Instead, much research is interested in factors that somehow seem to affect citations, ranging from the effects of authors to institutions.
Some also study aspects such as title length, number of pages and other characteristics, but I will ignore those studies here.
Most of the more quantitative studies do not explicitly use any citation model, but simply compare different articles with each other in one way or another, and try to draw conclusions from that comparison.
Other studies focus on the meaning of citations, and study the different types of ``influence'' that citations capture.

\citet{MacRoberts1989-er} provide a comprehensive overview of some of the problems in citation analysis.
Although their overview is over thirty years old by now, many of the identified problems are still playing a role, and continue to be discussed and studied.
I have already covered some common problems in section \ref{sec:citations}, namely varying citation patterns in different fields, years and document types.
Another category of problems concerns whether citations really capture the idea of ``influence'' or impact: not all influences are cited and some works are cited that have no influence (so-called perfunctory citations).
There are different types of citations \citep{Bornmann2008-ah}, which do not show an equal influence, with some citations for example being negative \citep{Lamers2021-xb}.
This does not necessarily mean that highly-cited publications are not influential.
For example, \citet{Teplitskiy2020-xv} find that highly-cited publications are actually more likely to have an intellectual influence on the work in which they are cited.
Another category of problems mentioned by \citet{MacRoberts1989-er} is more technical and relates to coverage issues \citep{Visser2021-ns}, problems of reference matching \citep{Olensky2016-yq} and problems of author disambiguation \citep{Caron2014-ai}.

Co-authored papers are cited more frequently, and this holds for multiple authors, multiple institutions and multiple countries \citep{Lariviere2015-uq}.
This seems not a result of self-citation, but really represents greater ``epistemic value'', as stated by \citet{Lariviere2015-uq}.
\citet{Wu2019-yv} have looked at this from a slightly different angle and found that larger teams typically produce less disruptive papers, but they are more likely to be more highly cited.

\citet{Cole1968-hs} find that the prestige of a department affects the visibility of authors.
\citet{Cole1970-nl} finds that the prestige of a department also affects (early) citation counts, especially for work that is of lower quality.
Similarly, \citet{Medoff2006-sh} finds that institutional prestige drives citations in economics, but only for elite universities.
\citet{Way2019-av} find evidence that research quality is driven by scholars' current work environment, and that it is not driven by selection of more highly cited scholar into more prestigious departments.

The role of journals in citations has been debated for a long time.
As I already discussed earlier, citation distributions of journals are roughly lognormal.
Correlations between the JIF of a journal, and the individual citations for each article is generally low~\citep{Seglen1997-sk}.
A more recent revisit of the work by \citet{Seglen1997-sk} again found that correlations between impact factors and citations are relatively low \citep{Zhang2017-yo}.
It was also shown that the correlation between the JIF and citations has weakened over the years \citep{Lozano2012-rg}, which was speculated to have been caused by digitalisation.
Electronic publication was observed to narrow the referencing, also to more recent literature \citep{Evans2008-wf}.
At the same time, where an article is published is one of the strongest single predictive factor of citations in several studies \citep{Stegehuis2015-tg, Callaham2002-po, Abramo2010-kd, Mingers2010-jz}.

As already stated earlier, the fundamental problem is that research quality is unobservable.
Clearly, citation distributions are highly skewed for each journal, and also overlap to a large extent, as I discussed earlier.
However, citations are only a proxy of quality, and are not equal to research quality.
Similarly, being published in a certain journal may be a proxy of quality.
The question is then: which is a better proxy?
Although many people may argue that citations are a more accurate proxy, this need not be the case, as \citet{Waltman2020-fh} demonstrate.
It is possible that all articles within the same journal have the same quality and that the broad distribution of citations is simply due to citations being a noisy proxy of this identical quality.
The average of these noisy citations can then be a more accurate representation of the underlying identical quality than the actual citations.
The extent to which journals publish similar quality articles is up for debate.
This for example will depend on reviewer uncertainty when scholars submit publications.
If there is substantial uncertainty, and reviewers try to assess the actual quality of the papers, then the resulting distributions of quality in journals may largely overlap \citep{Starbuck2005-ss}.

High-impact journals are more widely circulated, and hence have a higher readership \citep{Peritz1995-eo}.
There is a certain circularity here, and path dependency: higher impact journals have a higher readership, which attracts more interesting submissions, which in turn attracts more readers, which in turn attracts more citations.
\citet{Ellis1991-hc} found that current journal prestige is mostly determined by previous journal prestige and current impact, lending some support to this idea of path dependency and conservatism of journal prestige.

More generally, publicity has clear effects on citations.
\citet{Phillips1991-md} analysed what papers were being discussed in the New York Times, and how that influenced citations ten years later.
Using a three-month period during which the NYT did not appear, but the editorial process and selection remained, they studied the causal effect of publicity in the NYT.
They found a quite strong effect: featured papers received 73\% more citations.
At the same time, the newsworthiness itself also predicts the impact of the journal in which an article will appear \citep{Callaham2002-po}.

Citations to identical papers showed that versions that were published in more highly cited journals were cited more often \citep{Knothe2006-fd, Perneger2010-wp}, which was also coined as the Impact Factor's Matthew effect \citep{Lariviere2010-vw}.
\citet{Seglen1994-jo} questioned whether there was any causal relation between JIF and citations.
I will get back to this in section~\ref{sec:conclusion}.

\subsection{Comparing peer review and metrics}
\label{sec:comparing_review_metrics}

Metrics have been regularly compared to peer review outcomes.
Both are thought to be indicators of scientific ``quality'' or ``impact'', and both have been used in research evaluation.
One central difference is that metrics can only be used post-publication, while peer review is also used frequently pre-publication, for example when reviewing journal submissions.
Many national PBRFS I discussed earlier, such as the UK REF, the Italian VQR or the Norwegian system are post-publication evaluation systems, and some are based explicitly on metrics (such as the Norwegian model), while others are based on peer review (such as the UK REF) or a mixture of the two (such as the Italian VQR).
In the influential Metric Tide report \citep{Wilsdon2015-bd}, the use of metrics in the national research evaluation in the UK was extensively discussed.
They concluded that metrics could support but not supplant peer review, as also summarised by \citet{Wilsdon2015-gk}.

\citet{Aksnes2004-sf} compare peer review and metrics in Norway.
They find that normalised citations correlate best with peer review evaluations at the research group level, and report higher correlations for higher aggregate levels.
The average journal impact shows a similar level of correlation with peer review.
Interestingly, when considering citations relative to the journal (i.e. controlling for the journal impact) they find the lowest correlation.

\citet{Bornmann2013-tj} find that peer review, in the form of recommendations from F1000, is correlated with a number of citation-based indicators.
Noticeably, this again finds that when normalising based on the journal, there is barely any correlation between peer review and the journal-normalised citation-based indicator.

\citet{Radicchi2017-jk} asked respondents to compare pairs of papers, and asked them which paper had a higher influence on their own work.
Generally, they find a rather low correlation between citations and those pairwise preferences, but for respondents' own papers, more highly cited papers were more often said to have a higher influence on their own work.

\citet{Adams2008-ma} compared evaluation outcomes of papers in the RAE with journal-normalised citation scores, and found that they essentially did not correlate.
\citet{Eyre-Walker2013-ud} also showed that correlations between evaluation and citations are minimal when controlling for the journal, although some of their conclusions have been questioned by \citet{Eisen2013-yw}.

There are a few problems when comparing peer review and metrics.
Studies have reported wildly varying correlations, ranging from as low as $0.3$ to as high as $0.97$.
\citet{Traag2019-ul} argue that the comparison between peer review and metrics should take into consideration at least two aspects:
\begin{enumerate}
  \item Whether size-dependent (e.g. a sum) or size-independent (e.g. a mean) indicators are compared.
  \item What level of aggregation (e.g. individual papers, departments, entire universities) is being analysed.
\end{enumerate}
Depending on these choices, correlations can be high or low.
In addition, \citet{Traag2019-ul} argue that correlations should also be compared to a ``baseline'' of peer review uncertainty.

Analysis of data from the Italian VQR exercise shows that peer review is not very reliable \citep{Bertocchi2015-cs}, as I already discussed earlier.
Compared to correlations between two peer reviewers, correlations between peer review and metrics are found to be comparable.
This holds not only at the individual paper level \citep{Bertocchi2015-cs} but also at the aggregate institutional level \citep{Traag2020-vv}.
The correlations at the institutional level are typically higher, and this holds both for correlations between two peer reviewers and between peer review and metrics.
Of course, an average evaluation outcome can be estimated more accurately when using more peer reviewers.
Hence, in the case of many peer reviewers, repeating an evaluation exercise should give highly similar answers, but may still leave a difference between peer review and metrics.
However, the resources to do such a large scale peer review exercise are generally limited, so this is infeasible in practice.
A recent study by \citet{Forscher2019-iy} reported that in the context of NIH funding, one would need as many as 12 reviewers to obtain a modest reliability in funding decisions.

\section{Conclusion}

\label{sec:conclusion}

In this chapter, I briefly reviewed both citation models and some relevant aspects of research evaluation, including peer review and metrics.
Although they are treated separately, citation models and research evaluation are related, and the two literatures can be brought into closer conversation with each other.

First, citation models may help us draw inferences about certain effects.
As I reviewed in section~\ref{sec:metrics}, there are many questions about how various factors may or may not affect citations, such as author reputation, institutional reputation and journal reputation.
However, the inference of these effects is tricky.
Models such as the one by~\citet{Wang2013-tj} may help to disentangle such effects.
Essentially, the rate at which an article attracts citations can be formulated as $\lambda(t)$, where $\lambda(t)$ can be composed of multiple various factors, such as authorship, affiliation status, nationality, language, or the journal.
Following \citet{Wang2013-tj}, the number of citations at time $t$ can then in general be modelled as
\begin{equation}
  c(t) \sim \lambda(t) f(t) C(t)
\end{equation}
with $C(t) = \sum_{\tau = 0}^t c(\tau)$ the cumulative number of citations.
As said, $\lambda(t)$ can be composed of various factors, and we could for instance consider $\lambda(t) = \prod_k \phi_k(t)$ as a product of factors $\phi_k(t)$, which may include factors like author reputation, affiliation reputation, journal reputation, novelty, interdisciplinarity, methodological rigour, data quality, et cetera.
In general, this formulation would be highly degenerate: the overall rate $\lambda$ may be caused by a higher $\phi_1$ or a higher $\phi_2$ and it is not clear how we can properly identify and estimate the effects of these various factors separately.
With additional assumptions, some of these effects may sometimes be estimated.
For example, one can consider differences in citation rates when authors become affiliated with other institutions, as was done by~\citet{Way2019-av}, or one can compare preprints to their journal publications, as was done by~\citet{Traag2021-rq}.

Secondly, citation models provide more clarity about uncertainty.
They clarify that, even for a single paper, the number of citations is not uniquely pre-determined.
That is, for each observed outcome, a different outcome might have been observed, if the entire citation dynamics had been replayed.
Even considering a simple Poisson process, there is quite some variation in the realised citations, and so inferring the latent citation rate based on the observed citations will show quite some uncertainty.
Other factors, such as cumulative advantage process may increase the uncertainty even further.

It is good to explicitly consider uncertainty.
For example, for an early career researcher, we might observe only a few papers and a few citations.
When using empirical means, or other aggregate statistics, we might easily reach overly extreme conclusions when ignoring the uncertainty.
Explicitly considering the uncertainty in such statistics, for example using a Bayesian approach with informed priors, might provide much more reasonable estimates of performance, shrinking the observed number of citations towards more reasonable estimates.
As another example, \citet{Antonoyiannakis2018-cz} argues that smaller journals tend to have more extreme citation averages as a result of the law of large numbers: smaller samples will show more variation.
Explicitly modelling the uncertainty might help provide more reasonable estimates of journal performance.
Similar arguments could be made for estimates of citation impact of research groups, departments, or entire institutions.

Third, when building citation models, we should acknowledge the fundamental problem: \emph{research quality is unobservable}.
This means we cannot simply rely on citation models to draw inferences of research quality or ``academic success''.
However, citation models can help clarify how citations could potentially serve as an indicator for research quality.
Let us develop a preliminary notion of what an indicator is.
We could define an indicator as any variable that is causally affected by the variable for which it serves as an indicator.
So, if $X \rightarrow \ldots \rightarrow Y$, then $Y$ is an indicator for $X$, with the arrows representing a causal effect.
Typically, we do not know $X$ and we therefore use $Y$ to say something about $X$, and it is in this sense that $Y$ is an indicator for $X$.
However, what is typically the case is that some other factor $U$ also affects $Y$.
In this case, $Y$ might still be an indicator for $X$, but if $U$ is not exclusively affected by $X$, we could say that $Y$ is a \emph{biased} indicator for $X$.
After all, we use $Y$ to say something about $X$ in this context, but $Y$ is also affected by $U$, which is not relevant for saying something about $X$.
For example, if we have $Q$ the ``quality'' of an article and $C$ citations, where it is assumed that $Q \rightarrow \ldots \rightarrow C$, then citations are an indicator for quality.
However, if citations $C$ are also influenced by another factor $U$ that is deemed irrelevant, such as author affiliation, then using citations $C$ as an indicator for quality $Q$ would be biased by the affiliation $U$.

\begin{figure*}
  \includegraphics{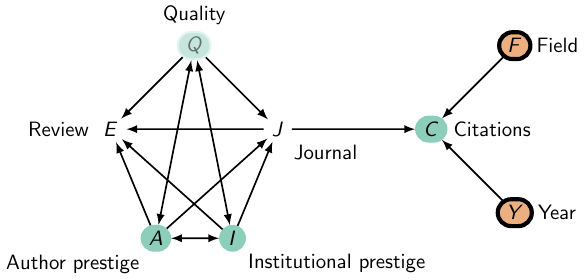}
  \caption{Possible causal model of how citations $C$ can act as an indicator for research quality $Q$.
  Here the field $F$ and the year $Y$ are assumed to be independent of quality $Q$, so that normalising citations $C$ by field $F$ and year $Y$ improves the accuracy of the normalised indicator for quality $Q$.
  }
  \label{fig:indicators}
\end{figure*}

We usually (implicitly) assume that the quality $Q$ is unrelated to some other factors that are related to citations, in particular the field $F$ and year $Y$.
That is, we usually assume that $F \rightarrow C$ and $Y \rightarrow C$, but that $F$ and $Y$ are independent of $Q$ otherwise (Fig.~\ref{fig:indicators}).
Under the assumption that $F$, $Y$ and $Q$ are independent, we can try to make $C$ a more accurate indicator for $Q$ by normalising citations $C$ based on $F$ and $Y$ \citep{Waltman2019-zt}, which amounts to conditioning on $F$ and $Y$.
Citation normalisation then makes sense from this point of view.

Sometimes, normalisation also considers the document type $D$, which implicitly assumes that $D \rightarrow C$ but that the quality is independent of the document type.
However, higher quality work might be more often made available as a research article, instead of for example as an editorial or a letter to the editor.
In that case, if we normalise citations by considering the document type, this pathway of quality $Q \rightarrow D \rightarrow C$ is blocked, and hence, this might actually deteriorate the accuracy of using normalised citations as an indicator for $Q$.

Normalising citations may make $C$ a more accurate indicator of $Q$, and if we could observe $Q$, we could study how the normalisation of citations makes $C$ a better predictor of $Q$.
Directly comparing $C$ to $Q$ is not possible, because of the fundamental problem: research quality $Q$ is unobservable.
Instead of comparing $C$ to $Q$, scholars regularly compare $C$ to another indicator of $Q$, namely peer review $E$, as I discussed in section \ref{sec:comparing_review_metrics}.
Let us assume that $Q \rightarrow \ldots \rightarrow E$, which seems a reasonable starting point.
By comparing $C$ and $E$, we hope to learn something about whether $C$ is an accurate indicator of $Q$.
Again, the fundamental problem in research evaluation is that we cannot observe $Q$, and so any correlation between $C$ and $E$ does not necessarily establish that they are accurate indicators of $Q$.
It merely establishes that $C$ and $E$ are correlated, but this can potentially also be caused by other causal factors.
For example, consider that the journal $J$ influences both citations $C$ and peer review $E$.
We would then observe a correlation between $C$ and $E$, but would be due to the ``confounding influence'' of journal $J$, and have nothing to do with $Q$.
In fact, as we discussed, there is empirical evidence that $C$ and $E$ are not correlated after controlling for the journal $J$.
This implies that if citations $C$ are any indicator for quality $Q$, then only because of the journal $J$.
Indeed, \citet{Waltman2020-fh} suggested that the journal $J$ might be a more accurate indicator of $Q$ than the citations $C$.

Now suppose that author prestige $A$ and institutional prestige $I$ affect acceptance for publication in a journal, so that $A \rightarrow J$ and $I \rightarrow J$, for which there is some evidence, as we saw in section~\ref{sec:peer_review}.
Author and institutional prestige are most likely associated with quality $Q$ (and perhaps mutually reinforce each other).
Most likely, $A$ and $I$ also affect the peer evaluation $E$.
However, they cannot directly affect citations $C$ as well, since we have empirical evidence that entire correlation between $E$ and $C$ is due to the journal $J$.

All in all, this thinking exercise suggests a couple of prestige feedback loops: author prestige, institutional prestige and journal prestige.
These prestige cycles are not independent of the underlying quality of the science, and author prestige, institutional prestige, and journal prestige are all related to quality.
However, they do seem to obfuscate and confound much of the measurement of research quality by citations, such that citation-based indicators may better be seen as indicators of academic prestige than as scientific impact.

In a sense, these prestige feedback loops may be similar to what \citet{ONeil2016-sa} referred to as pernicious cycles.
By not considering the effects of predictions when people act upon predictions, the predictions themselves become ill-informed, and may potentially have serious consequences.
For example, if we use citations to predict institutional scientific performance, scientists may leave certain departments or institutions because of this.
On the face of it, citations may then seem to have some predictive value, but it is exactly \emph{because} citations were used to predict scientific performance that resulted in this behaviour.
If we had correctly considered the potential effect of citations on this behaviour, we would perhaps have concluded that on the contrary, citations do \emph{not} have any predictive value.
This calls attention to clearer considerations of causality \citep{klebel_introduction_2024}.
Whenever an indicator, or a prediction, is used in practice, that is, we \emph{act} on it, we enter causal territory.
Even when an indicator initially might have been valid, through its very use, the consequences of its use may invalidate it.
This is perhaps what could be called a causal understanding of constitutive effects.
By combining citation models with proper causal reasoning and acknowledging the fundamental problem about unobservable research quality, we may hope to make some progress.

\section{Further reading}

There are a great number of books and reviews that cover quantitative science studies.
An older overview of scientometrics is provided by \citet{Hood2001-le}, providing also a history of the origins and various terms related to this field, such as bibliometrics and informetrics.
A useful overview of informetrics is provided by \citet{Bar-Ilan2008-oy}.
\citet{De_Bellis2009-nh} provides a comprehensive overview of the field, and also includes some of the more theoretical frameworks that underpin some of the research.
\citet{Sugimoto2018-yr} cover the essentials of measuring research, and makes for a great introductory read.
The science of science approach was briefly reviewed by \citet{Fortunato2018-ly}, and more recently, was covered in a more accessible form by \citet{Wang2021-aa}.
Some of the literature was also reviewed by \citet{Zeng2017-tg} who took a complex network and complex systems approach.
A related, but different perspective was offered by \citet{Evans2011-gh}.
An overview of some of the literature concerning research evaluation was written by \citet{De_Rijcke2016-kf}.

\bibliography{bibliography}

\begin{thebibliography}{174}%
\makeatletter
\providecommand \@ifxundefined [1]{%
 \@ifx{#1\undefined}
}%
\providecommand \@ifnum [1]{%
 \ifnum #1\expandafter \@firstoftwo
 \else \expandafter \@secondoftwo
 \fi
}%
\providecommand \@ifx [1]{%
 \ifx #1\expandafter \@firstoftwo
 \else \expandafter \@secondoftwo
 \fi
}%
\providecommand \natexlab [1]{#1}%
\providecommand \enquote  [1]{``#1''}%
\providecommand \bibnamefont  [1]{#1}%
\providecommand \bibfnamefont [1]{#1}%
\providecommand \citenamefont [1]{#1}%
\providecommand \href@noop [0]{\@secondoftwo}%
\providecommand \href [0]{\begingroup \@sanitize@url \@href}%
\providecommand \@href[1]{\@@startlink{#1}\@@href}%
\providecommand \@@href[1]{\endgroup#1\@@endlink}%
\providecommand \@sanitize@url [0]{\catcode `\\12\catcode `\$12\catcode `\&12\catcode `\#12\catcode `\^12\catcode `\_12\catcode `\%12\relax}%
\providecommand \@@startlink[1]{}%
\providecommand \@@endlink[0]{}%
\providecommand \url  [0]{\begingroup\@sanitize@url \@url }%
\providecommand \@url [1]{\endgroup\@href {#1}{\urlprefix }}%
\providecommand \urlprefix  [0]{URL }%
\providecommand \Eprint [0]{\href }%
\providecommand \doibase [0]{https://doi.org/}%
\providecommand \selectlanguage [0]{\@gobble}%
\providecommand \bibinfo  [0]{\@secondoftwo}%
\providecommand \bibfield  [0]{\@secondoftwo}%
\providecommand \translation [1]{[#1]}%
\providecommand \BibitemOpen [0]{}%
\providecommand \bibitemStop [0]{}%
\providecommand \bibitemNoStop [0]{.\EOS\space}%
\providecommand \EOS [0]{\spacefactor3000\relax}%
\providecommand \BibitemShut  [1]{\csname bibitem#1\endcsname}%
\let\auto@bib@innerbib\@empty
\bibitem [{\citenamefont {Aagaard}\ and\ \citenamefont {Schneider}(2017)}]{Aagaard2017-fh}%
  \BibitemOpen
  \bibfield  {author} {\bibinfo {author} {\bibnamefont {Aagaard}, \bibfnamefont {K.}}and\ \bibinfo {author} {\bibnamefont {Schneider}, \bibfnamefont {J.~W.}},\ }\bibfield  {title} {\enquote {\bibinfo {title} {Some considerations about causes and effects in studies of performance-based research funding systems},}\ }\href {https://doi.org/10.1016/j.joi.2017.05.018} {\bibfield  {journal} {\bibinfo  {journal} {J. Informetr.}\ }\textbf {\bibinfo {volume} {11}},\ \bibinfo {pages} {923--926} (\bibinfo {year} {2017})}\BibitemShut {NoStop}%
\bibitem [{\citenamefont {Abramo}, \citenamefont {D'Angelo},\ and\ \citenamefont {Di~Costa}(2010)}]{Abramo2010-kd}%
  \BibitemOpen
  \bibfield  {author} {\bibinfo {author} {\bibnamefont {Abramo}, \bibfnamefont {G.}}, \bibinfo {author} {\bibnamefont {D'Angelo}, \bibfnamefont {C.~A.}}, and\ \bibinfo {author} {\bibnamefont {Di~Costa}, \bibfnamefont {F.}},\ }\bibfield  {title} {\enquote {\bibinfo {title} {Citations versus journal impact factor as proxy of quality: could the latter ever be preferable?}}\ }\href {https://doi.org/10.1007/s11192-010-0200-1} {\bibfield  {journal} {\bibinfo  {journal} {Scientometrics}\ }\textbf {\bibinfo {volume} {84}},\ \bibinfo {pages} {821--833} (\bibinfo {year} {2010})}\BibitemShut {NoStop}%
\bibitem [{\citenamefont {Adams}, \citenamefont {Gurney},\ and\ \citenamefont {Jackson}(2008)}]{Adams2008-ma}%
  \BibitemOpen
  \bibfield  {author} {\bibinfo {author} {\bibnamefont {Adams}, \bibfnamefont {J.}}, \bibinfo {author} {\bibnamefont {Gurney}, \bibfnamefont {K.}}, and\ \bibinfo {author} {\bibnamefont {Jackson}, \bibfnamefont {L.}},\ }\bibfield  {title} {\enquote {\bibinfo {title} {Calibrating the zoom - a test of zitt's hypothesis},}\ }\href {https://doi.org/10.1007/s11192-007-1832-7} {\bibfield  {journal} {\bibinfo  {journal} {Scientometrics}\ }\textbf {\bibinfo {volume} {75}},\ \bibinfo {pages} {81--95} (\bibinfo {year} {2008})}\BibitemShut {NoStop}%
\bibitem [{\citenamefont {Aksnes}, \citenamefont {Langfeldt},\ and\ \citenamefont {Wouters}(2019)}]{Aksnes2019-kl}%
  \BibitemOpen
  \bibfield  {author} {\bibinfo {author} {\bibnamefont {Aksnes}, \bibfnamefont {D.~W.}}, \bibinfo {author} {\bibnamefont {Langfeldt}, \bibfnamefont {L.}}, and\ \bibinfo {author} {\bibnamefont {Wouters}, \bibfnamefont {P.}},\ }\bibfield  {title} {\enquote {\bibinfo {title} {Citations, citation indicators, and research quality: An overview of basic concepts and theories:},}\ }\href {https://doi.org/10.1177/2158244019829575} {\bibfield  {journal} {\bibinfo  {journal} {https://doi.org/10.1177/2158244019829575}\ }\textbf {\bibinfo {volume} {9}},\ \bibinfo {pages} {215824401982957} (\bibinfo {year} {2019})}\BibitemShut {NoStop}%
\bibitem [{\citenamefont {Aksnes}\ and\ \citenamefont {Taxt}(2004)}]{Aksnes2004-sf}%
  \BibitemOpen
  \bibfield  {author} {\bibinfo {author} {\bibnamefont {Aksnes}, \bibfnamefont {D.~W.}}and\ \bibinfo {author} {\bibnamefont {Taxt}, \bibfnamefont {R.~E.}},\ }\bibfield  {title} {\enquote {\bibinfo {title} {Peer reviews and bibliometric indicators: a comparative study at a norwegian university},}\ }\href {https://doi.org/10.3152/147154404781776563} {\bibfield  {journal} {\bibinfo  {journal} {Res. Eval.}\ }\textbf {\bibinfo {volume} {13}},\ \bibinfo {pages} {33--41} (\bibinfo {year} {2004})}\BibitemShut {NoStop}%
\bibitem [{\citenamefont {Antonoyiannakis}(2018)}]{Antonoyiannakis2018-cz}%
  \BibitemOpen
  \bibfield  {author} {\bibinfo {author} {\bibnamefont {Antonoyiannakis}, \bibfnamefont {M.}},\ }\bibfield  {title} {\enquote {\bibinfo {title} {Impact factors and the central limit theorem: Why citation averages are scale dependent},}\ }\href {https://doi.org/10.1016/j.joi.2018.08.011} {\bibfield  {journal} {\bibinfo  {journal} {J. Informetr.}\ }\textbf {\bibinfo {volume} {12}},\ \bibinfo {pages} {1072--1088} (\bibinfo {year} {2018})}\BibitemShut {NoStop}%
\bibitem [{\citenamefont {Antonoyiannakis}(2021)}]{Antonoyiannakis2021-od}%
  \BibitemOpen
  \bibfield  {author} {\bibinfo {author} {\bibnamefont {Antonoyiannakis}, \bibfnamefont {M.}},\ }\bibfield  {title} {\enquote {\bibinfo {title} {Does publicity in the science press drive citations? a vindication of peer review},}\ }\href@noop {} {\  (\bibinfo {year} {2021})},\ \Eprint {https://arxiv.org/abs/2105.08118} {arXiv:2105.08118 [cs.DL]} \BibitemShut {NoStop}%
\bibitem [{\citenamefont {Avramescu}(1979)}]{Avramescu1979-sr}%
  \BibitemOpen
  \bibfield  {author} {\bibinfo {author} {\bibnamefont {Avramescu}, \bibfnamefont {A.}},\ }\bibfield  {title} {\enquote {\bibinfo {title} {Actuality and obsolescence of scientific literature},}\ }\href {https://doi.org/10.1002/asi.4630300509} {\bibfield  {journal} {\bibinfo  {journal} {Journal of the American Society for Information Science}\ }\textbf {\bibinfo {volume} {30}},\ \bibinfo {pages} {296--303} (\bibinfo {year} {1979})}\BibitemShut {NoStop}%
\bibitem [{\citenamefont {Baccini}, \citenamefont {De~Nicolao},\ and\ \citenamefont {Petrovich}(2019)}]{Baccini2019-is}%
  \BibitemOpen
  \bibfield  {author} {\bibinfo {author} {\bibnamefont {Baccini}, \bibfnamefont {A.}}, \bibinfo {author} {\bibnamefont {De~Nicolao}, \bibfnamefont {G.}}, and\ \bibinfo {author} {\bibnamefont {Petrovich}, \bibfnamefont {E.}},\ }\bibfield  {title} {\enquote {\bibinfo {title} {Citation gaming induced by bibliometric evaluation: A country-level comparative analysis},}\ }\href {https://doi.org/10.1371/journal.pone.0221212} {\bibfield  {journal} {\bibinfo  {journal} {PLoS One}\ }\textbf {\bibinfo {volume} {14}},\ \bibinfo {pages} {e0221212} (\bibinfo {year} {2019})}\BibitemShut {NoStop}%
\bibitem [{\citenamefont {Bakare}\ and\ \citenamefont {Lewison}(2017)}]{Bakare2017-xe}%
  \BibitemOpen
  \bibfield  {author} {\bibinfo {author} {\bibnamefont {Bakare}, \bibfnamefont {V.}}and\ \bibinfo {author} {\bibnamefont {Lewison}, \bibfnamefont {G.}},\ }\bibfield  {title} {\enquote {\bibinfo {title} {Country over-citation ratios},}\ }\href {https://doi.org/10.1007/s11192-017-2490-z} {\bibfield  {journal} {\bibinfo  {journal} {Scientometrics}\ }\textbf {\bibinfo {volume} {113}},\ \bibinfo {pages} {1199--1207} (\bibinfo {year} {2017})}\BibitemShut {NoStop}%
\bibitem [{\citenamefont {Bar-Ilan}(2008)}]{Bar-Ilan2008-oy}%
  \BibitemOpen
  \bibfield  {author} {\bibinfo {author} {\bibnamefont {Bar-Ilan}, \bibfnamefont {J.}},\ }\bibfield  {title} {\enquote {\bibinfo {title} {Informetrics at the beginning of the 21st century---a review},}\ }\href {https://doi.org/10.1016/j.joi.2007.11.001} {\bibfield  {journal} {\bibinfo  {journal} {J. Informetr.}\ }\textbf {\bibinfo {volume} {2}},\ \bibinfo {pages} {1--52} (\bibinfo {year} {2008})}\BibitemShut {NoStop}%
\bibitem [{\citenamefont {Barab{\'a}si}\ and\ \citenamefont {Albert}(1999)}]{Barabasi1999-dz}%
  \BibitemOpen
  \bibfield  {author} {\bibinfo {author} {\bibnamefont {Barab{\'a}si}, \bibfnamefont {A.-L.}}and\ \bibinfo {author} {\bibnamefont {Albert}, \bibfnamefont {R.}},\ }\bibfield  {title} {\enquote {\bibinfo {title} {Emergence of scaling in random networks},}\ }\href {https://doi.org/10.1126/science.286.5439.509} {\bibfield  {journal} {\bibinfo  {journal} {Science}\ }\textbf {\bibinfo {volume} {286}},\ \bibinfo {pages} {509--512} (\bibinfo {year} {1999})}\BibitemShut {NoStop}%
\bibitem [{\citenamefont {Bertin}\ \emph {et~al.}(2016)\citenamefont {Bertin}, \citenamefont {Atanassova}, \citenamefont {Gingras},\ and\ \citenamefont {Larivi{\`e}re}}]{Bertin2016-mp}%
  \BibitemOpen
  \bibfield  {author} {\bibinfo {author} {\bibnamefont {Bertin}, \bibfnamefont {M.}}, \bibinfo {author} {\bibnamefont {Atanassova}, \bibfnamefont {I.}}, \bibinfo {author} {\bibnamefont {Gingras}, \bibfnamefont {Y.}}, and\ \bibinfo {author} {\bibnamefont {Larivi{\`e}re}, \bibfnamefont {V.}},\ }\bibfield  {title} {\enquote {\bibinfo {title} {The invariant distribution of references in scientific articles},}\ }\href {https://doi.org/10.1002/asi.23367} {\bibfield  {journal} {\bibinfo  {journal} {J. Assoc. Inf. Sci. Technol.}\ }\textbf {\bibinfo {volume} {67}},\ \bibinfo {pages} {164--177} (\bibinfo {year} {2016})}\BibitemShut {NoStop}%
\bibitem [{\citenamefont {Bertocchi}\ \emph {et~al.}(2015)\citenamefont {Bertocchi}, \citenamefont {Gambardella}, \citenamefont {Jappelli}, \citenamefont {Nappi},\ and\ \citenamefont {Peracchi}}]{Bertocchi2015-cs}%
  \BibitemOpen
  \bibfield  {author} {\bibinfo {author} {\bibnamefont {Bertocchi}, \bibfnamefont {G.}}, \bibinfo {author} {\bibnamefont {Gambardella}, \bibfnamefont {A.}}, \bibinfo {author} {\bibnamefont {Jappelli}, \bibfnamefont {T.}}, \bibinfo {author} {\bibnamefont {Nappi}, \bibfnamefont {C.~A.}}, and\ \bibinfo {author} {\bibnamefont {Peracchi}, \bibfnamefont {F.}},\ }\bibfield  {title} {\enquote {\bibinfo {title} {Bibliometric evaluation vs. informed peer review: Evidence from italy},}\ }\href {https://doi.org/10.1016/j.respol.2014.08.004} {\bibfield  {journal} {\bibinfo  {journal} {Res. Policy}\ }\textbf {\bibinfo {volume} {44}},\ \bibinfo {pages} {451--466} (\bibinfo {year} {2015})}\BibitemShut {NoStop}%
\bibitem [{\citenamefont {Van~den Besselaar}, \citenamefont {Heyman},\ and\ \citenamefont {Sandstr{\"o}m}(2017)}]{Van_den_Besselaar2017-yr}%
  \BibitemOpen
  \bibfield  {author} {\bibinfo {author} {\bibnamefont {Van~den Besselaar}, \bibfnamefont {P.}}, \bibinfo {author} {\bibnamefont {Heyman}, \bibfnamefont {U.}}, and\ \bibinfo {author} {\bibnamefont {Sandstr{\"o}m}, \bibfnamefont {U.}},\ }\bibfield  {title} {\enquote {\bibinfo {title} {Perverse effects of output-based research funding? butler's australian case revisited},}\ }\href {https://doi.org/10.1016/j.joi.2017.05.016} {\bibfield  {journal} {\bibinfo  {journal} {J. Informetr.}\ }\textbf {\bibinfo {volume} {11}},\ \bibinfo {pages} {905--918} (\bibinfo {year} {2017})}\BibitemShut {NoStop}%
\bibitem [{\citenamefont {Van~den Besselaar}\ and\ \citenamefont {Sandstr{\"o}m}(2015)}]{Van_den_Besselaar2015-qp}%
  \BibitemOpen
  \bibfield  {author} {\bibinfo {author} {\bibnamefont {Van~den Besselaar}, \bibfnamefont {P.}}and\ \bibinfo {author} {\bibnamefont {Sandstr{\"o}m}, \bibfnamefont {U.}},\ }\bibfield  {title} {\enquote {\bibinfo {title} {Early career grants, performance, and careers: A study on predictive validity of grant decisions},}\ }\href {https://doi.org/10.1016/j.joi.2015.07.011} {\bibfield  {journal} {\bibinfo  {journal} {J. Informetr.}\ }\textbf {\bibinfo {volume} {9}},\ \bibinfo {pages} {826--838} (\bibinfo {year} {2015})}\BibitemShut {NoStop}%
\bibitem [{\citenamefont {Bhattacharya}\ and\ \citenamefont {Packalen}(2020)}]{Bhattacharya2020-ya}%
  \BibitemOpen
  \bibfield  {author} {\bibinfo {author} {\bibnamefont {Bhattacharya}, \bibfnamefont {J.}}and\ \bibinfo {author} {\bibnamefont {Packalen}, \bibfnamefont {M.}},\ }\href {https://doi.org/10.3386/w26752} {\enquote {\bibinfo {title} {Stagnation and scientific incentives},}\ }\bibinfo {type} {Tech. Rep.}\ \bibinfo {number} {26752}\ (\bibinfo  {institution} {National Bureau of Economic Research},\ \bibinfo {address} {Cambridge, MA},\ \bibinfo {year} {2020})\BibitemShut {NoStop}%
\bibitem [{\citenamefont {Bloch}\ and\ \citenamefont {Schneider}(2016)}]{Bloch2016-dy}%
  \BibitemOpen
  \bibfield  {author} {\bibinfo {author} {\bibnamefont {Bloch}, \bibfnamefont {C.}}and\ \bibinfo {author} {\bibnamefont {Schneider}, \bibfnamefont {J.~W.}},\ }\bibfield  {title} {\enquote {\bibinfo {title} {Performance-based funding models and researcher behavior: An analysis of the influence of the norwegian publication indicator at the individual level},}\ }\href {https://doi.org/10.1093/reseval/rvv047} {\bibfield  {journal} {\bibinfo  {journal} {Res. Eval.}\ }\textbf {\bibinfo {volume} {25}},\ \bibinfo {pages} {371--382} (\bibinfo {year} {2016})}\BibitemShut {NoStop}%
\bibitem [{\citenamefont {Bookstein}\ and\ \citenamefont {Yitzhaki}(1999)}]{Bookstein1999-rb}%
  \BibitemOpen
  \bibfield  {author} {\bibinfo {author} {\bibnamefont {Bookstein}, \bibfnamefont {A.}}and\ \bibinfo {author} {\bibnamefont {Yitzhaki}, \bibfnamefont {M.}},\ }\bibfield  {title} {\enquote {\bibinfo {title} {Own-language preference: A new measure of ``relative language self-citation''},}\ }\href {https://doi.org/10.1007/BF02464782} {\bibfield  {journal} {\bibinfo  {journal} {Scientometrics}\ }\textbf {\bibinfo {volume} {46}},\ \bibinfo {pages} {337--348} (\bibinfo {year} {1999})}\BibitemShut {NoStop}%
\bibitem [{\citenamefont {Bornemann-Cimenti}, \citenamefont {Szilagyi},\ and\ \citenamefont {Sandner-Kiesling}(2016)}]{Bornemann-Cimenti2016-nf}%
  \BibitemOpen
  \bibfield  {author} {\bibinfo {author} {\bibnamefont {Bornemann-Cimenti}, \bibfnamefont {H.}}, \bibinfo {author} {\bibnamefont {Szilagyi}, \bibfnamefont {I.~S.}}, and\ \bibinfo {author} {\bibnamefont {Sandner-Kiesling}, \bibfnamefont {A.}},\ }\bibfield  {title} {\enquote {\bibinfo {title} {Perpetuation of retracted publications using the example of the scott s. reuben case: Incidences, reasons and possible improvements},}\ }\href {https://doi.org/10.1007/s11948-015-9680-y} {\bibfield  {journal} {\bibinfo  {journal} {Sci. Eng. Ethics}\ }\textbf {\bibinfo {volume} {22}},\ \bibinfo {pages} {1063--1072} (\bibinfo {year} {2016})}\BibitemShut {NoStop}%
\bibitem [{\citenamefont {Bornmann}(2011)}]{Bornmann2011-eg}%
  \BibitemOpen
  \bibfield  {author} {\bibinfo {author} {\bibnamefont {Bornmann}, \bibfnamefont {L.}},\ }\bibfield  {title} {\enquote {\bibinfo {title} {Scientific peer review},}\ }\href@noop {} {\bibfield  {journal} {\bibinfo  {journal} {Annual Rev. Info. Sci \& Technol.}\ }\textbf {\bibinfo {volume} {45}},\ \bibinfo {pages} {197--245} (\bibinfo {year} {2011})}\BibitemShut {NoStop}%
\bibitem [{\citenamefont {Bornmann}\ and\ \citenamefont {Daniel}(2008{\natexlab{a}})}]{Bornmann2008-ah}%
  \BibitemOpen
  \bibfield  {author} {\bibinfo {author} {\bibnamefont {Bornmann}, \bibfnamefont {L.}}and\ \bibinfo {author} {\bibnamefont {Daniel}, \bibfnamefont {H.}},\ }\bibfield  {title} {\enquote {\bibinfo {title} {What do citation counts measure? a review of studies on citing behavior},}\ }\href {https://doi.org/10.1108/00220410810844150} {\bibfield  {journal} {\bibinfo  {journal} {Journal of Documentation}\ }\textbf {\bibinfo {volume} {64}},\ \bibinfo {pages} {45--80} (\bibinfo {year} {2008}{\natexlab{a}})}\BibitemShut {NoStop}%
\bibitem [{\citenamefont {Bornmann}\ and\ \citenamefont {Daniel}(2008{\natexlab{b}})}]{Bornmann2008-kk}%
  \BibitemOpen
  \bibfield  {author} {\bibinfo {author} {\bibnamefont {Bornmann}, \bibfnamefont {L.}}and\ \bibinfo {author} {\bibnamefont {Daniel}, \bibfnamefont {H.-D.}},\ }\bibfield  {title} {\enquote {\bibinfo {title} {Selecting manuscripts for a high-impact journal through peer review: A citation analysis of communications that were accepted by angewandte chemie international edition, or rejected but published elsewhere},}\ }\href {https://doi.org/10.1002/asi.20901} {\bibfield  {journal} {\bibinfo  {journal} {J. Am. Soc. Inf. Sci. Technol.}\ }\textbf {\bibinfo {volume} {59}},\ \bibinfo {pages} {1841--1852} (\bibinfo {year} {2008}{\natexlab{b}})}\BibitemShut {NoStop}%
\bibitem [{\citenamefont {Bornmann}\ and\ \citenamefont {Leydesdorff}(2013)}]{Bornmann2013-tj}%
  \BibitemOpen
  \bibfield  {author} {\bibinfo {author} {\bibnamefont {Bornmann}, \bibfnamefont {L.}}and\ \bibinfo {author} {\bibnamefont {Leydesdorff}, \bibfnamefont {L.}},\ }\bibfield  {title} {\enquote {\bibinfo {title} {The validation of (advanced) bibliometric indicators through peer assessments: A comparative study using data from {InCites} and {F1000}},}\ }\href {https://doi.org/10.1016/j.joi.2012.12.003} {\bibfield  {journal} {\bibinfo  {journal} {J. Informetr.}\ }\textbf {\bibinfo {volume} {7}},\ \bibinfo {pages} {286--291} (\bibinfo {year} {2013})}\BibitemShut {NoStop}%
\bibitem [{\citenamefont {Brookes}(1970)}]{Brookes1970-lk}%
  \BibitemOpen
  \bibfield  {author} {\bibinfo {author} {\bibnamefont {Brookes}, \bibfnamefont {B.~C.}},\ }\bibfield  {title} {\enquote {\bibinfo {title} {The growth, utility, and obsolescence of scientific periodical literature},}\ }\href {https://doi.org/10.1108/eb026500} {\bibfield  {journal} {\bibinfo  {journal} {Journal of Documentation}\ }\textbf {\bibinfo {volume} {26}},\ \bibinfo {pages} {283--294} (\bibinfo {year} {1970})}\BibitemShut {NoStop}%
\bibitem [{\citenamefont {Burrel}(2001)}]{Burrel2001-de}%
  \BibitemOpen
  \bibfield  {author} {\bibinfo {author} {\bibnamefont {Burrel}, \bibfnamefont {Q.~L.}},\ }\bibfield  {title} {\enquote {\bibinfo {title} {Stochastic modelling of the first-citation distribution},}\ }\href {https://doi.org/10.1023/A:1012751509975} {\bibfield  {journal} {\bibinfo  {journal} {Scientometrics}\ }\textbf {\bibinfo {volume} {52}},\ \bibinfo {pages} {3--12} (\bibinfo {year} {2001})}\BibitemShut {NoStop}%
\bibitem [{\citenamefont {Burrell}(2002)}]{Burrell2002-cu}%
  \BibitemOpen
  \bibfield  {author} {\bibinfo {author} {\bibnamefont {Burrell}, \bibfnamefont {Q.~L.}},\ }\bibfield  {title} {\enquote {\bibinfo {title} {The nth-citation distribution and obsolescence},}\ }\href {https://doi.org/10.1023/A:1014816911511} {\bibfield  {journal} {\bibinfo  {journal} {Scientometrics}\ }\textbf {\bibinfo {volume} {53}},\ \bibinfo {pages} {309--323} (\bibinfo {year} {2002})}\BibitemShut {NoStop}%
\bibitem [{\citenamefont {Burton}\ and\ \citenamefont {Kebler}(1960)}]{Burton1960-xm}%
  \BibitemOpen
  \bibfield  {author} {\bibinfo {author} {\bibnamefont {Burton}, \bibfnamefont {R.~E.}}and\ \bibinfo {author} {\bibnamefont {Kebler}, \bibfnamefont {R.~W.}},\ }\bibfield  {title} {\enquote {\bibinfo {title} {The ``half-life'' of some scientific and technical literatures},}\ }\href {https://doi.org/10.1002/asi.5090110105} {\bibfield  {journal} {\bibinfo  {journal} {Am. doc.}\ }\textbf {\bibinfo {volume} {11}},\ \bibinfo {pages} {18--22} (\bibinfo {year} {1960})}\BibitemShut {NoStop}%
\bibitem [{\citenamefont {Butler}(2003)}]{Butler2003-em}%
  \BibitemOpen
  \bibfield  {author} {\bibinfo {author} {\bibnamefont {Butler}, \bibfnamefont {L.}},\ }\bibfield  {title} {\enquote {\bibinfo {title} {Explaining australia's increased share of {ISI} publications---the effects of a funding formula based on publication counts},}\ }\href {https://doi.org/10.1016/S0048-7333(02)00007-0} {\bibfield  {journal} {\bibinfo  {journal} {Res. Policy}\ }\textbf {\bibinfo {volume} {32}},\ \bibinfo {pages} {143--155} (\bibinfo {year} {2003})}\BibitemShut {NoStop}%
\bibitem [{\citenamefont {Callaham}(2002)}]{Callaham2002-po}%
  \BibitemOpen
  \bibfield  {author} {\bibinfo {author} {\bibnamefont {Callaham}, \bibfnamefont {M.}},\ }\bibfield  {title} {\enquote {\bibinfo {title} {Journal prestige, publication bias, and other characteristics associated with citation of published studies in {Peer-Reviewed} journals},}\ }\href {https://doi.org/10.1001/jama.287.21.2847} {\bibfield  {journal} {\bibinfo  {journal} {JAMA}\ }\textbf {\bibinfo {volume} {287}},\ \bibinfo {pages} {2847} (\bibinfo {year} {2002})}\BibitemShut {NoStop}%
\bibitem [{\citenamefont {Caron}\ and\ \citenamefont {Van~Eck}(2014)}]{Caron2014-ai}%
  \BibitemOpen
  \bibfield  {author} {\bibinfo {author} {\bibnamefont {Caron}, \bibfnamefont {E.}}and\ \bibinfo {author} {\bibnamefont {Van~Eck}, \bibfnamefont {N.~J.}},\ }\bibfield  {title} {\enquote {\bibinfo {title} {Large scale author name disambiguation using rule-based scoring and clustering},}\ }in\ \href@noop {} {\emph {\bibinfo {booktitle} {Proceedings of the {STI}}}}\ (\bibinfo {year} {2014})\ pp.\ \bibinfo {pages} {79--86}\BibitemShut {NoStop}%
\bibitem [{\citenamefont {Casadevall}\ and\ \citenamefont {Fang}(2014)}]{Casadevall2014-nx}%
  \BibitemOpen
  \bibfield  {author} {\bibinfo {author} {\bibnamefont {Casadevall}, \bibfnamefont {A.}}and\ \bibinfo {author} {\bibnamefont {Fang}, \bibfnamefont {F.~C.}},\ }\bibfield  {title} {\enquote {\bibinfo {title} {Causes for the persistence of impact factor mania},}\ }\href {https://doi.org/10.1128/mBio.00064-14} {\bibfield  {journal} {\bibinfo  {journal} {MBio}\ }\textbf {\bibinfo {volume} {5}},\ \bibinfo {pages} {e00064--14} (\bibinfo {year} {2014})}\BibitemShut {NoStop}%
\bibitem [{\citenamefont {Chatterjee}, \citenamefont {Ghosh},\ and\ \citenamefont {Chakrabarti}(2016)}]{Chatterjee2016-ih}%
  \BibitemOpen
  \bibfield  {author} {\bibinfo {author} {\bibnamefont {Chatterjee}, \bibfnamefont {A.}}, \bibinfo {author} {\bibnamefont {Ghosh}, \bibfnamefont {A.}}, and\ \bibinfo {author} {\bibnamefont {Chakrabarti}, \bibfnamefont {B.~K.}},\ }\bibfield  {title} {\enquote {\bibinfo {title} {Universality of citation distributions for academic institutions and journals},}\ }\href {https://doi.org/10.1371/journal.pone.0146762} {\bibfield  {journal} {\bibinfo  {journal} {PLoS One}\ }\textbf {\bibinfo {volume} {11}},\ \bibinfo {pages} {e0146762} (\bibinfo {year} {2016})}\BibitemShut {NoStop}%
\bibitem [{\citenamefont {Cole}(1970)}]{Cole1970-nl}%
  \BibitemOpen
  \bibfield  {author} {\bibinfo {author} {\bibnamefont {Cole}, \bibfnamefont {S.}},\ }\bibfield  {title} {\enquote {\bibinfo {title} {Professional standing and the reception of scientific discoveries},}\ }\href {https://doi.org/10.1086/224934} {\bibfield  {journal} {\bibinfo  {journal} {Am. J. Sociol.}\ }\textbf {\bibinfo {volume} {76}},\ \bibinfo {pages} {286--306} (\bibinfo {year} {1970})}\BibitemShut {NoStop}%
\bibitem [{\citenamefont {Cole}\ and\ \citenamefont {Cole}(1968)}]{Cole1968-hs}%
  \BibitemOpen
  \bibfield  {author} {\bibinfo {author} {\bibnamefont {Cole}, \bibfnamefont {S.}}and\ \bibinfo {author} {\bibnamefont {Cole}, \bibfnamefont {J.~R.}},\ }\bibfield  {title} {\enquote {\bibinfo {title} {Visibility and the structural bases of awareness of scientific research},}\ }\href {https://doi.org/10.2307/2091914} {\bibfield  {journal} {\bibinfo  {journal} {Am. Sociol. Rev.}\ }\textbf {\bibinfo {volume} {33}},\ \bibinfo {pages} {397--413} (\bibinfo {year} {1968})}\BibitemShut {NoStop}%
\bibitem [{\citenamefont {Cole}, \citenamefont {Cole},\ and\ \citenamefont {Simon}(1981)}]{Cole1981-dt}%
  \BibitemOpen
  \bibfield  {author} {\bibinfo {author} {\bibnamefont {Cole}, \bibfnamefont {S.}}, \bibinfo {author} {\bibnamefont {Cole}, \bibfnamefont {J.~R.}}, and\ \bibinfo {author} {\bibnamefont {Simon}, \bibfnamefont {G.~A.}},\ }\bibfield  {title} {\enquote {\bibinfo {title} {Chance and consensus in peer review},}\ }\href {https://doi.org/10.1126/science.7302566} {\bibfield  {journal} {\bibinfo  {journal} {Science}\ }\textbf {\bibinfo {volume} {214}},\ \bibinfo {pages} {881--886} (\bibinfo {year} {1981})}\BibitemShut {NoStop}%
\bibitem [{\citenamefont {Dahler-Larsen}(2014)}]{Dahler-Larsen2014-kl}%
  \BibitemOpen
  \bibfield  {author} {\bibinfo {author} {\bibnamefont {Dahler-Larsen}, \bibfnamefont {P.}},\ }\bibfield  {title} {\enquote {\bibinfo {title} {Constitutive effects of performance indicators: Getting beyond unintended consequences},}\ }\href {https://doi.org/10.1080/14719037.2013.770058} {\bibfield  {journal} {\bibinfo  {journal} {Public Management Review}\ }\textbf {\bibinfo {volume} {16}},\ \bibinfo {pages} {969--986} (\bibinfo {year} {2014})}\BibitemShut {NoStop}%
\bibitem [{\citenamefont {De~Bellis}(2009)}]{De_Bellis2009-nh}%
  \BibitemOpen
  \bibfield  {author} {\bibinfo {author} {\bibnamefont {De~Bellis}, \bibfnamefont {N.}},\ }\href@noop {} {\emph {\bibinfo {title} {Bibliometrics and Citation Analysis: From the Science Citation Index to Cybermetrics}}}\ (\bibinfo  {publisher} {Scarecrow Press},\ \bibinfo {year} {2009})\BibitemShut {NoStop}%
\bibitem [{\citenamefont {{DORA}}(2013)}]{Dora2013-kt}%
  \BibitemOpen
  \bibfield  {author} {\bibinfo {author} {\bibnamefont {{DORA}},},\ }\href@noop {} {\enquote {\bibinfo {title} {San francisco declaration on research assessment ({DORA})},}\ }\bibinfo {howpublished} {\url{https://sfdora.org/}} (\bibinfo {year} {2013}),\ \bibinfo {note} {accessed: 2013-NA-NA}\BibitemShut {NoStop}%
\bibitem [{\citenamefont {Dorogovtsev}\ and\ \citenamefont {Mendes}(2000)}]{Dorogovtsev2000-jd}%
  \BibitemOpen
  \bibfield  {author} {\bibinfo {author} {\bibnamefont {Dorogovtsev}, \bibfnamefont {S.~N.}}and\ \bibinfo {author} {\bibnamefont {Mendes}, \bibfnamefont {J.~F.}},\ }\bibfield  {title} {\enquote {\bibinfo {title} {Evolution of networks with aging of sites},}\ }\href {https://doi.org/10.1103/physreve.62.1842} {\bibfield  {journal} {\bibinfo  {journal} {Phys. Rev. E Stat. Phys. Plasmas Fluids Relat. Interdiscip. Topics}\ }\textbf {\bibinfo {volume} {62}},\ \bibinfo {pages} {1842--1845} (\bibinfo {year} {2000})}\BibitemShut {NoStop}%
\bibitem [{\citenamefont {Egghe}(1994)}]{Egghe1994-rm}%
  \BibitemOpen
  \bibfield  {author} {\bibinfo {author} {\bibnamefont {Egghe}, \bibfnamefont {L.}},\ }\bibfield  {title} {\enquote {\bibinfo {title} {A theory of continuous rates and applications to the theory of growth and obsolescence rates},}\ }\href {https://doi.org/10.1016/0306-4573(94)90070-1} {\bibfield  {journal} {\bibinfo  {journal} {Inf. Process. Manag.}\ }\textbf {\bibinfo {volume} {30}},\ \bibinfo {pages} {279--292} (\bibinfo {year} {1994})}\BibitemShut {NoStop}%
\bibitem [{\citenamefont {Egghe}(2005)}]{Egghe2005-pa}%
  \BibitemOpen
  \bibfield  {author} {\bibinfo {author} {\bibnamefont {Egghe}, \bibfnamefont {L.}},\ }\href@noop {} {\emph {\bibinfo {title} {Power Laws in the Information Production Process: Lotkaian Informetrics}}}\ (\bibinfo  {publisher} {Elsevier Academic Press},\ \bibinfo {address} {Amsterdam},\ \bibinfo {year} {2005})\ p.\ \bibinfo {pages} {427}\BibitemShut {NoStop}%
\bibitem [{\citenamefont {Egghe}(2010)}]{Egghe2010-cx}%
  \BibitemOpen
  \bibfield  {author} {\bibinfo {author} {\bibnamefont {Egghe}, \bibfnamefont {L.}},\ }\bibfield  {title} {\enquote {\bibinfo {title} {A model showing the increase in time of the average and median reference age and the decrease in time of the price index},}\ }\href {https://doi.org/10.1007/s11192-009-0057-3} {\bibfield  {journal} {\bibinfo  {journal} {Scientometrics}\ }\textbf {\bibinfo {volume} {82}},\ \bibinfo {pages} {243--248} (\bibinfo {year} {2010})}\BibitemShut {NoStop}%
\bibitem [{\citenamefont {Egghe}\ and\ \citenamefont {Ravichandra~rao}(1992)}]{Egghe1992-uo}%
  \BibitemOpen
  \bibfield  {author} {\bibinfo {author} {\bibnamefont {Egghe}, \bibfnamefont {L.}}and\ \bibinfo {author} {\bibnamefont {Ravichandra~rao}, \bibfnamefont {I.~K.}},\ }\bibfield  {title} {\enquote {\bibinfo {title} {Citation age data and the obsolescence function: Fits and explanations},}\ }\href {https://doi.org/10.1016/0306-4573(92)90046-3} {\bibfield  {journal} {\bibinfo  {journal} {Inf. Process. Manag.}\ }\textbf {\bibinfo {volume} {28}},\ \bibinfo {pages} {201--217} (\bibinfo {year} {1992})}\BibitemShut {NoStop}%
\bibitem [{\citenamefont {Egghe}\ and\ \citenamefont {Rousseau}(2000)}]{Egghe2000-td}%
  \BibitemOpen
  \bibfield  {author} {\bibinfo {author} {\bibnamefont {Egghe}, \bibfnamefont {L.}}and\ \bibinfo {author} {\bibnamefont {Rousseau}, \bibfnamefont {R.}},\ }\href {https://doi.org/10.1002/1097-4571(2000)9999:9999<::aid-asi1003>3.0.co;2-8} {\enquote {\bibinfo {title} {Aging, obsolescence, impact, growth, and utilization: Definitions and relations},}\ } (\bibinfo {year} {2000})\BibitemShut {NoStop}%
\bibitem [{\citenamefont {Eisen}\ \emph {et~al.}(2013)\citenamefont {Eisen}, \citenamefont {MacCallum}, \citenamefont {Neylon}, \citenamefont {Sugimoto},\ and\ \citenamefont {Walport}}]{Eisen2013-yw}%
  \BibitemOpen
  \bibfield  {author} {\bibinfo {author} {\bibnamefont {Eisen}, \bibfnamefont {J.~A.}}, \bibinfo {author} {\bibnamefont {MacCallum}, \bibfnamefont {C.~J.}}, \bibinfo {author} {\bibnamefont {Neylon}, \bibfnamefont {C.}}, \bibinfo {author} {\bibnamefont {Sugimoto}, \bibfnamefont {C.~R.}}, and\ \bibinfo {author} {\bibnamefont {Walport}, \bibfnamefont {M.}},\ }\bibfield  {title} {\enquote {\bibinfo {title} {Expert failure: Re-evaluating research assessment},}\ }\href {https://doi.org/10.1371/journal.pbio.1001677} {\bibfield  {journal} {\bibinfo  {journal} {PLoS Biol.}\ }\textbf {\bibinfo {volume} {11}},\ \bibinfo {pages} {e1001677} (\bibinfo {year} {2013})}\BibitemShut {NoStop}%
\bibitem [{\citenamefont {Ellis}\ and\ \citenamefont {Durden}(1991)}]{Ellis1991-hc}%
  \BibitemOpen
  \bibfield  {author} {\bibinfo {author} {\bibnamefont {Ellis}, \bibfnamefont {L.~V.}}and\ \bibinfo {author} {\bibnamefont {Durden}, \bibfnamefont {G.~C.}},\ }\bibfield  {title} {\enquote {\bibinfo {title} {Why economists rank their journals the way they do},}\ }\href {https://doi.org/10.1016/0148-6195(91)90024-Q} {\bibfield  {journal} {\bibinfo  {journal} {J. Econ. Bus.}\ }\textbf {\bibinfo {volume} {43}},\ \bibinfo {pages} {265--270} (\bibinfo {year} {1991})}\BibitemShut {NoStop}%
\bibitem [{\citenamefont {Eom}\ and\ \citenamefont {Fortunato}(2011)}]{Eom2011-jy}%
  \BibitemOpen
  \bibfield  {author} {\bibinfo {author} {\bibnamefont {Eom}, \bibfnamefont {Y.-H.}}and\ \bibinfo {author} {\bibnamefont {Fortunato}, \bibfnamefont {S.}},\ }\bibfield  {title} {\enquote {\bibinfo {title} {Characterizing and modeling citation dynamics},}\ }\href {https://doi.org/10.1371/journal.pone.0024926} {\bibfield  {journal} {\bibinfo  {journal} {PLoS One}\ }\textbf {\bibinfo {volume} {6}},\ \bibinfo {pages} {e24926} (\bibinfo {year} {2011})}\BibitemShut {NoStop}%
\bibitem [{\citenamefont {Ernst}, \citenamefont {Saradeth},\ and\ \citenamefont {Resch}(1993)}]{Ernst1993-jo}%
  \BibitemOpen
  \bibfield  {author} {\bibinfo {author} {\bibnamefont {Ernst}, \bibfnamefont {E.}}, \bibinfo {author} {\bibnamefont {Saradeth}, \bibfnamefont {T.}}, and\ \bibinfo {author} {\bibnamefont {Resch}, \bibfnamefont {K.~L.}},\ }\bibfield  {title} {\enquote {\bibinfo {title} {Drawbacks of peer review},}\ }\href {https://doi.org/10.1038/363296a0} {\bibfield  {journal} {\bibinfo  {journal} {Nature}\ }\textbf {\bibinfo {volume} {363}},\ \bibinfo {pages} {296} (\bibinfo {year} {1993})}\BibitemShut {NoStop}%
\bibitem [{\citenamefont {Evans}(2008)}]{Evans2008-wf}%
  \BibitemOpen
  \bibfield  {author} {\bibinfo {author} {\bibnamefont {Evans}, \bibfnamefont {J.~a.}},\ }\bibfield  {title} {\enquote {\bibinfo {title} {Electronic publication and the narrowing of science and scholarship},}\ }\href {https://doi.org/10.1126/science.1150473} {\bibfield  {journal} {\bibinfo  {journal} {Science}\ }\textbf {\bibinfo {volume} {321}},\ \bibinfo {pages} {395--399} (\bibinfo {year} {2008})}\BibitemShut {NoStop}%
\bibitem [{\citenamefont {Evans}\ and\ \citenamefont {Foster}(2011)}]{Evans2011-gh}%
  \BibitemOpen
  \bibfield  {author} {\bibinfo {author} {\bibnamefont {Evans}, \bibfnamefont {J.~A.}}and\ \bibinfo {author} {\bibnamefont {Foster}, \bibfnamefont {J.~G.}},\ }\bibfield  {title} {\enquote {\bibinfo {title} {Metaknowledge},}\ }\href {https://doi.org/10.1126/science.1201765} {\bibfield  {journal} {\bibinfo  {journal} {Science}\ }\textbf {\bibinfo {volume} {331}},\ \bibinfo {pages} {721--725} (\bibinfo {year} {2011})}\BibitemShut {NoStop}%
\bibitem [{\citenamefont {Eyre-Walker}\ and\ \citenamefont {Stoletzki}(2013)}]{Eyre-Walker2013-ud}%
  \BibitemOpen
  \bibfield  {author} {\bibinfo {author} {\bibnamefont {Eyre-Walker}, \bibfnamefont {A.}}and\ \bibinfo {author} {\bibnamefont {Stoletzki}, \bibfnamefont {N.}},\ }\bibfield  {title} {\enquote {\bibinfo {title} {The assessment of science: The relative merits of {Post-Publication} review, the impact factor, and the number of citations},}\ }\href {https://doi.org/10.1371/journal.pbio.1001675} {\bibfield  {journal} {\bibinfo  {journal} {PLoS Biol.}\ }\textbf {\bibinfo {volume} {11}},\ \bibinfo {pages} {e1001675} (\bibinfo {year} {2013})}\BibitemShut {NoStop}%
\bibitem [{\citenamefont {Forscher}\ \emph {et~al.}(2019)\citenamefont {Forscher}, \citenamefont {Brauer}, \citenamefont {Azevedo}, \citenamefont {Cox},\ and\ \citenamefont {Devine}}]{Forscher2019-iy}%
  \BibitemOpen
  \bibfield  {author} {\bibinfo {author} {\bibnamefont {Forscher}, \bibfnamefont {P.~S.}}, \bibinfo {author} {\bibnamefont {Brauer}, \bibfnamefont {M.}}, \bibinfo {author} {\bibnamefont {Azevedo}, \bibfnamefont {F.}}, \bibinfo {author} {\bibnamefont {Cox}, \bibfnamefont {W.~T.~L.}}, and\ \bibinfo {author} {\bibnamefont {Devine}, \bibfnamefont {P.~G.}},\ }\href {https://doi.org/10.31234/osf.io/483zj} {\enquote {\bibinfo {title} {How many reviewers are required to obtain reliable evaluations of {NIH} {R01} grant proposals?}}\ } (\bibinfo {year} {2019})\BibitemShut {NoStop}%
\bibitem [{\citenamefont {Fortunato}\ \emph {et~al.}(2018)\citenamefont {Fortunato}, \citenamefont {Bergstrom}, \citenamefont {B{\"o}rner}, \citenamefont {Evans}, \citenamefont {Helbing}, \citenamefont {Milojevi{\'c}}, \citenamefont {Petersen}, \citenamefont {Radicchi}, \citenamefont {Sinatra}, \citenamefont {Uzzi}, \citenamefont {Vespignani}, \citenamefont {Waltman}, \citenamefont {Wang},\ and\ \citenamefont {Barab{\'a}si}}]{Fortunato2018-ly}%
  \BibitemOpen
  \bibfield  {author} {\bibinfo {author} {\bibnamefont {Fortunato}, \bibfnamefont {S.}}, \bibinfo {author} {\bibnamefont {Bergstrom}, \bibfnamefont {C.~T.}}, \bibinfo {author} {\bibnamefont {B{\"o}rner}, \bibfnamefont {K.}}, \bibinfo {author} {\bibnamefont {Evans}, \bibfnamefont {J.~A.}}, \bibinfo {author} {\bibnamefont {Helbing}, \bibfnamefont {D.}}, \bibinfo {author} {\bibnamefont {Milojevi{\'c}}, \bibfnamefont {S.}}, \bibinfo {author} {\bibnamefont {Petersen}, \bibfnamefont {A.~M.}}, \bibinfo {author} {\bibnamefont {Radicchi}, \bibfnamefont {F.}}, \bibinfo {author} {\bibnamefont {Sinatra}, \bibfnamefont {R.}}, \bibinfo {author} {\bibnamefont {Uzzi}, \bibfnamefont {B.}}, \bibinfo {author} {\bibnamefont {Vespignani}, \bibfnamefont {A.}}, \bibinfo {author} {\bibnamefont {Waltman}, \bibfnamefont {L.}}, \bibinfo {author} {\bibnamefont {Wang}, \bibfnamefont {D.}}, and\ \bibinfo {author} {\bibnamefont {Barab{\'a}si}, \bibfnamefont {A.-L.}},\ }\bibfield  {title} {\enquote {\bibinfo {title} {Science of science},}\ }\href {https://doi.org/10.1126/science.aao0185} {\bibfield  {journal} {\bibinfo  {journal} {Science}\ }\textbf {\bibinfo {volume} {359}} (\bibinfo {year} {2018}),\ 10.1126/science.aao0185}\BibitemShut {NoStop}%
\bibitem [{\citenamefont {Furman}, \citenamefont {Jensen},\ and\ \citenamefont {Murray}(2012)}]{Furman2012-uq}%
  \BibitemOpen
  \bibfield  {author} {\bibinfo {author} {\bibnamefont {Furman}, \bibfnamefont {J.~L.}}, \bibinfo {author} {\bibnamefont {Jensen}, \bibfnamefont {K.}}, and\ \bibinfo {author} {\bibnamefont {Murray}, \bibfnamefont {F.}},\ }\bibfield  {title} {\enquote {\bibinfo {title} {Governing knowledge in the scientific community: Exploring the role of retractions in biomedicine},}\ }\href {https://doi.org/10.1016/j.respol.2011.11.001} {\bibfield  {journal} {\bibinfo  {journal} {Res. Policy}\ }\textbf {\bibinfo {volume} {41}},\ \bibinfo {pages} {276--290} (\bibinfo {year} {2012})}\BibitemShut {NoStop}%
\bibitem [{\citenamefont {Garfield}(1957)}]{Garfield1957-qo}%
  \BibitemOpen
  \bibfield  {author} {\bibinfo {author} {\bibnamefont {Garfield}, \bibfnamefont {E.}},\ }\bibfield  {title} {\enquote {\bibinfo {title} {The obliteration phenomenon in science, and the advantage of being obliterated},}\ }\href@noop {} {\bibfield  {journal} {\bibinfo  {journal} {Current Contents}\ ,\ \bibinfo {pages} {5--7}} (\bibinfo {year} {1957})}\BibitemShut {NoStop}%
\bibitem [{\citenamefont {Gingras}\ \emph {et~al.}(2008)\citenamefont {Gingras}, \citenamefont {Larivi{\`e}re}, \citenamefont {Macaluso},\ and\ \citenamefont {Robitaille}}]{Gingras2008-zr}%
  \BibitemOpen
  \bibfield  {author} {\bibinfo {author} {\bibnamefont {Gingras}, \bibfnamefont {Y.}}, \bibinfo {author} {\bibnamefont {Larivi{\`e}re}, \bibfnamefont {V.}}, \bibinfo {author} {\bibnamefont {Macaluso}, \bibfnamefont {B.}}, and\ \bibinfo {author} {\bibnamefont {Robitaille}, \bibfnamefont {J.-P.}},\ }\bibfield  {title} {\enquote {\bibinfo {title} {The effects of aging on researchers' publication and citation patterns},}\ }\href {https://doi.org/10.1371/journal.pone.0004048} {\bibfield  {journal} {\bibinfo  {journal} {PLoS One}\ }\textbf {\bibinfo {volume} {3}},\ \bibinfo {pages} {e4048} (\bibinfo {year} {2008})}\BibitemShut {NoStop}%
\bibitem [{\citenamefont {Gl{\"a}ser}(2007)}]{Glaser2007-aq}%
  \BibitemOpen
  \bibfield  {author} {\bibinfo {author} {\bibnamefont {Gl{\"a}ser}, \bibfnamefont {J.}},\ }\bibfield  {title} {\enquote {\bibinfo {title} {The social orders of research evaluation systems},}\ }in\ \href {https://doi.org/10.1007/978-1-4020-6746-4\_12} {\emph {\bibinfo {booktitle} {The Changing Governance of the Sciences: The Advent of Research Evaluation Systems}}},\ \bibinfo {editor} {edited by\ \bibinfo {editor} {\bibfnamefont {R.}~\bibnamefont {Whitley}}\ and\ \bibinfo {editor} {\bibfnamefont {J.}~\bibnamefont {Gl{\"a}ser}}}\ (\bibinfo  {publisher} {Springer Netherlands},\ \bibinfo {address} {Dordrecht},\ \bibinfo {year} {2007})\ pp.\ \bibinfo {pages} {245--266}\BibitemShut {NoStop}%
\bibitem [{\citenamefont {Gl{\"a}ser}\ and\ \citenamefont {Laudel}(2016)}]{Glaser2016-uk}%
  \BibitemOpen
  \bibfield  {author} {\bibinfo {author} {\bibnamefont {Gl{\"a}ser}, \bibfnamefont {J.}}and\ \bibinfo {author} {\bibnamefont {Laudel}, \bibfnamefont {G.}},\ }\bibfield  {title} {\enquote {\bibinfo {title} {Governing science},}\ }\href {https://doi.org/10.1017/s0003975616000047} {\bibfield  {journal} {\bibinfo  {journal} {Arch. Eur. Sociol.}\ }\textbf {\bibinfo {volume} {57}},\ \bibinfo {pages} {117--168} (\bibinfo {year} {2016})}\BibitemShut {NoStop}%
\bibitem [{\citenamefont {Goldberg}, \citenamefont {Anthony},\ and\ \citenamefont {Evans}(2015)}]{Goldberg2015-fl}%
  \BibitemOpen
  \bibfield  {author} {\bibinfo {author} {\bibnamefont {Goldberg}, \bibfnamefont {S.~R.}}, \bibinfo {author} {\bibnamefont {Anthony}, \bibfnamefont {H.}}, and\ \bibinfo {author} {\bibnamefont {Evans}, \bibfnamefont {T.~S.}},\ }\bibfield  {title} {\enquote {\bibinfo {title} {Modelling citation networks},}\ }\href {https://doi.org/10.1007/s11192-015-1737-9} {\bibfield  {journal} {\bibinfo  {journal} {Scientometrics}\ }\textbf {\bibinfo {volume} {105}},\ \bibinfo {pages} {1577--1604} (\bibinfo {year} {2015})}\BibitemShut {NoStop}%
\bibitem [{\citenamefont {Goodman}\ \emph {et~al.}(1994)\citenamefont {Goodman}, \citenamefont {Berlin}, \citenamefont {Fletcher},\ and\ \citenamefont {Fletcher}}]{Goodman1994-yw}%
  \BibitemOpen
  \bibfield  {author} {\bibinfo {author} {\bibnamefont {Goodman}, \bibfnamefont {S.~N.}}, \bibinfo {author} {\bibnamefont {Berlin}, \bibfnamefont {J.}}, \bibinfo {author} {\bibnamefont {Fletcher}, \bibfnamefont {S.~W.}}, and\ \bibinfo {author} {\bibnamefont {Fletcher}, \bibfnamefont {R.~H.}},\ }\bibfield  {title} {\enquote {\bibinfo {title} {Manuscript quality before and after peer review and editing at annals of internal medicine},}\ }\href {https://doi.org/10.7326/0003-4819-121-1-199407010-00003} {\bibfield  {journal} {\bibinfo  {journal} {Ann. Intern. Med.}\ }\textbf {\bibinfo {volume} {121}},\ \bibinfo {pages} {11} (\bibinfo {year} {1994})}\BibitemShut {NoStop}%
\bibitem [{\citenamefont {Gross}\ and\ \citenamefont {Gross}(1927)}]{Gross1927-yr}%
  \BibitemOpen
  \bibfield  {author} {\bibinfo {author} {\bibnamefont {Gross}, \bibfnamefont {P.~L.}}and\ \bibinfo {author} {\bibnamefont {Gross}, \bibfnamefont {E.~M.}},\ }\bibfield  {title} {\enquote {\bibinfo {title} {{COLLEGE} {LIBRARIES} {AND} {CHEMICAL} {EDUCATION}},}\ }\href {https://doi.org/10.1126/science.66.1713.385} {\bibfield  {journal} {\bibinfo  {journal} {Science}\ }\textbf {\bibinfo {volume} {66}},\ \bibinfo {pages} {385--389} (\bibinfo {year} {1927})}\BibitemShut {NoStop}%
\bibitem [{\citenamefont {Hajra}\ and\ \citenamefont {Sen}(2006)}]{Hajra2006-po}%
  \BibitemOpen
  \bibfield  {author} {\bibinfo {author} {\bibnamefont {Hajra}, \bibfnamefont {K.~B.}}and\ \bibinfo {author} {\bibnamefont {Sen}, \bibfnamefont {P.}},\ }\bibfield  {title} {\enquote {\bibinfo {title} {Modelling aging characteristics in citation networks},}\ }\href {https://doi.org/10.1016/j.physa.2005.12.044} {\bibfield  {journal} {\bibinfo  {journal} {Physica A: Statistical Mechanics and its Applications}\ }\textbf {\bibinfo {volume} {368}},\ \bibinfo {pages} {575--582} (\bibinfo {year} {2006})}\BibitemShut {NoStop}%
\bibitem [{\citenamefont {Hargens}\ and\ \citenamefont {Felmlee}(1984)}]{Hargens1984-gl}%
  \BibitemOpen
  \bibfield  {author} {\bibinfo {author} {\bibnamefont {Hargens}, \bibfnamefont {L.~L.}}and\ \bibinfo {author} {\bibnamefont {Felmlee}, \bibfnamefont {D.~H.}},\ }\bibfield  {title} {\enquote {\bibinfo {title} {Structural determinants of stratification in science},}\ }\href {https://doi.org/10.2307/2095425} {\bibfield  {journal} {\bibinfo  {journal} {Am. Sociol. Rev.}\ }\textbf {\bibinfo {volume} {49}},\ \bibinfo {pages} {685} (\bibinfo {year} {1984})}\BibitemShut {NoStop}%
\bibitem [{\citenamefont {Herman}(2004{\natexlab{a}})}]{Herman2004-sm}%
  \BibitemOpen
  \bibfield  {author} {\bibinfo {author} {\bibnamefont {Herman}, \bibfnamefont {E.}},\ }\bibfield  {title} {\enquote {\bibinfo {title} {Research in progress. part 2 -- some preliminary insights into the information needs of the contemporary academic researcher},}\ }\href {https://doi.org/10.1108/00012530410529495} {\bibfield  {journal} {\bibinfo  {journal} {Aslib Proc.}\ }\textbf {\bibinfo {volume} {56}},\ \bibinfo {pages} {118--131} (\bibinfo {year} {2004}{\natexlab{a}})}\BibitemShut {NoStop}%
\bibitem [{\citenamefont {Herman}(2004{\natexlab{b}})}]{Herman2004-bq}%
  \BibitemOpen
  \bibfield  {author} {\bibinfo {author} {\bibnamefont {Herman}, \bibfnamefont {E.}},\ }\bibfield  {title} {\enquote {\bibinfo {title} {Research in progress: some preliminary and key insights into the information needs of the contemporary academic researcher. part 1},}\ }\href {https://doi.org/10.1108/00012530410516859} {\bibfield  {journal} {\bibinfo  {journal} {Aslib Proc.}\ }\textbf {\bibinfo {volume} {56}},\ \bibinfo {pages} {34--47} (\bibinfo {year} {2004}{\natexlab{b}})}\BibitemShut {NoStop}%
\bibitem [{\citenamefont {Hicks}(2012)}]{Hicks2012-tb}%
  \BibitemOpen
  \bibfield  {author} {\bibinfo {author} {\bibnamefont {Hicks}, \bibfnamefont {D.}},\ }\bibfield  {title} {\enquote {\bibinfo {title} {Performance-based university research funding systems},}\ }\href {https://doi.org/10.1016/j.respol.2011.09.007} {\bibfield  {journal} {\bibinfo  {journal} {Res. Policy}\ }\textbf {\bibinfo {volume} {41}},\ \bibinfo {pages} {251--261} (\bibinfo {year} {2012})}\BibitemShut {NoStop}%
\bibitem [{\citenamefont {Hicks}\ \emph {et~al.}(2015)\citenamefont {Hicks}, \citenamefont {Wouters}, \citenamefont {Waltman}, \citenamefont {Rijcke},\ and\ \citenamefont {Rafols}}]{Hicks2015-nj}%
  \BibitemOpen
  \bibfield  {author} {\bibinfo {author} {\bibnamefont {Hicks}, \bibfnamefont {D.}}, \bibinfo {author} {\bibnamefont {Wouters}, \bibfnamefont {P.}}, \bibinfo {author} {\bibnamefont {Waltman}, \bibfnamefont {L.}}, \bibinfo {author} {\bibnamefont {Rijcke}, \bibfnamefont {S.~d.}}, and\ \bibinfo {author} {\bibnamefont {Rafols}, \bibfnamefont {I.}},\ }\bibfield  {title} {\enquote {\bibinfo {title} {The leiden manifesto for research metrics},}\ }\href {https://doi.org/10.1038/520429a} {\bibfield  {journal} {\bibinfo  {journal} {Nature}\ }\textbf {\bibinfo {volume} {520}},\ \bibinfo {pages} {429--431} (\bibinfo {year} {2015})}\BibitemShut {NoStop}%
\bibitem [{\citenamefont {Higham}\ \emph {et~al.}(2017)\citenamefont {Higham}, \citenamefont {Governale}, \citenamefont {Jaffe},\ and\ \citenamefont {Z{\"u}licke}}]{Higham2017-ho}%
  \BibitemOpen
  \bibfield  {author} {\bibinfo {author} {\bibnamefont {Higham}, \bibfnamefont {K.~W.}}, \bibinfo {author} {\bibnamefont {Governale}, \bibfnamefont {M.}}, \bibinfo {author} {\bibnamefont {Jaffe}, \bibfnamefont {A.~B.}}, and\ \bibinfo {author} {\bibnamefont {Z{\"u}licke}, \bibfnamefont {U.}},\ }\bibfield  {title} {\enquote {\bibinfo {title} {Unraveling the dynamics of growth, aging and inflation for citations to scientific articles from specific research fields},}\ }\href {https://doi.org/10.1016/j.joi.2017.10.004} {\bibfield  {journal} {\bibinfo  {journal} {Journal of Informetrics}\ }\textbf {\bibinfo {volume} {11}},\ \bibinfo {pages} {1190--1200} (\bibinfo {year} {2017})},\ \Eprint {https://arxiv.org/abs/1708.08335} {arXiv:1708.08335 [cs.DL]} \BibitemShut {NoStop}%
\bibitem [{\citenamefont {Hood}\ and\ \citenamefont {Wilson}(2001)}]{Hood2001-le}%
  \BibitemOpen
  \bibfield  {author} {\bibinfo {author} {\bibnamefont {Hood}, \bibfnamefont {W.~W.}}and\ \bibinfo {author} {\bibnamefont {Wilson}, \bibfnamefont {C.~S.}},\ }\bibfield  {title} {\enquote {\bibinfo {title} {The literature of bibliometrics, scientometrics, and informetrics},}\ }\href {https://doi.org/10.1023/A:1017919924342} {\bibfield  {journal} {\bibinfo  {journal} {Scientometrics}\ }\textbf {\bibinfo {volume} {52}},\ \bibinfo {pages} {291--314} (\bibinfo {year} {2001})}\BibitemShut {NoStop}%
\bibitem [{\citenamefont {King}(2017)}]{King2017-gr}%
  \BibitemOpen
  \bibfield  {author} {\bibinfo {author} {\bibnamefont {King}, \bibfnamefont {S.~R.~F.}},\ }\bibfield  {title} {\enquote {\bibinfo {title} {Consultative review is worth the wait},}\ }\href {https://doi.org/10.7554/eLife.32012} {\bibfield  {journal} {\bibinfo  {journal} {Elife}\ }\textbf {\bibinfo {volume} {6}},\ \bibinfo {pages} {e32012} (\bibinfo {year} {2017})}\BibitemShut {NoStop}%
\bibitem [{\citenamefont {Klebel}\ and\ \citenamefont {Traag}(2024)}]{klebel_introduction_2024}%
  \BibitemOpen
  \bibfield  {author} {\bibinfo {author} {\bibnamefont {Klebel}, \bibfnamefont {T.}}and\ \bibinfo {author} {\bibnamefont {Traag}, \bibfnamefont {V.}},\ }\bibfield  {title} {\enquote {\bibinfo {title} {Introduction to causality in science studies},}\ }\href {https://doi.org/10.31235/osf.io/4bw9e} {\bibfield  {journal} {\bibinfo  {journal} {SocArXiv}\ } (\bibinfo {year} {2024}),\ 10.31235/osf.io/4bw9e}\BibitemShut {NoStop}%
\bibitem [{\citenamefont {Klein}\ \emph {et~al.}(2016)\citenamefont {Klein}, \citenamefont {Broadwell}, \citenamefont {Farb},\ and\ \citenamefont {Grappone}}]{Klein2016-sx}%
  \BibitemOpen
  \bibfield  {author} {\bibinfo {author} {\bibnamefont {Klein}, \bibfnamefont {M.}}, \bibinfo {author} {\bibnamefont {Broadwell}, \bibfnamefont {P.}}, \bibinfo {author} {\bibnamefont {Farb}, \bibfnamefont {S.~E.}}, and\ \bibinfo {author} {\bibnamefont {Grappone}, \bibfnamefont {T.}},\ }\bibfield  {title} {\enquote {\bibinfo {title} {Comparing published scientific journal articles to their pre-print versions},}\ }in\ \href {https://doi.org/10.1145/2910896.2910909} {\emph {\bibinfo {booktitle} {Proceedings of the 16th {ACM/IEEE-CS} on Joint Conference on Digital Libraries - {JCDL} '16}}}\ (\bibinfo  {publisher} {ACM Press},\ \bibinfo {address} {New York, New York, USA},\ \bibinfo {year} {2016})\ pp.\ \bibinfo {pages} {153--162}\BibitemShut {NoStop}%
\bibitem [{\citenamefont {Knothe}(2006)}]{Knothe2006-fd}%
  \BibitemOpen
  \bibfield  {author} {\bibinfo {author} {\bibnamefont {Knothe}, \bibfnamefont {G.}},\ }\bibfield  {title} {\enquote {\bibinfo {title} {Comparative citation analysis of duplicate or highly related publications},}\ }\href {https://doi.org/10.1002/asi.20409} {\bibfield  {journal} {\bibinfo  {journal} {J. Am. Soc. Inf. Sci. Technol.}\ }\textbf {\bibinfo {volume} {57}},\ \bibinfo {pages} {1830--1839} (\bibinfo {year} {2006})}\BibitemShut {NoStop}%
\bibitem [{\citenamefont {Kuhn}(2012)}]{Kuhn2012-in}%
  \BibitemOpen
  \bibfield  {author} {\bibinfo {author} {\bibnamefont {Kuhn}, \bibfnamefont {T.~S.}},\ }\href@noop {} {\emph {\bibinfo {title} {The structure of scientific revolutions}}}\ (\bibinfo  {publisher} {The University of Chicago Press},\ \bibinfo {address} {Chicago},\ \bibinfo {year} {2012})\BibitemShut {NoStop}%
\bibitem [{\citenamefont {Laherr{\`e}re}\ and\ \citenamefont {Sornette}(1998)}]{Laherrere1998-vo}%
  \BibitemOpen
  \bibfield  {author} {\bibinfo {author} {\bibnamefont {Laherr{\`e}re}, \bibfnamefont {J.}}and\ \bibinfo {author} {\bibnamefont {Sornette}, \bibfnamefont {D.}},\ }\bibfield  {title} {\enquote {\bibinfo {title} {Stretched exponential distributions in nature and economy: ``fat tails'' with characteristic scales},}\ }\href {https://doi.org/10.1007/s100510050276} {\bibfield  {journal} {\bibinfo  {journal} {The European Physical Journal B - Condensed Matter and Complex Systems}\ }\textbf {\bibinfo {volume} {2}},\ \bibinfo {pages} {525--539} (\bibinfo {year} {1998})}\BibitemShut {NoStop}%
\bibitem [{\citenamefont {Lai}, \citenamefont {Traag},\ and\ \citenamefont {Waltman}(2020)}]{Lai2020-bi}%
  \BibitemOpen
  \bibfield  {author} {\bibinfo {author} {\bibnamefont {Lai}, \bibfnamefont {K.~H.}}, \bibinfo {author} {\bibnamefont {Traag}, \bibfnamefont {V.}}, and\ \bibinfo {author} {\bibnamefont {Waltman}, \bibfnamefont {L.}},\ }\bibfield  {title} {\enquote {\bibinfo {title} {Challenges in using bibliometric indicators to assess peer review decisions: A simulation model},}\ }in\ \href@noop {} {\emph {\bibinfo {booktitle} {{PEERE} Conference}}}\ (\bibinfo {year} {2020})\BibitemShut {NoStop}%
\bibitem [{\citenamefont {Lamers}\ \emph {et~al.}(2021)\citenamefont {Lamers}, \citenamefont {Boyack}, \citenamefont {Larivi{\`e}re}, \citenamefont {Sugimoto},\ and\ \citenamefont {{others}}}]{Lamers2021-xb}%
  \BibitemOpen
  \bibfield  {author} {\bibinfo {author} {\bibnamefont {Lamers}, \bibfnamefont {W.~S.}}, \bibinfo {author} {\bibnamefont {Boyack}, \bibfnamefont {K.}}, \bibinfo {author} {\bibnamefont {Larivi{\`e}re}, \bibfnamefont {V.}}, \bibinfo {author} {\bibnamefont {Sugimoto}, \bibfnamefont {C.~R.}}, and\ \bibinfo {author} {\bibnamefont {{others}},},\ }\bibfield  {title} {\enquote {\bibinfo {title} {{Meta-Research}: Investigating disagreement in the scientific literature},}\ }\href@noop {} {\bibfield  {journal} {\bibinfo  {journal} {Elife}\ } (\bibinfo {year} {2021})}\BibitemShut {NoStop}%
\bibitem [{\citenamefont {Larivi{\`e}re}, \citenamefont {Archambault},\ and\ \citenamefont {Gingras}(2008)}]{Lariviere2008-zg}%
  \BibitemOpen
  \bibfield  {author} {\bibinfo {author} {\bibnamefont {Larivi{\`e}re}, \bibfnamefont {V.}}, \bibinfo {author} {\bibnamefont {Archambault}, \bibfnamefont {{\'E}.}}, and\ \bibinfo {author} {\bibnamefont {Gingras}, \bibfnamefont {Y.}},\ }\bibfield  {title} {\enquote {\bibinfo {title} {Long-term variations in the aging of scientific literature: From exponential growth to steady-state science (1900--2004)},}\ }\href {https://doi.org/10.1002/asi.20744} {\bibfield  {journal} {\bibinfo  {journal} {J. Am. Soc. Inf. Sci. Technol.}\ }\textbf {\bibinfo {volume} {59}},\ \bibinfo {pages} {288--296} (\bibinfo {year} {2008})}\BibitemShut {NoStop}%
\bibitem [{\citenamefont {Larivi{\`e}re}\ and\ \citenamefont {Gingras}(2010)}]{Lariviere2010-vw}%
  \BibitemOpen
  \bibfield  {author} {\bibinfo {author} {\bibnamefont {Larivi{\`e}re}, \bibfnamefont {V.}}and\ \bibinfo {author} {\bibnamefont {Gingras}, \bibfnamefont {Y.}},\ }\bibfield  {title} {\enquote {\bibinfo {title} {The impact factor's matthew effect: A natural experiment in bibliometrics},}\ }\href {https://doi.org/10.1002/asi.21232} {\bibfield  {journal} {\bibinfo  {journal} {J. Am. Soc. Inf. Sci. Technol.}\ }\textbf {\bibinfo {volume} {61}},\ \bibinfo {pages} {424--427} (\bibinfo {year} {2010})}\BibitemShut {NoStop}%
\bibitem [{\citenamefont {Larivi{\`e}re}, \citenamefont {Gingras},\ and\ \citenamefont {Archambault}(2009)}]{Lariviere2009-ap}%
  \BibitemOpen
  \bibfield  {author} {\bibinfo {author} {\bibnamefont {Larivi{\`e}re}, \bibfnamefont {V.}}, \bibinfo {author} {\bibnamefont {Gingras}, \bibfnamefont {Y.}}, and\ \bibinfo {author} {\bibnamefont {Archambault}, \bibfnamefont {{\'E}.}},\ }\bibfield  {title} {\enquote {\bibinfo {title} {The decline in the concentration of citations, 1900-2007},}\ }\href {https://doi.org/10.1002/asi.21011} {\bibfield  {journal} {\bibinfo  {journal} {J. Am. Soc. Inf. Sci. Technol.}\ }\textbf {\bibinfo {volume} {60}},\ \bibinfo {pages} {858--862} (\bibinfo {year} {2009})}\BibitemShut {NoStop}%
\bibitem [{\citenamefont {Larivi{\`e}re}\ \emph {et~al.}(2015)\citenamefont {Larivi{\`e}re}, \citenamefont {Gingras}, \citenamefont {Sugimoto},\ and\ \citenamefont {Tsou}}]{Lariviere2015-uq}%
  \BibitemOpen
  \bibfield  {author} {\bibinfo {author} {\bibnamefont {Larivi{\`e}re}, \bibfnamefont {V.}}, \bibinfo {author} {\bibnamefont {Gingras}, \bibfnamefont {Y.}}, \bibinfo {author} {\bibnamefont {Sugimoto}, \bibfnamefont {C.~R.}}, and\ \bibinfo {author} {\bibnamefont {Tsou}, \bibfnamefont {A.}},\ }\bibfield  {title} {\enquote {\bibinfo {title} {Team size matters: Collaboration and scientific impact since 1900: On the relationship between collaboration and scientific impact since 1900},}\ }\href {https://doi.org/10.1002/asi.23266} {\bibfield  {journal} {\bibinfo  {journal} {J. Assoc. Inf. Sci. Technol.}\ }\textbf {\bibinfo {volume} {66}},\ \bibinfo {pages} {1323--1332} (\bibinfo {year} {2015})}\BibitemShut {NoStop}%
\bibitem [{\citenamefont {Lariviere}\ \emph {et~al.}(2016)\citenamefont {Lariviere}, \citenamefont {Kiermer}, \citenamefont {MacCallum}, \citenamefont {McNutt}, \citenamefont {Patterson}, \citenamefont {Pulverer}, \citenamefont {Swaminathan}, \citenamefont {Taylor},\ and\ \citenamefont {Curry}}]{Lariviere2016-ws}%
  \BibitemOpen
  \bibfield  {author} {\bibinfo {author} {\bibnamefont {Lariviere}, \bibfnamefont {V.}}, \bibinfo {author} {\bibnamefont {Kiermer}, \bibfnamefont {V.}}, \bibinfo {author} {\bibnamefont {MacCallum}, \bibfnamefont {C.~J.}}, \bibinfo {author} {\bibnamefont {McNutt}, \bibfnamefont {M.}}, \bibinfo {author} {\bibnamefont {Patterson}, \bibfnamefont {M.}}, \bibinfo {author} {\bibnamefont {Pulverer}, \bibfnamefont {B.}}, \bibinfo {author} {\bibnamefont {Swaminathan}, \bibfnamefont {S.}}, \bibinfo {author} {\bibnamefont {Taylor}, \bibfnamefont {S.}}, and\ \bibinfo {author} {\bibnamefont {Curry}, \bibfnamefont {S.}},\ }\bibfield  {title} {\enquote {\bibinfo {title} {A simple proposal for the publication of journal citation distributions},}\ }\href {https://doi.org/10.1101/062109} {\bibfield  {journal} {\bibinfo  {journal} {bioRxiv}\ ,\ \bibinfo {pages} {062109}} (\bibinfo {year} {2016})}\BibitemShut {NoStop}%
\bibitem [{\citenamefont {Larivi{\`e}re}\ and\ \citenamefont {Sugimoto}(2019)}]{Lariviere2019-ib}%
  \BibitemOpen
  \bibfield  {author} {\bibinfo {author} {\bibnamefont {Larivi{\`e}re}, \bibfnamefont {V.}}and\ \bibinfo {author} {\bibnamefont {Sugimoto}, \bibfnamefont {C.~R.}},\ }\bibfield  {title} {\enquote {\bibinfo {title} {The journal impact factor: A brief history, critique, and discussion of adverse effects},}\ }in\ \href {https://doi.org/10.1007/978-3-030-02511-3\_1} {\emph {\bibinfo {booktitle} {Springer Handbook of Science and Technology Indicators}}},\ \bibinfo {editor} {edited by\ \bibinfo {editor} {\bibfnamefont {W.~G. F. M.~S.}\ \bibnamefont {Thelwall}}}\ (\bibinfo  {publisher} {Springer},\ \bibinfo {address} {Cham},\ \bibinfo {year} {2019})\ pp.\ \bibinfo {pages} {3--24}\BibitemShut {NoStop}%
\bibitem [{\citenamefont {Lee}\ \emph {et~al.}(2013)\citenamefont {Lee}, \citenamefont {Sugimoto}, \citenamefont {Zhang},\ and\ \citenamefont {Cronin}}]{Lee2013-rb}%
  \BibitemOpen
  \bibfield  {author} {\bibinfo {author} {\bibnamefont {Lee}, \bibfnamefont {C.~J.}}, \bibinfo {author} {\bibnamefont {Sugimoto}, \bibfnamefont {C.~R.}}, \bibinfo {author} {\bibnamefont {Zhang}, \bibfnamefont {G.}}, and\ \bibinfo {author} {\bibnamefont {Cronin}, \bibfnamefont {B.}},\ }\bibfield  {title} {\enquote {\bibinfo {title} {Bias in peer review},}\ }\href {https://doi.org/10.1002/asi.22784} {\bibfield  {journal} {\bibinfo  {journal} {J. Am. Soc. Inf. Sci. Technol.}\ }\textbf {\bibinfo {volume} {64}},\ \bibinfo {pages} {2--17} (\bibinfo {year} {2013})}\BibitemShut {NoStop}%
\bibitem [{\citenamefont {Line}(1970)}]{Line1970-gl}%
  \BibitemOpen
  \bibfield  {author} {\bibinfo {author} {\bibnamefont {Line}, \bibfnamefont {M.~B.}},\ }\bibfield  {title} {\enquote {\bibinfo {title} {{THE} {'HALF-LIFE'} {OF} {PERIODICAL} {LITERATURE}: {APPARENT} {AND} {REAL} {OBSOLESCENCE}},}\ }\href {https://doi.org/10.1108/eb026486} {\bibfield  {journal} {\bibinfo  {journal} {Journal of Documentation}\ }\textbf {\bibinfo {volume} {26}},\ \bibinfo {pages} {46--54} (\bibinfo {year} {1970})}\BibitemShut {NoStop}%
\bibitem [{\citenamefont {Line}\ and\ \citenamefont {Sandison}(1974)}]{Line1974-tt}%
  \BibitemOpen
  \bibfield  {author} {\bibinfo {author} {\bibnamefont {Line}, \bibfnamefont {M.~B.}}and\ \bibinfo {author} {\bibnamefont {Sandison}, \bibfnamefont {A.}},\ }\bibfield  {title} {\enquote {\bibinfo {title} {{PROGRESS} {IN} {DOCUMENTATION}: `obsolescence' and changes in the use of literature with time},}\ }\href {https://doi.org/10.1108/eb026583} {\bibfield  {journal} {\bibinfo  {journal} {Journal of Documentation}\ }\textbf {\bibinfo {volume} {30}},\ \bibinfo {pages} {283--350} (\bibinfo {year} {1974})}\BibitemShut {NoStop}%
\bibitem [{\citenamefont {Lozano}, \citenamefont {Larivi{\`e}re},\ and\ \citenamefont {Gingras}(2012)}]{Lozano2012-rg}%
  \BibitemOpen
  \bibfield  {author} {\bibinfo {author} {\bibnamefont {Lozano}, \bibfnamefont {G.~A.}}, \bibinfo {author} {\bibnamefont {Larivi{\`e}re}, \bibfnamefont {V.}}, and\ \bibinfo {author} {\bibnamefont {Gingras}, \bibfnamefont {Y.}},\ }\bibfield  {title} {\enquote {\bibinfo {title} {The weakening relationship between the impact factor and papers' citations in the digital age},}\ }\href {https://doi.org/10.1002/asi.22731} {\bibfield  {journal} {\bibinfo  {journal} {J. Am. Soc. Inf. Sci. Technol.}\ } (\bibinfo {year} {2012}),\ 10.1002/asi.22731}\BibitemShut {NoStop}%
\bibitem [{\citenamefont {MacRoberts}\ and\ \citenamefont {MacRoberts}(1989)}]{MacRoberts1989-er}%
  \BibitemOpen
  \bibfield  {author} {\bibinfo {author} {\bibnamefont {MacRoberts}, \bibfnamefont {M.~H.}}and\ \bibinfo {author} {\bibnamefont {MacRoberts}, \bibfnamefont {B.~R.}},\ }\bibfield  {title} {\enquote {\bibinfo {title} {Problems of citation analysis: A critical review},}\ }\href@noop {} {\bibfield  {journal} {\bibinfo  {journal} {Journal of the American Society for Information Science (1986-1998)}\ }\textbf {\bibinfo {volume} {40}},\ \bibinfo {pages} {342} (\bibinfo {year} {1989})}\BibitemShut {NoStop}%
\bibitem [{\citenamefont {McCain}(2011)}]{McCain2011-fr}%
  \BibitemOpen
  \bibfield  {author} {\bibinfo {author} {\bibnamefont {McCain}, \bibfnamefont {K.~W.}},\ }\bibfield  {title} {\enquote {\bibinfo {title} {Eponymy and obliteration by incorporation: The case of the ``nash equilibrium''},}\ }\href {https://doi.org/10.1002/asi.21536} {\bibfield  {journal} {\bibinfo  {journal} {J. Am. Soc. Inf. Sci. Technol.}\ }\textbf {\bibinfo {volume} {62}},\ \bibinfo {pages} {1412--1424} (\bibinfo {year} {2011})}\BibitemShut {NoStop}%
\bibitem [{\citenamefont {McKiernan}\ \emph {et~al.}(2019)\citenamefont {McKiernan}, \citenamefont {Schimanski}, \citenamefont {Mu{\~n}oz~Nieves}, \citenamefont {Matthias}, \citenamefont {Niles},\ and\ \citenamefont {Alperin}}]{McKiernan2019-nu}%
  \BibitemOpen
  \bibfield  {author} {\bibinfo {author} {\bibnamefont {McKiernan}, \bibfnamefont {E.~C.}}, \bibinfo {author} {\bibnamefont {Schimanski}, \bibfnamefont {L.~A.}}, \bibinfo {author} {\bibnamefont {Mu{\~n}oz~Nieves}, \bibfnamefont {C.}}, \bibinfo {author} {\bibnamefont {Matthias}, \bibfnamefont {L.}}, \bibinfo {author} {\bibnamefont {Niles}, \bibfnamefont {M.~T.}}, and\ \bibinfo {author} {\bibnamefont {Alperin}, \bibfnamefont {J.~P.}},\ }\bibfield  {title} {\enquote {\bibinfo {title} {Use of the journal impact factor in academic review, promotion, and tenure evaluations},}\ }\href {https://doi.org/10.7554/eLife.47338} {\bibfield  {journal} {\bibinfo  {journal} {Elife}\ }\textbf {\bibinfo {volume} {8}} (\bibinfo {year} {2019}),\ 10.7554/eLife.47338}\BibitemShut {NoStop}%
\bibitem [{\citenamefont {Medoff}(2006)}]{Medoff2006-sh}%
  \BibitemOpen
  \bibfield  {author} {\bibinfo {author} {\bibnamefont {Medoff}, \bibfnamefont {M.~H.}},\ }\bibfield  {title} {\enquote {\bibinfo {title} {Evidence of a harvard and chicago matthew effect},}\ }\href {https://doi.org/10.1080/13501780601049079} {\bibfield  {journal} {\bibinfo  {journal} {Journal of Economic Methodology}\ }\textbf {\bibinfo {volume} {13}},\ \bibinfo {pages} {485--506} (\bibinfo {year} {2006})}\BibitemShut {NoStop}%
\bibitem [{\citenamefont {Merton}(1968)}]{Merton1968-qu}%
  \BibitemOpen
  \bibfield  {author} {\bibinfo {author} {\bibnamefont {Merton}, \bibfnamefont {R.~K.}},\ }\bibfield  {title} {\enquote {\bibinfo {title} {The matthew effect in science},}\ }\href@noop {} {\bibfield  {journal} {\bibinfo  {journal} {Science}\ }\textbf {\bibinfo {volume} {159}},\ \bibinfo {pages} {56--63} (\bibinfo {year} {1968})}\BibitemShut {NoStop}%
\bibitem [{\citenamefont {Milojevi{\'c}}, \citenamefont {Radicchi},\ and\ \citenamefont {Bar-Ilan}(2016)}]{Milojevic2016-yb}%
  \BibitemOpen
  \bibfield  {author} {\bibinfo {author} {\bibnamefont {Milojevi{\'c}}, \bibfnamefont {S.}}, \bibinfo {author} {\bibnamefont {Radicchi}, \bibfnamefont {F.}}, and\ \bibinfo {author} {\bibnamefont {Bar-Ilan}, \bibfnamefont {J.}},\ }\bibfield  {title} {\enquote {\bibinfo {title} {Citation success index - an intuitive pair-wise journal comparison metric},}\ }\href {https://doi.org/10.1016/j.joi.2016.12.006} {\bibfield  {journal} {\bibinfo  {journal} {Journal of Informetrics}\ }\textbf {\bibinfo {volume} {11}},\ \bibinfo {pages} {223--231} (\bibinfo {year} {2016})},\ \Eprint {https://arxiv.org/abs/1607.03179} {arXiv:1607.03179 [cs.DL]} \BibitemShut {NoStop}%
\bibitem [{\citenamefont {Mingers}\ and\ \citenamefont {Burrell}(2006)}]{Mingers2006-my}%
  \BibitemOpen
  \bibfield  {author} {\bibinfo {author} {\bibnamefont {Mingers}, \bibfnamefont {J.}}and\ \bibinfo {author} {\bibnamefont {Burrell}, \bibfnamefont {Q.~L.}},\ }\bibfield  {title} {\enquote {\bibinfo {title} {Modeling citation behavior in management science journals},}\ }\href {https://doi.org/10.1016/j.ipm.2006.03.012} {\bibfield  {journal} {\bibinfo  {journal} {Inf. Process. Manag.}\ }\textbf {\bibinfo {volume} {42}},\ \bibinfo {pages} {1451--1464} (\bibinfo {year} {2006})}\BibitemShut {NoStop}%
\bibitem [{\citenamefont {Mingers}\ and\ \citenamefont {Xu}(2010)}]{Mingers2010-jz}%
  \BibitemOpen
  \bibfield  {author} {\bibinfo {author} {\bibnamefont {Mingers}, \bibfnamefont {J.}}and\ \bibinfo {author} {\bibnamefont {Xu}, \bibfnamefont {F.}},\ }\bibfield  {title} {\enquote {\bibinfo {title} {The drivers of citations in management science journals},}\ }\href {https://doi.org/10.1016/j.ejor.2009.12.008} {\bibfield  {journal} {\bibinfo  {journal} {Eur. J. Oper. Res.}\ }\textbf {\bibinfo {volume} {205}},\ \bibinfo {pages} {422--430} (\bibinfo {year} {2010})}\BibitemShut {NoStop}%
\bibitem [{\citenamefont {Moed}(2008)}]{Moed2008-ni}%
  \BibitemOpen
  \bibfield  {author} {\bibinfo {author} {\bibnamefont {Moed}, \bibfnamefont {H.~F.}},\ }\bibfield  {title} {\enquote {\bibinfo {title} {{UK} research assessment exercises: Informed judgments on research quality or quantity?}}\ }\href {https://doi.org/10.1007/s11192-008-0108-1} {\bibfield  {journal} {\bibinfo  {journal} {Scientometrics}\ }\textbf {\bibinfo {volume} {74}},\ \bibinfo {pages} {153--161} (\bibinfo {year} {2008})}\BibitemShut {NoStop}%
\bibitem [{\citenamefont {Molas-Gallart}\ and\ \citenamefont {Rafols}(2018)}]{Molas-Gallart2018-qw}%
  \BibitemOpen
  \bibfield  {author} {\bibinfo {author} {\bibnamefont {Molas-Gallart}, \bibfnamefont {J.}}and\ \bibinfo {author} {\bibnamefont {Rafols}, \bibfnamefont {I.}},\ }\href {https://doi.org/10.2139/ssrn.3174954} {\enquote {\bibinfo {title} {Why bibliometric indicators break down: Unstable parameters, incorrect models and irrelevant properties},}\ } (\bibinfo {year} {2018})\BibitemShut {NoStop}%
\bibitem [{\citenamefont {Moreira}, \citenamefont {Zeng},\ and\ \citenamefont {Amaral}(2015)}]{Moreira2015-ww}%
  \BibitemOpen
  \bibfield  {author} {\bibinfo {author} {\bibnamefont {Moreira}, \bibfnamefont {J.~A.~G.}}, \bibinfo {author} {\bibnamefont {Zeng}, \bibfnamefont {X.~H.~T.}}, and\ \bibinfo {author} {\bibnamefont {Amaral}, \bibfnamefont {L.~A.~N.}},\ }\bibfield  {title} {\enquote {\bibinfo {title} {The distribution of the asymptotic number of citations to sets of publications by a researcher or from an academic department are consistent with a discrete lognormal model},}\ }\href {https://doi.org/10.1371/journal.pone.0143108} {\bibfield  {journal} {\bibinfo  {journal} {PLoS One}\ }\textbf {\bibinfo {volume} {10}},\ \bibinfo {pages} {e0143108} (\bibinfo {year} {2015})}\BibitemShut {NoStop}%
\bibitem [{\citenamefont {M{\"u}ller}\ and\ \citenamefont {de~Rijcke}(2017)}]{Muller2017-bq}%
  \BibitemOpen
  \bibfield  {author} {\bibinfo {author} {\bibnamefont {M{\"u}ller}, \bibfnamefont {R.}}and\ \bibinfo {author} {\bibnamefont {de~Rijcke}, \bibfnamefont {S.}},\ }\bibfield  {title} {\enquote {\bibinfo {title} {Thinking with indicators. exploring the epistemic impacts of academic performance indicators in the life sciences},}\ }\href {https://doi.org/10.1093/reseval/rvx023} {\bibfield  {journal} {\bibinfo  {journal} {Res. Eval.}\ }\textbf {\bibinfo {volume} {26}},\ \bibinfo {pages} {157--168} (\bibinfo {year} {2017})}\BibitemShut {NoStop}%
\bibitem [{\citenamefont {Nakamoto}(1988)}]{Nakamoto1988-nh}%
  \BibitemOpen
  \bibfield  {author} {\bibinfo {author} {\bibnamefont {Nakamoto}, \bibfnamefont {H.}},\ }\bibfield  {title} {\enquote {\bibinfo {title} {Synchronous and diachronous citation distributions},}\ }\href@noop {} {\bibfield  {journal} {\bibinfo  {journal} {Informetrics}\ } (\bibinfo {year} {1988})}\BibitemShut {NoStop}%
\bibitem [{\citenamefont {Okike}\ \emph {et~al.}(2016)\citenamefont {Okike}, \citenamefont {Hug}, \citenamefont {Kocher},\ and\ \citenamefont {Leopold}}]{Okike2016-cv}%
  \BibitemOpen
  \bibfield  {author} {\bibinfo {author} {\bibnamefont {Okike}, \bibfnamefont {K.}}, \bibinfo {author} {\bibnamefont {Hug}, \bibfnamefont {K.~T.}}, \bibinfo {author} {\bibnamefont {Kocher}, \bibfnamefont {M.~S.}}, and\ \bibinfo {author} {\bibnamefont {Leopold}, \bibfnamefont {S.~S.}},\ }\bibfield  {title} {\enquote {\bibinfo {title} {Single-blind vs double-blind peer review in the setting of author prestige},}\ }\href {https://doi.org/10.1001/jama.2016.11014} {\bibfield  {journal} {\bibinfo  {journal} {JAMA}\ }\textbf {\bibinfo {volume} {316}},\ \bibinfo {pages} {1315--1316} (\bibinfo {year} {2016})}\BibitemShut {NoStop}%
\bibitem [{\citenamefont {Olensky}, \citenamefont {Schmidt},\ and\ \citenamefont {Van~Eck}(2016)}]{Olensky2016-yq}%
  \BibitemOpen
  \bibfield  {author} {\bibinfo {author} {\bibnamefont {Olensky}, \bibfnamefont {M.}}, \bibinfo {author} {\bibnamefont {Schmidt}, \bibfnamefont {M.}}, and\ \bibinfo {author} {\bibnamefont {Van~Eck}, \bibfnamefont {N.~J.}},\ }\bibfield  {title} {\enquote {\bibinfo {title} {Evaluation of the citation matching algorithms of {CWTS} and {iFQ} in comparison to the web of science},}\ }\href {https://doi.org/10.1002/asi.23590} {\bibfield  {journal} {\bibinfo  {journal} {Journal of the Association for Information Science and Technology}\ }\textbf {\bibinfo {volume} {67}},\ \bibinfo {pages} {2550--2564} (\bibinfo {year} {2016})}\BibitemShut {NoStop}%
\bibitem [{\citenamefont {O'Neil}(2016)}]{ONeil2016-sa}%
  \BibitemOpen
  \bibfield  {author} {\bibinfo {author} {\bibnamefont {O'Neil}, \bibfnamefont {C.}},\ }\href@noop {} {\emph {\bibinfo {title} {Weapons of Math Destruction: How Big Data Increases Inequality and Threatens Democracy}}}\ (\bibinfo  {publisher} {Crown},\ \bibinfo {year} {2016})\BibitemShut {NoStop}%
\bibitem [{\citenamefont {Osuna}, \citenamefont {Cruz-Castro},\ and\ \citenamefont {Sanz-Men{\'e}ndez}(2011)}]{Osuna2011-se}%
  \BibitemOpen
  \bibfield  {author} {\bibinfo {author} {\bibnamefont {Osuna}, \bibfnamefont {C.}}, \bibinfo {author} {\bibnamefont {Cruz-Castro}, \bibfnamefont {L.}}, and\ \bibinfo {author} {\bibnamefont {Sanz-Men{\'e}ndez}, \bibfnamefont {L.}},\ }\bibfield  {title} {\enquote {\bibinfo {title} {Overturning some assumptions about the effects of evaluation systems on publication performance},}\ }\href {https://doi.org/10.1007/s11192-010-0312-7} {\bibfield  {journal} {\bibinfo  {journal} {Scientometrics}\ }\textbf {\bibinfo {volume} {86}},\ \bibinfo {pages} {575--592} (\bibinfo {year} {2011})}\BibitemShut {NoStop}%
\bibitem [{\citenamefont {Owens}(2013)}]{Owens2013-ku}%
  \BibitemOpen
  \bibfield  {author} {\bibinfo {author} {\bibnamefont {Owens}, \bibfnamefont {B.}},\ }\bibfield  {title} {\enquote {\bibinfo {title} {Research assessments: Judgement day},}\ }\href {https://doi.org/10.1038/502288a} {\bibfield  {journal} {\bibinfo  {journal} {Nature}\ }\textbf {\bibinfo {volume} {502}},\ \bibinfo {pages} {288--290} (\bibinfo {year} {2013})}\BibitemShut {NoStop}%
\bibitem [{\citenamefont {Pan}, \citenamefont {Kaski},\ and\ \citenamefont {Fortunato}(2012)}]{Pan2012-da}%
  \BibitemOpen
  \bibfield  {author} {\bibinfo {author} {\bibnamefont {Pan}, \bibfnamefont {R.~K.}}, \bibinfo {author} {\bibnamefont {Kaski}, \bibfnamefont {K.}}, and\ \bibinfo {author} {\bibnamefont {Fortunato}, \bibfnamefont {S.}},\ }\bibfield  {title} {\enquote {\bibinfo {title} {World citation and collaboration networks: uncovering the role of geography in science},}\ }\href {https://doi.org/10.1038/srep00902} {\bibfield  {journal} {\bibinfo  {journal} {Scientific Reports 2012 2}\ }\textbf {\bibinfo {volume} {2}},\ \bibinfo {pages} {902} (\bibinfo {year} {2012})}\BibitemShut {NoStop}%
\bibitem [{\citenamefont {Pan}\ \emph {et~al.}(2018)\citenamefont {Pan}, \citenamefont {Petersen}, \citenamefont {Pammolli},\ and\ \citenamefont {Fortunato}}]{Pan2018-yc}%
  \BibitemOpen
  \bibfield  {author} {\bibinfo {author} {\bibnamefont {Pan}, \bibfnamefont {R.~K.}}, \bibinfo {author} {\bibnamefont {Petersen}, \bibfnamefont {A.~M.}}, \bibinfo {author} {\bibnamefont {Pammolli}, \bibfnamefont {F.}}, and\ \bibinfo {author} {\bibnamefont {Fortunato}, \bibfnamefont {S.}},\ }\bibfield  {title} {\enquote {\bibinfo {title} {The memory of science: Inflation, myopia, and the knowledge network},}\ }\href {https://doi.org/10.1016/j.joi.2018.06.005} {\bibfield  {journal} {\bibinfo  {journal} {J. Informetr.}\ }\textbf {\bibinfo {volume} {12}},\ \bibinfo {pages} {656--678} (\bibinfo {year} {2018})}\BibitemShut {NoStop}%
\bibitem [{\citenamefont {Parolo}\ \emph {et~al.}(2015)\citenamefont {Parolo}, \citenamefont {Pan}, \citenamefont {Ghosh}, \citenamefont {Huberman}, \citenamefont {Kaski},\ and\ \citenamefont {Fortunato}}]{Parolo2015-iq}%
  \BibitemOpen
  \bibfield  {author} {\bibinfo {author} {\bibnamefont {Parolo}, \bibfnamefont {P.~D.~B.}}, \bibinfo {author} {\bibnamefont {Pan}, \bibfnamefont {R.~K.}}, \bibinfo {author} {\bibnamefont {Ghosh}, \bibfnamefont {R.}}, \bibinfo {author} {\bibnamefont {Huberman}, \bibfnamefont {B.~A.}}, \bibinfo {author} {\bibnamefont {Kaski}, \bibfnamefont {K.}}, and\ \bibinfo {author} {\bibnamefont {Fortunato}, \bibfnamefont {S.}},\ }\bibfield  {title} {\enquote {\bibinfo {title} {Attention decay in science},}\ }\href {https://doi.org/10.1016/j.joi.2015.07.006} {\bibfield  {journal} {\bibinfo  {journal} {J. Informetr.}\ }\textbf {\bibinfo {volume} {9}},\ \bibinfo {pages} {734--745} (\bibinfo {year} {2015})}\BibitemShut {NoStop}%
\bibitem [{\citenamefont {Penner}\ \emph {et~al.}(2013)\citenamefont {Penner}, \citenamefont {Pan}, \citenamefont {Petersen}, \citenamefont {Kaski},\ and\ \citenamefont {Fortunato}}]{Penner2013-dt}%
  \BibitemOpen
  \bibfield  {author} {\bibinfo {author} {\bibnamefont {Penner}, \bibfnamefont {O.}}, \bibinfo {author} {\bibnamefont {Pan}, \bibfnamefont {R.~K.}}, \bibinfo {author} {\bibnamefont {Petersen}, \bibfnamefont {A.~M.}}, \bibinfo {author} {\bibnamefont {Kaski}, \bibfnamefont {K.}}, and\ \bibinfo {author} {\bibnamefont {Fortunato}, \bibfnamefont {S.}},\ }\bibfield  {title} {\enquote {\bibinfo {title} {On the predictability of future impact in science},}\ }\href {https://doi.org/10.1038/srep03052} {\bibfield  {journal} {\bibinfo  {journal} {Sci. Rep.}\ }\textbf {\bibinfo {volume} {3}} (\bibinfo {year} {2013}),\ 10.1038/srep03052}\BibitemShut {NoStop}%
\bibitem [{\citenamefont {Peritz}(1995)}]{Peritz1995-eo}%
  \BibitemOpen
  \bibfield  {author} {\bibinfo {author} {\bibnamefont {Peritz}, \bibfnamefont {B.~C.}},\ }\bibfield  {title} {\enquote {\bibinfo {title} {On the association between journal circulation and impact factor},}\ }\href {https://doi.org/10.1177/016555159502100108} {\bibfield  {journal} {\bibinfo  {journal} {J. Inf. Sci. Eng.}\ }\textbf {\bibinfo {volume} {21}},\ \bibinfo {pages} {63--67} (\bibinfo {year} {1995})}\BibitemShut {NoStop}%
\bibitem [{\citenamefont {Perneger}(2010)}]{Perneger2010-wp}%
  \BibitemOpen
  \bibfield  {author} {\bibinfo {author} {\bibnamefont {Perneger}, \bibfnamefont {T.~V.}},\ }\bibfield  {title} {\enquote {\bibinfo {title} {Citation analysis of identical consensus statements revealed journal-related bias},}\ }\href {https://doi.org/10.1016/j.jclinepi.2009.09.012} {\bibfield  {journal} {\bibinfo  {journal} {J. Clin. Epidemiol.}\ }\textbf {\bibinfo {volume} {63}},\ \bibinfo {pages} {660--664} (\bibinfo {year} {2010})}\BibitemShut {NoStop}%
\bibitem [{\citenamefont {Peterson}, \citenamefont {Press{\'e}},\ and\ \citenamefont {Dill}(2010)}]{Peterson2010-lx}%
  \BibitemOpen
  \bibfield  {author} {\bibinfo {author} {\bibnamefont {Peterson}, \bibfnamefont {G.~J.}}, \bibinfo {author} {\bibnamefont {Press{\'e}}, \bibfnamefont {S.}}, and\ \bibinfo {author} {\bibnamefont {Dill}, \bibfnamefont {K.~A.}},\ }\bibfield  {title} {\enquote {\bibinfo {title} {Nonuniversal power law scaling in the probability distribution of scientific citations},}\ }\href {https://doi.org/10.1073/pnas.1010757107} {\bibfield  {journal} {\bibinfo  {journal} {Proc. Natl. Acad. Sci. U. S. A.}\ }\textbf {\bibinfo {volume} {107}},\ \bibinfo {pages} {16023--16027} (\bibinfo {year} {2010})}\BibitemShut {NoStop}%
\bibitem [{\citenamefont {Phillips}\ \emph {et~al.}(1991)\citenamefont {Phillips}, \citenamefont {Kanter}, \citenamefont {Bednarczyk},\ and\ \citenamefont {Tastad}}]{Phillips1991-md}%
  \BibitemOpen
  \bibfield  {author} {\bibinfo {author} {\bibnamefont {Phillips}, \bibfnamefont {D.~P.}}, \bibinfo {author} {\bibnamefont {Kanter}, \bibfnamefont {E.~J.}}, \bibinfo {author} {\bibnamefont {Bednarczyk}, \bibfnamefont {B.}}, and\ \bibinfo {author} {\bibnamefont {Tastad}, \bibfnamefont {P.~L.}},\ }\bibfield  {title} {\enquote {\bibinfo {title} {Importance of the lay press in the transmission of medical knowledge to the scientific community},}\ }\href {https://doi.org/10.1056/NEJM199110173251620} {\bibfield  {journal} {\bibinfo  {journal} {N. Engl. J. Med.}\ }\textbf {\bibinfo {volume} {325}},\ \bibinfo {pages} {1180--1183} (\bibinfo {year} {1991})}\BibitemShut {NoStop}%
\bibitem [{\citenamefont {Pier}\ \emph {et~al.}(2018)\citenamefont {Pier}, \citenamefont {Brauer}, \citenamefont {Filut}, \citenamefont {Kaatz}, \citenamefont {Raclaw}, \citenamefont {Nathan}, \citenamefont {Ford},\ and\ \citenamefont {Carnes}}]{Pier2018-rj}%
  \BibitemOpen
  \bibfield  {author} {\bibinfo {author} {\bibnamefont {Pier}, \bibfnamefont {E.~L.}}, \bibinfo {author} {\bibnamefont {Brauer}, \bibfnamefont {M.}}, \bibinfo {author} {\bibnamefont {Filut}, \bibfnamefont {A.}}, \bibinfo {author} {\bibnamefont {Kaatz}, \bibfnamefont {A.}}, \bibinfo {author} {\bibnamefont {Raclaw}, \bibfnamefont {J.}}, \bibinfo {author} {\bibnamefont {Nathan}, \bibfnamefont {M.~J.}}, \bibinfo {author} {\bibnamefont {Ford}, \bibfnamefont {C.~E.}}, and\ \bibinfo {author} {\bibnamefont {Carnes}, \bibfnamefont {M.}},\ }\bibfield  {title} {\enquote {\bibinfo {title} {Low agreement among reviewers evaluating the same {NIH} grant applications},}\ }\href {https://doi.org/10.1073/pnas.1714379115} {\bibfield  {journal} {\bibinfo  {journal} {Proc. Natl. Acad. Sci. U. S. A.}\ }\textbf {\bibinfo {volume} {115}},\ \bibinfo {pages} {2952--2957} (\bibinfo {year} {2018})}\BibitemShut {NoStop}%
\bibitem [{\citenamefont {Poncela-Casasnovas}\ \emph {et~al.}(2019)\citenamefont {Poncela-Casasnovas}, \citenamefont {Gerlach}, \citenamefont {Aguirre},\ and\ \citenamefont {Amaral}}]{Poncela-Casasnovas2019-gf}%
  \BibitemOpen
  \bibfield  {author} {\bibinfo {author} {\bibnamefont {Poncela-Casasnovas}, \bibfnamefont {J.}}, \bibinfo {author} {\bibnamefont {Gerlach}, \bibfnamefont {M.}}, \bibinfo {author} {\bibnamefont {Aguirre}, \bibfnamefont {N.}}, and\ \bibinfo {author} {\bibnamefont {Amaral}, \bibfnamefont {L.~A.~N.}},\ }\bibfield  {title} {\enquote {\bibinfo {title} {Large-scale analysis of micro-level citation patterns reveals nuanced selection criteria},}\ }\href {https://doi.org/10.1038/s41562-019-0585-7} {\bibfield  {journal} {\bibinfo  {journal} {Nature Human Behaviour}\ }\textbf {\bibinfo {volume} {3}},\ \bibinfo {pages} {568--575} (\bibinfo {year} {2019})}\BibitemShut {NoStop}%
\bibitem [{\citenamefont {Price}(1976)}]{Price1976-rb}%
  \BibitemOpen
  \bibfield  {author} {\bibinfo {author} {\bibnamefont {Price}, \bibfnamefont {D.~D.~S.}},\ }\bibfield  {title} {\enquote {\bibinfo {title} {A general theory of bibliometric and other cumulative advantage processes},}\ }\href {https://doi.org/10.1002/asi.4630270505} {\bibfield  {journal} {\bibinfo  {journal} {Journal of the American Society for Information Science}\ }\textbf {\bibinfo {volume} {27}},\ \bibinfo {pages} {292--306} (\bibinfo {year} {1976})}\BibitemShut {NoStop}%
\bibitem [{\citenamefont {Price}(1965)}]{Price1965-gg}%
  \BibitemOpen
  \bibfield  {author} {\bibinfo {author} {\bibnamefont {Price}, \bibfnamefont {D.~J.}},\ }\bibfield  {title} {\enquote {\bibinfo {title} {{NETWORKS} {OF} {SCIENTIFIC} {PAPERS}},}\ }\href {https://doi.org/10.1126/science.149.3683.510} {\bibfield  {journal} {\bibinfo  {journal} {Science}\ }\textbf {\bibinfo {volume} {149}},\ \bibinfo {pages} {510--515} (\bibinfo {year} {1965})}\BibitemShut {NoStop}%
\bibitem [{\citenamefont {Radicchi}\ and\ \citenamefont {Castellano}(2011)}]{Radicchi2011-wx}%
  \BibitemOpen
  \bibfield  {author} {\bibinfo {author} {\bibnamefont {Radicchi}, \bibfnamefont {F.}}and\ \bibinfo {author} {\bibnamefont {Castellano}, \bibfnamefont {C.}},\ }\bibfield  {title} {\enquote {\bibinfo {title} {Rescaling citations of publications in physics},}\ }\href {https://doi.org/10.1103/PhysRevE.83.046116} {\bibfield  {journal} {\bibinfo  {journal} {Physical Review E}\ }\textbf {\bibinfo {volume} {83}},\ \bibinfo {pages} {046116} (\bibinfo {year} {2011})}\BibitemShut {NoStop}%
\bibitem [{\citenamefont {Radicchi}\ and\ \citenamefont {Castellano}(2012)}]{Radicchi2012-vo}%
  \BibitemOpen
  \bibfield  {author} {\bibinfo {author} {\bibnamefont {Radicchi}, \bibfnamefont {F.}}and\ \bibinfo {author} {\bibnamefont {Castellano}, \bibfnamefont {C.}},\ }\bibfield  {title} {\enquote {\bibinfo {title} {A reverse engineering approach to the suppression of citation biases reveals universal properties of citation distributions},}\ }\href {https://doi.org/10.1371/journal.pone.0033833} {\bibfield  {journal} {\bibinfo  {journal} {PLoS One}\ }\textbf {\bibinfo {volume} {7}},\ \bibinfo {pages} {e33833} (\bibinfo {year} {2012})}\BibitemShut {NoStop}%
\bibitem [{\citenamefont {Radicchi}, \citenamefont {Fortunato},\ and\ \citenamefont {Castellano}(2008)}]{Radicchi2008-ju}%
  \BibitemOpen
  \bibfield  {author} {\bibinfo {author} {\bibnamefont {Radicchi}, \bibfnamefont {F.}}, \bibinfo {author} {\bibnamefont {Fortunato}, \bibfnamefont {S.}}, and\ \bibinfo {author} {\bibnamefont {Castellano}, \bibfnamefont {C.}},\ }\bibfield  {title} {\enquote {\bibinfo {title} {Universality of citation distributions: toward an objective measure of scientific impact},}\ }\href {https://doi.org/10.1073/pnas.0806977105} {\bibfield  {journal} {\bibinfo  {journal} {Proc. Natl. Acad. Sci. U. S. A.}\ }\textbf {\bibinfo {volume} {105}},\ \bibinfo {pages} {17268--17272} (\bibinfo {year} {2008})}\BibitemShut {NoStop}%
\bibitem [{\citenamefont {Radicchi}, \citenamefont {Weissman},\ and\ \citenamefont {Bollen}(2017)}]{Radicchi2017-jk}%
  \BibitemOpen
  \bibfield  {author} {\bibinfo {author} {\bibnamefont {Radicchi}, \bibfnamefont {F.}}, \bibinfo {author} {\bibnamefont {Weissman}, \bibfnamefont {A.}}, and\ \bibinfo {author} {\bibnamefont {Bollen}, \bibfnamefont {J.}},\ }\bibfield  {title} {\enquote {\bibinfo {title} {Quantifying perceived impact of scientific publications},}\ }\href {https://doi.org/10.1016/j.joi.2017.05.010} {\bibfield  {journal} {\bibinfo  {journal} {J. Informetr.}\ }\textbf {\bibinfo {volume} {11}},\ \bibinfo {pages} {704--712} (\bibinfo {year} {2017})}\BibitemShut {NoStop}%
\bibitem [{\citenamefont {Redner}(1998)}]{Redner1998-wy}%
  \BibitemOpen
  \bibfield  {author} {\bibinfo {author} {\bibnamefont {Redner}, \bibfnamefont {S.}},\ }\bibfield  {title} {\enquote {\bibinfo {title} {How popular is your paper? an empirical study of the citation distribution},}\ }\href {https://doi.org/10.1007/s100510050359} {\bibfield  {journal} {\bibinfo  {journal} {Eur. Phys. J. B}\ }\textbf {\bibinfo {volume} {4}},\ \bibinfo {pages} {131--134} (\bibinfo {year} {1998})}\BibitemShut {NoStop}%
\bibitem [{\citenamefont {Redner}(2005)}]{Redner2005-qk}%
  \BibitemOpen
  \bibfield  {author} {\bibinfo {author} {\bibnamefont {Redner}, \bibfnamefont {S.}},\ }\bibfield  {title} {\enquote {\bibinfo {title} {Citation statistics from 110 years of physical review},}\ }\href {https://doi.org/10.1063/1.1996475} {\bibfield  {journal} {\bibinfo  {journal} {Phys. Today}\ }\textbf {\bibinfo {volume} {58}},\ \bibinfo {pages} {49--54} (\bibinfo {year} {2005})}\BibitemShut {NoStop}%
\bibitem [{\citenamefont {de~Rijcke}\ \emph {et~al.}(2016)\citenamefont {de~Rijcke}, \citenamefont {Wouters}, \citenamefont {Rushforth}, \citenamefont {Franssen},\ and\ \citenamefont {Hammarfelt}}]{De_Rijcke2016-kf}%
  \BibitemOpen
  \bibfield  {author} {\bibinfo {author} {\bibnamefont {de~Rijcke}, \bibfnamefont {S.}}, \bibinfo {author} {\bibnamefont {Wouters}, \bibfnamefont {P.~F.}}, \bibinfo {author} {\bibnamefont {Rushforth}, \bibfnamefont {A.~D.}}, \bibinfo {author} {\bibnamefont {Franssen}, \bibfnamefont {T.~P.}}, and\ \bibinfo {author} {\bibnamefont {Hammarfelt}, \bibfnamefont {B.}},\ }\bibfield  {title} {\enquote {\bibinfo {title} {Evaluation practices and effects of indicator use---a literature review},}\ }\href {https://doi.org/10.1093/reseval/rvv038} {\bibfield  {journal} {\bibinfo  {journal} {Res. Eval.}\ }\textbf {\bibinfo {volume} {25}},\ \bibinfo {pages} {161--169} (\bibinfo {year} {2016})}\BibitemShut {NoStop}%
\bibitem [{\citenamefont {Van~de Rijt}\ \emph {et~al.}(2014)\citenamefont {Van~de Rijt}, \citenamefont {Kang}, \citenamefont {Restivo},\ and\ \citenamefont {Patil}}]{Van_de_Rijt2014-yx}%
  \BibitemOpen
  \bibfield  {author} {\bibinfo {author} {\bibnamefont {Van~de Rijt}, \bibfnamefont {A.}}, \bibinfo {author} {\bibnamefont {Kang}, \bibfnamefont {S.~M.}}, \bibinfo {author} {\bibnamefont {Restivo}, \bibfnamefont {M.}}, and\ \bibinfo {author} {\bibnamefont {Patil}, \bibfnamefont {A.}},\ }\bibfield  {title} {\enquote {\bibinfo {title} {Field experiments of success-breeds-success dynamics},}\ }\href {https://doi.org/10.1073/pnas.1316836111} {\bibfield  {journal} {\bibinfo  {journal} {Proc. Natl. Acad. Sci. U. S. A.}\ }\textbf {\bibinfo {volume} {111}},\ \bibinfo {pages} {6934--6939} (\bibinfo {year} {2014})}\BibitemShut {NoStop}%
\bibitem [{\citenamefont {Rothwell}\ and\ \citenamefont {Martyn}(2000)}]{Rothwell2000-oe}%
  \BibitemOpen
  \bibfield  {author} {\bibinfo {author} {\bibnamefont {Rothwell}, \bibfnamefont {P.~M.}}and\ \bibinfo {author} {\bibnamefont {Martyn}, \bibfnamefont {C.~N.}},\ }\bibfield  {title} {\enquote {\bibinfo {title} {Reproducibility of peer review in clinical neuroscience. is agreement between reviewers any greater than would be expected by chance alone?}}\ }\href {https://doi.org/10.1093/brain/123.9.1964} {\bibfield  {journal} {\bibinfo  {journal} {Brain}\ }\textbf {\bibinfo {volume} {123 ( Pt 9)}},\ \bibinfo {pages} {1964--1969} (\bibinfo {year} {2000})}\BibitemShut {NoStop}%
\bibitem [{\citenamefont {Rushforth}\ and\ \citenamefont {de~Rijcke}(2015)}]{Rushforth2015-vg}%
  \BibitemOpen
  \bibfield  {author} {\bibinfo {author} {\bibnamefont {Rushforth}, \bibfnamefont {A.}}and\ \bibinfo {author} {\bibnamefont {de~Rijcke}, \bibfnamefont {S.}},\ }\bibfield  {title} {\enquote {\bibinfo {title} {Accounting for impact? the journal impact factor and the making of biomedical research in the netherlands},}\ }\href {https://doi.org/10.1007/s11024-015-9274-5} {\bibfield  {journal} {\bibinfo  {journal} {Minerva}\ }\textbf {\bibinfo {volume} {53}},\ \bibinfo {pages} {117--139} (\bibinfo {year} {2015})}\BibitemShut {NoStop}%
\bibitem [{\citenamefont {Sandstr{\"o}m}\ and\ \citenamefont {Van~den Besselaar}(2018)}]{Sandstrom2018-mp}%
  \BibitemOpen
  \bibfield  {author} {\bibinfo {author} {\bibnamefont {Sandstr{\"o}m}, \bibfnamefont {U.}}and\ \bibinfo {author} {\bibnamefont {Van~den Besselaar}, \bibfnamefont {P.}},\ }\bibfield  {title} {\enquote {\bibinfo {title} {Funding, evaluation, and the performance of national research systems},}\ }\href {https://doi.org/10.1016/j.joi.2018.01.007} {\bibfield  {journal} {\bibinfo  {journal} {J. Informetr.}\ }\textbf {\bibinfo {volume} {12}},\ \bibinfo {pages} {365--384} (\bibinfo {year} {2018})}\BibitemShut {NoStop}%
\bibitem [{\citenamefont {Schneider}, \citenamefont {Aagaard},\ and\ \citenamefont {Bloch}(2016)}]{Schneider2016-fs}%
  \BibitemOpen
  \bibfield  {author} {\bibinfo {author} {\bibnamefont {Schneider}, \bibfnamefont {J.~W.}}, \bibinfo {author} {\bibnamefont {Aagaard}, \bibfnamefont {K.}}, and\ \bibinfo {author} {\bibnamefont {Bloch}, \bibfnamefont {C.~W.}},\ }\bibfield  {title} {\enquote {\bibinfo {title} {What happens when national research funding is linked to differentiated publication counts? a comparison of the australian and norwegian publication-based funding models},}\ }\href {https://doi.org/10.1093/reseval/rvv036} {\bibfield  {journal} {\bibinfo  {journal} {Res. Eval.}\ }\textbf {\bibinfo {volume} {25}},\ \bibinfo {pages} {244--256} (\bibinfo {year} {2016})}\BibitemShut {NoStop}%
\bibitem [{\citenamefont {Schubert}\ and\ \citenamefont {Gl{\"a}nzel}(2006)}]{Schubert2006-ng}%
  \BibitemOpen
  \bibfield  {author} {\bibinfo {author} {\bibnamefont {Schubert}, \bibfnamefont {A.}}and\ \bibinfo {author} {\bibnamefont {Gl{\"a}nzel}, \bibfnamefont {W.}},\ }\bibfield  {title} {\enquote {\bibinfo {title} {Cross-national preference in co-authorship, references and citations},}\ }\href {https://doi.org/10.1007/s11192-006-0160-7} {\bibfield  {journal} {\bibinfo  {journal} {Scientometrics}\ }\textbf {\bibinfo {volume} {69}},\ \bibinfo {pages} {409--428} (\bibinfo {year} {2006})}\BibitemShut {NoStop}%
\bibitem [{\citenamefont {Seeber}\ \emph {et~al.}(2017)\citenamefont {Seeber}, \citenamefont {Cattaneo}, \citenamefont {Meoli},\ and\ \citenamefont {Malighetti}}]{Seeber2017-wo}%
  \BibitemOpen
  \bibfield  {author} {\bibinfo {author} {\bibnamefont {Seeber}, \bibfnamefont {M.}}, \bibinfo {author} {\bibnamefont {Cattaneo}, \bibfnamefont {M.}}, \bibinfo {author} {\bibnamefont {Meoli}, \bibfnamefont {M.}}, and\ \bibinfo {author} {\bibnamefont {Malighetti}, \bibfnamefont {P.}},\ }\bibfield  {title} {\enquote {\bibinfo {title} {Self-citations as strategic response to the use of metrics for career decisions},}\ }\href {https://doi.org/10.1016/j.respol.2017.12.004} {\bibfield  {journal} {\bibinfo  {journal} {Res. Policy}\ } (\bibinfo {year} {2017}),\ 10.1016/j.respol.2017.12.004}\BibitemShut {NoStop}%
\bibitem [{\citenamefont {Seglen}(1994)}]{Seglen1994-jo}%
  \BibitemOpen
  \bibfield  {author} {\bibinfo {author} {\bibnamefont {Seglen}, \bibfnamefont {P.~O.}},\ }\bibfield  {title} {\enquote {\bibinfo {title} {Causal relationship between article citedness and journal impact},}\ }\href {https://doi.org/10.1002/(SICI)1097-4571(199401)45:1<1::AID-ASI1>3.0.CO;2-Y} {\bibfield  {journal} {\bibinfo  {journal} {Journal of the American Society for Information Science}\ }\textbf {\bibinfo {volume} {45}},\ \bibinfo {pages} {1--11} (\bibinfo {year} {1994})}\BibitemShut {NoStop}%
\bibitem [{\citenamefont {Seglen}(1997)}]{Seglen1997-sk}%
  \BibitemOpen
  \bibfield  {author} {\bibinfo {author} {\bibnamefont {Seglen}, \bibfnamefont {P.~O.}},\ }\bibfield  {title} {\enquote {\bibinfo {title} {Why the impact factor of journals should not be used for evaluating research},}\ }\href {https://doi.org/10.1136/bmj.314.7079.497} {\bibfield  {journal} {\bibinfo  {journal} {BMJ}\ }\textbf {\bibinfo {volume} {314}},\ \bibinfo {pages} {498--502} (\bibinfo {year} {1997})}\BibitemShut {NoStop}%
\bibitem [{\citenamefont {Simkin}\ and\ \citenamefont {Roychowdhury}(2007)}]{Simkin2007-jy}%
  \BibitemOpen
  \bibfield  {author} {\bibinfo {author} {\bibnamefont {Simkin}, \bibfnamefont {M.~V.}}and\ \bibinfo {author} {\bibnamefont {Roychowdhury}, \bibfnamefont {V.~P.}},\ }\bibfield  {title} {\enquote {\bibinfo {title} {A mathematical theory of citing},}\ }\href {https://doi.org/10.1002/asi.20653} {\bibfield  {journal} {\bibinfo  {journal} {J. Am. Soc. Inf. Sci. Technol.}\ }\textbf {\bibinfo {volume} {58}},\ \bibinfo {pages} {1661--1673} (\bibinfo {year} {2007})}\BibitemShut {NoStop}%
\bibitem [{\citenamefont {Sinatra}\ \emph {et~al.}(2015)\citenamefont {Sinatra}, \citenamefont {Deville}, \citenamefont {Szell}, \citenamefont {Wang},\ and\ \citenamefont {Barab{\'a}si}}]{Sinatra2015-ag}%
  \BibitemOpen
  \bibfield  {author} {\bibinfo {author} {\bibnamefont {Sinatra}, \bibfnamefont {R.}}, \bibinfo {author} {\bibnamefont {Deville}, \bibfnamefont {P.}}, \bibinfo {author} {\bibnamefont {Szell}, \bibfnamefont {M.}}, \bibinfo {author} {\bibnamefont {Wang}, \bibfnamefont {D.}}, and\ \bibinfo {author} {\bibnamefont {Barab{\'a}si}, \bibfnamefont {A.~L.}},\ }\bibfield  {title} {\enquote {\bibinfo {title} {A century of physics},}\ }\href {https://doi.org/10.1038/nphys3494} {\bibfield  {journal} {\bibinfo  {journal} {Nat. Phys.}\ }\textbf {\bibinfo {volume} {11}},\ \bibinfo {pages} {791--796} (\bibinfo {year} {2015})}\BibitemShut {NoStop}%
\bibitem [{\citenamefont {Sinatra}\ \emph {et~al.}(2016)\citenamefont {Sinatra}, \citenamefont {Wang}, \citenamefont {Deville}, \citenamefont {Song},\ and\ \citenamefont {Barab{\'a}si}}]{Sinatra2016-up}%
  \BibitemOpen
  \bibfield  {author} {\bibinfo {author} {\bibnamefont {Sinatra}, \bibfnamefont {R.}}, \bibinfo {author} {\bibnamefont {Wang}, \bibfnamefont {D.}}, \bibinfo {author} {\bibnamefont {Deville}, \bibfnamefont {P.}}, \bibinfo {author} {\bibnamefont {Song}, \bibfnamefont {C.}}, and\ \bibinfo {author} {\bibnamefont {Barab{\'a}si}, \bibfnamefont {A.-L.}},\ }\bibfield  {title} {\enquote {\bibinfo {title} {Quantifying the evolution of individual scientific impact},}\ }\href {https://doi.org/10.1126/science.aaf5239} {\bibfield  {journal} {\bibinfo  {journal} {Science}\ }\textbf {\bibinfo {volume} {354}} (\bibinfo {year} {2016}),\ 10.1126/science.aaf5239}\BibitemShut {NoStop}%
\bibitem [{\citenamefont {Smaldino}\ and\ \citenamefont {McElreath}(2016)}]{Smaldino2016-io}%
  \BibitemOpen
  \bibfield  {author} {\bibinfo {author} {\bibnamefont {Smaldino}, \bibfnamefont {P.~E.}}and\ \bibinfo {author} {\bibnamefont {McElreath}, \bibfnamefont {R.}},\ }\bibfield  {title} {\enquote {\bibinfo {title} {The natural selection of bad science},}\ }\href {https://doi.org/10.1098/rsos.160384} {\bibfield  {journal} {\bibinfo  {journal} {Royal Society Open Science}\ }\textbf {\bibinfo {volume} {3}},\ \bibinfo {pages} {160384} (\bibinfo {year} {2016})}\BibitemShut {NoStop}%
\bibitem [{\citenamefont {Starbuck}(2005)}]{Starbuck2005-ss}%
  \BibitemOpen
  \bibfield  {author} {\bibinfo {author} {\bibnamefont {Starbuck}, \bibfnamefont {W.~H.}},\ }\bibfield  {title} {\enquote {\bibinfo {title} {How much better are the {Most-Prestigious} journals? the statistics of academic publication},}\ }\href {https://doi.org/10.1287/orsc.1040.0107} {\bibfield  {journal} {\bibinfo  {journal} {Organization Science}\ }\textbf {\bibinfo {volume} {16}},\ \bibinfo {pages} {180--200} (\bibinfo {year} {2005})}\BibitemShut {NoStop}%
\bibitem [{\citenamefont {Stegehuis}, \citenamefont {Litvak},\ and\ \citenamefont {Waltman}(2015)}]{Stegehuis2015-tg}%
  \BibitemOpen
  \bibfield  {author} {\bibinfo {author} {\bibnamefont {Stegehuis}, \bibfnamefont {C.}}, \bibinfo {author} {\bibnamefont {Litvak}, \bibfnamefont {N.}}, and\ \bibinfo {author} {\bibnamefont {Waltman}, \bibfnamefont {L.}},\ }\bibfield  {title} {\enquote {\bibinfo {title} {Predicting the long-term citation impact of recent publications},}\ }\href {https://doi.org/10.1016/j.joi.2015.06.005} {\bibfield  {journal} {\bibinfo  {journal} {J. Informetr.}\ }\textbf {\bibinfo {volume} {9}},\ \bibinfo {pages} {642--657} (\bibinfo {year} {2015})}\BibitemShut {NoStop}%
\bibitem [{\citenamefont {Stinson}\ and\ \citenamefont {Lancaster}(1987)}]{Stinson1987-wk}%
  \BibitemOpen
  \bibfield  {author} {\bibinfo {author} {\bibnamefont {Stinson}, \bibfnamefont {E.~R.}}and\ \bibinfo {author} {\bibnamefont {Lancaster}, \bibfnamefont {F.~W.}},\ }\bibfield  {title} {\enquote {\bibinfo {title} {Synchronous versus diachronous methods in the measurement of obsolescence by citation studies},}\ }\href {https://doi.org/10.1177/016555158701300201} {\bibfield  {journal} {\bibinfo  {journal} {J. Inf. Sci. Eng.}\ }\textbf {\bibinfo {volume} {13}},\ \bibinfo {pages} {65--74} (\bibinfo {year} {1987})}\BibitemShut {NoStop}%
\bibitem [{\citenamefont {Stringer}, \citenamefont {Sales-Pardo},\ and\ \citenamefont {Amaral}(2005)}]{Stringer2005-em}%
  \BibitemOpen
  \bibfield  {author} {\bibinfo {author} {\bibnamefont {Stringer}, \bibfnamefont {M.~J.}}, \bibinfo {author} {\bibnamefont {Sales-Pardo}, \bibfnamefont {M.}}, and\ \bibinfo {author} {\bibnamefont {Amaral}, \bibfnamefont {L.~A.~N.}},\ }\bibfield  {title} {\enquote {\bibinfo {title} {Effectiveness of journal ranking schemes as a tool for locating information},}\ }\href {https://doi.org/10.1371/journal.pone.0001683} {\bibfield  {journal} {\bibinfo  {journal} {PLoS One}\ }\textbf {\bibinfo {volume} {3}},\ \bibinfo {pages} {e1683} (\bibinfo {year} {2005})}\BibitemShut {NoStop}%
\bibitem [{\citenamefont {{\v S}ubelj}\ and\ \citenamefont {Fiala}(2017)}]{Subelj2017-jr}%
  \BibitemOpen
  \bibfield  {author} {\bibinfo {author} {\bibnamefont {{\v S}ubelj}, \bibfnamefont {L.}}and\ \bibinfo {author} {\bibnamefont {Fiala}, \bibfnamefont {D.}},\ }\bibfield  {title} {\enquote {\bibinfo {title} {Publication boost in web of science journals and its effect on citation distributions},}\ }\href {https://doi.org/10.1002/asi.23718} {\bibfield  {journal} {\bibinfo  {journal} {J. Assoc. Inf. Sci. Technol.}\ }\textbf {\bibinfo {volume} {68}},\ \bibinfo {pages} {1018--1023} (\bibinfo {year} {2017})}\BibitemShut {NoStop}%
\bibitem [{\citenamefont {Sugimoto}\ and\ \citenamefont {Larivi{\`e}re}(2018)}]{Sugimoto2018-yr}%
  \BibitemOpen
  \bibfield  {author} {\bibinfo {author} {\bibnamefont {Sugimoto}, \bibfnamefont {C.~R.}}and\ \bibinfo {author} {\bibnamefont {Larivi{\`e}re}, \bibfnamefont {V.}},\ }\href@noop {} {\emph {\bibinfo {title} {Measuring Research: What Everyone Needs to Know}}}\ (\bibinfo  {publisher} {Oxford University Press},\ \bibinfo {year} {2018})\ pp.\ \bibinfo {pages} {1--143}\BibitemShut {NoStop}%
\bibitem [{\citenamefont {Teplitskiy}\ \emph {et~al.}(2020)\citenamefont {Teplitskiy}, \citenamefont {Duede}, \citenamefont {Menietti},\ and\ \citenamefont {Lakhani}}]{Teplitskiy2020-xv}%
  \BibitemOpen
  \bibfield  {author} {\bibinfo {author} {\bibnamefont {Teplitskiy}, \bibfnamefont {M.}}, \bibinfo {author} {\bibnamefont {Duede}, \bibfnamefont {E.}}, \bibinfo {author} {\bibnamefont {Menietti}, \bibfnamefont {M.}}, and\ \bibinfo {author} {\bibnamefont {Lakhani}, \bibfnamefont {K.~R.}},\ }\bibfield  {title} {\enquote {\bibinfo {title} {Status drives how we cite: Evidence from thousands of authors},}\ }\href@noop {} {\ ,\ \bibinfo {pages} {20} (\bibinfo {year} {2020})},\ \Eprint {https://arxiv.org/abs/2002.10033} {arXiv:2002.10033 [cs.SI]} \BibitemShut {NoStop}%
\bibitem [{\citenamefont {Thelwall}\ and\ \citenamefont {Wilson}(2014)}]{Thelwall2014-gk}%
  \BibitemOpen
  \bibfield  {author} {\bibinfo {author} {\bibnamefont {Thelwall}, \bibfnamefont {M.}}and\ \bibinfo {author} {\bibnamefont {Wilson}, \bibfnamefont {P.}},\ }\bibfield  {title} {\enquote {\bibinfo {title} {Regression for citation data: An evaluation of different methods},}\ }\href {https://doi.org/10.1016/j.joi.2014.09.011} {\bibfield  {journal} {\bibinfo  {journal} {J. Informetr.}\ }\textbf {\bibinfo {volume} {8}},\ \bibinfo {pages} {963--971} (\bibinfo {year} {2014})}\BibitemShut {NoStop}%
\bibitem [{\citenamefont {Tiokhin}\ \emph {et~al.}(2021)\citenamefont {Tiokhin}, \citenamefont {Panchanathan}, \citenamefont {Smaldino},\ and\ \citenamefont {Lakens}}]{Tiokhin2021-ch}%
  \BibitemOpen
  \bibfield  {author} {\bibinfo {author} {\bibnamefont {Tiokhin}, \bibfnamefont {L.}}, \bibinfo {author} {\bibnamefont {Panchanathan}, \bibfnamefont {K.}}, \bibinfo {author} {\bibnamefont {Smaldino}, \bibfnamefont {P.~E.}}, and\ \bibinfo {author} {\bibnamefont {Lakens}, \bibfnamefont {D.}},\ }\href {https://doi.org/10.31222/osf.io/juwck} {\enquote {\bibinfo {title} {Shifting the level of selection in science},}\ } (\bibinfo {year} {2021})\BibitemShut {NoStop}%
\bibitem [{\citenamefont {Tomkins}, \citenamefont {Zhang},\ and\ \citenamefont {Heavlin}(2017)}]{Tomkins2017-lb}%
  \BibitemOpen
  \bibfield  {author} {\bibinfo {author} {\bibnamefont {Tomkins}, \bibfnamefont {A.}}, \bibinfo {author} {\bibnamefont {Zhang}, \bibfnamefont {M.}}, and\ \bibinfo {author} {\bibnamefont {Heavlin}, \bibfnamefont {W.~D.}},\ }\bibfield  {title} {\enquote {\bibinfo {title} {Reviewer bias in single- versus double-blind peer review},}\ }\href {https://doi.org/10.1073/pnas.1707323114} {\bibfield  {journal} {\bibinfo  {journal} {Proc. Natl. Acad. Sci. U. S. A.}\ }\textbf {\bibinfo {volume} {114}},\ \bibinfo {pages} {12708--12713} (\bibinfo {year} {2017})}\BibitemShut {NoStop}%
\bibitem [{\citenamefont {Traag}(2021)}]{Traag2021-rq}%
  \BibitemOpen
  \bibfield  {author} {\bibinfo {author} {\bibnamefont {Traag}, \bibfnamefont {V.~A.}},\ }\bibfield  {title} {\enquote {\bibinfo {title} {Inferring the causal effect of journals on citations},}\ }\href {https://doi.org/10.1162/qss\_a\_00128} {\bibfield  {journal} {\bibinfo  {journal} {Quantitative Science Studies}\ ,\ \bibinfo {pages} {1--9}} (\bibinfo {year} {2021})}\BibitemShut {NoStop}%
\bibitem [{\citenamefont {Traag}, \citenamefont {Malgarini},\ and\ \citenamefont {Sarlo}(2020)}]{Traag2020-vv}%
  \BibitemOpen
  \bibfield  {author} {\bibinfo {author} {\bibnamefont {Traag}, \bibfnamefont {V.~A.}}, \bibinfo {author} {\bibnamefont {Malgarini}, \bibfnamefont {M.}}, and\ \bibinfo {author} {\bibnamefont {Sarlo}, \bibfnamefont {S.}},\ }\bibfield  {title} {\enquote {\bibinfo {title} {Metrics and peer review agreement at the institutional level},}\ }\href@noop {} {\  (\bibinfo {year} {2020})},\ \Eprint {https://arxiv.org/abs/2006.14830} {arXiv:2006.14830 [cs.DL]} \BibitemShut {NoStop}%
\bibitem [{\citenamefont {Traag}\ and\ \citenamefont {Waltman}(2019)}]{Traag2019-ul}%
  \BibitemOpen
  \bibfield  {author} {\bibinfo {author} {\bibnamefont {Traag}, \bibfnamefont {V.~A.}}and\ \bibinfo {author} {\bibnamefont {Waltman}, \bibfnamefont {L.}},\ }\bibfield  {title} {\enquote {\bibinfo {title} {Systematic analysis of agreement between metrics and peer review in the {UK} {REF}},}\ }\href {https://doi.org/10.1057/s41599-019-0233-x} {\bibfield  {journal} {\bibinfo  {journal} {Palgrave Communications}\ }\textbf {\bibinfo {volume} {5}},\ \bibinfo {pages} {29} (\bibinfo {year} {2019})}\BibitemShut {NoStop}%
\bibitem [{\citenamefont {Traag}\ and\ \citenamefont {Waltman}(2022)}]{traag_causal_2022}%
  \BibitemOpen
  \bibfield  {author} {\bibinfo {author} {\bibnamefont {Traag}, \bibfnamefont {V.~A.}}and\ \bibinfo {author} {\bibnamefont {Waltman}, \bibfnamefont {L.}},\ }\bibfield  {title} {\enquote {\bibinfo {title} {Causal foundations of bias, disparity and fairness},}\ }\href {https://doi.org/10.48550/arXiv.2207.13665} {\bibfield  {journal} {\bibinfo  {journal} {arXiv}\ } (\bibinfo {year} {2022}),\ 10.48550/arXiv.2207.13665}\BibitemShut {NoStop}%
\bibitem [{\citenamefont {Tsallis}\ and\ \citenamefont {de~Albuquerque}(2000)}]{Tsallis2000-qu}%
  \BibitemOpen
  \bibfield  {author} {\bibinfo {author} {\bibnamefont {Tsallis}, \bibfnamefont {C.}}and\ \bibinfo {author} {\bibnamefont {de~Albuquerque}, \bibfnamefont {M.~P.}},\ }\bibfield  {title} {\enquote {\bibinfo {title} {Are citations of scientific papers a case of nonextensivity?}}\ }\href {https://doi.org/10.1007/s100510050097} {\bibfield  {journal} {\bibinfo  {journal} {Eur. Phys. J. B}\ }\textbf {\bibinfo {volume} {13}},\ \bibinfo {pages} {777--780} (\bibinfo {year} {2000})}\BibitemShut {NoStop}%
\bibitem [{\citenamefont {Verstak}\ \emph {et~al.}(2014)\citenamefont {Verstak}, \citenamefont {Acharya}, \citenamefont {Suzuki}, \citenamefont {Henderson}, \citenamefont {Iakhiaev}, \citenamefont {Lin},\ and\ \citenamefont {Shetty}}]{Verstak2014-hc}%
  \BibitemOpen
  \bibfield  {author} {\bibinfo {author} {\bibnamefont {Verstak}, \bibfnamefont {A.}}, \bibinfo {author} {\bibnamefont {Acharya}, \bibfnamefont {A.}}, \bibinfo {author} {\bibnamefont {Suzuki}, \bibfnamefont {H.}}, \bibinfo {author} {\bibnamefont {Henderson}, \bibfnamefont {S.}}, \bibinfo {author} {\bibnamefont {Iakhiaev}, \bibfnamefont {M.}}, \bibinfo {author} {\bibnamefont {Lin}, \bibfnamefont {C.~C.~Y.}}, and\ \bibinfo {author} {\bibnamefont {Shetty}, \bibfnamefont {N.}},\ }\bibfield  {title} {\enquote {\bibinfo {title} {On the shoulders of giants: The growing impact of older articles},}\ }\href@noop {} {\  (\bibinfo {year} {2014})},\ \Eprint {https://arxiv.org/abs/1411.0275} {arXiv:1411.0275 [cs.DL]} \BibitemShut {NoStop}%
\bibitem [{\citenamefont {Vinkler}(1996)}]{Vinkler1996-eb}%
  \BibitemOpen
  \bibfield  {author} {\bibinfo {author} {\bibnamefont {Vinkler}, \bibfnamefont {P.}},\ }\bibfield  {title} {\enquote {\bibinfo {title} {Relationships between the rate of scientific development and citations. the chance for citedness model},}\ }\href {https://doi.org/10.1007/BF02016908} {\bibfield  {journal} {\bibinfo  {journal} {Scientometrics}\ }\textbf {\bibinfo {volume} {35}},\ \bibinfo {pages} {375--386} (\bibinfo {year} {1996})}\BibitemShut {NoStop}%
\bibitem [{\citenamefont {Visser}, \citenamefont {Van~Eck},\ and\ \citenamefont {Waltman}(2021)}]{Visser2021-ns}%
  \BibitemOpen
  \bibfield  {author} {\bibinfo {author} {\bibnamefont {Visser}, \bibfnamefont {M.}}, \bibinfo {author} {\bibnamefont {Van~Eck}, \bibfnamefont {N.~J.}}, and\ \bibinfo {author} {\bibnamefont {Waltman}, \bibfnamefont {L.}},\ }\bibfield  {title} {\enquote {\bibinfo {title} {Large-scale comparison of bibliographic data sources: Scopus, web of science, dimensions, crossref, and microsoft academic},}\ }\href {https://doi.org/10.1162/qss\_a\_00112} {\bibfield  {journal} {\bibinfo  {journal} {Quantitative Science Studies}\ }\textbf {\bibinfo {volume} {2}},\ \bibinfo {pages} {20--41} (\bibinfo {year} {2021})}\BibitemShut {NoStop}%
\bibitem [{\citenamefont {Wallace}, \citenamefont {Larivi{\`e}re},\ and\ \citenamefont {Gingras}(2009)}]{Wallace2009-lc}%
  \BibitemOpen
  \bibfield  {author} {\bibinfo {author} {\bibnamefont {Wallace}, \bibfnamefont {M.~L.}}, \bibinfo {author} {\bibnamefont {Larivi{\`e}re}, \bibfnamefont {V.}}, and\ \bibinfo {author} {\bibnamefont {Gingras}, \bibfnamefont {Y.}},\ }\bibfield  {title} {\enquote {\bibinfo {title} {Modeling a century of citation distributions},}\ }\href {https://doi.org/10.1016/j.joi.2009.03.010} {\bibfield  {journal} {\bibinfo  {journal} {J. Informetr.}\ }\textbf {\bibinfo {volume} {3}},\ \bibinfo {pages} {296--303} (\bibinfo {year} {2009})}\BibitemShut {NoStop}%
\bibitem [{\citenamefont {Waltman}\ and\ \citenamefont {Traag}(2020)}]{Waltman2020-fh}%
  \BibitemOpen
  \bibfield  {author} {\bibinfo {author} {\bibnamefont {Waltman}, \bibfnamefont {L.}}and\ \bibinfo {author} {\bibnamefont {Traag}, \bibfnamefont {V.~A.}},\ }\bibfield  {title} {\enquote {\bibinfo {title} {Use of the journal impact factor for assessing individual articles: Statistically flawed or not?}}\ }\href {https://doi.org/10.12688/f1000research.23418.2} {\bibfield  {journal} {\bibinfo  {journal} {F1000Res.}\ }\textbf {\bibinfo {volume} {9}},\ \bibinfo {pages} {366} (\bibinfo {year} {2020})},\ \Eprint {https://arxiv.org/abs/1703.02334} {arXiv:1703.02334 [cs.DL]} \BibitemShut {NoStop}%
\bibitem [{\citenamefont {Waltman}\ and\ \citenamefont {Van~Eck}(2019)}]{Waltman2019-zt}%
  \BibitemOpen
  \bibfield  {author} {\bibinfo {author} {\bibnamefont {Waltman}, \bibfnamefont {L.}}and\ \bibinfo {author} {\bibnamefont {Van~Eck}, \bibfnamefont {N.~J.}},\ }\bibfield  {title} {\enquote {\bibinfo {title} {Field normalization of scientometric indicators},}\ }\href {https://doi.org/10.1007/978-3-030-02511-3\_11} {\bibfield  {journal} {\bibinfo  {journal} {Springer Handbook of Science and Technology Indicators}\ ,\ \bibinfo {pages} {281--300}} (\bibinfo {year} {2019})}\BibitemShut {NoStop}%
\bibitem [{\citenamefont {Waltman}, \citenamefont {Van~Eck},\ and\ \citenamefont {Van~Raan}(2011)}]{Waltman2011-jw}%
  \BibitemOpen
  \bibfield  {author} {\bibinfo {author} {\bibnamefont {Waltman}, \bibfnamefont {L.}}, \bibinfo {author} {\bibnamefont {Van~Eck}, \bibfnamefont {N.~J.}}, and\ \bibinfo {author} {\bibnamefont {Van~Raan}, \bibfnamefont {A.~F.~J.}},\ }\bibfield  {title} {\enquote {\bibinfo {title} {Universality of citation distributions revisited},}\ }\href {https://doi.org/10.1002/asi.21671} {\bibfield  {journal} {\bibinfo  {journal} {J. Am. Soc. Inf. Sci. Technol.}\ }\textbf {\bibinfo {volume} {63}},\ \bibinfo {pages} {72--77} (\bibinfo {year} {2011})}\BibitemShut {NoStop}%
\bibitem [{\citenamefont {Wang}\ and\ \citenamefont {Barab{\'a}si}(2021)}]{Wang2021-aa}%
  \BibitemOpen
  \bibfield  {author} {\bibinfo {author} {\bibnamefont {Wang}, \bibfnamefont {D.}}and\ \bibinfo {author} {\bibnamefont {Barab{\'a}si}, \bibfnamefont {A.-L.}},\ }\href@noop {} {\emph {\bibinfo {title} {The Science of Science}}},\ \bibinfo {edition} {1st}\ ed.\ (\bibinfo  {publisher} {Cambridge University Press},\ \bibinfo {year} {2021})\BibitemShut {NoStop}%
\bibitem [{\citenamefont {Wang}, \citenamefont {Song},\ and\ \citenamefont {Barab{\'a}si}(2013)}]{Wang2013-tj}%
  \BibitemOpen
  \bibfield  {author} {\bibinfo {author} {\bibnamefont {Wang}, \bibfnamefont {D.}}, \bibinfo {author} {\bibnamefont {Song}, \bibfnamefont {C.}}, and\ \bibinfo {author} {\bibnamefont {Barab{\'a}si}, \bibfnamefont {A.-L.}},\ }\bibfield  {title} {\enquote {\bibinfo {title} {Quantifying {Long-Term} scientific impact},}\ }\href {https://doi.org/10.1126/science.1237825} {\bibfield  {journal} {\bibinfo  {journal} {Science}\ }\textbf {\bibinfo {volume} {342}},\ \bibinfo {pages} {127--132} (\bibinfo {year} {2013})}\BibitemShut {NoStop}%
\bibitem [{\citenamefont {Wang}\ \emph {et~al.}(2014)\citenamefont {Wang}, \citenamefont {Song}, \citenamefont {Shen},\ and\ \citenamefont {Barab{\'a}si}}]{Wang2014-sh}%
  \BibitemOpen
  \bibfield  {author} {\bibinfo {author} {\bibnamefont {Wang}, \bibfnamefont {D.}}, \bibinfo {author} {\bibnamefont {Song}, \bibfnamefont {C.}}, \bibinfo {author} {\bibnamefont {Shen}, \bibfnamefont {H.-W.}}, and\ \bibinfo {author} {\bibnamefont {Barab{\'a}si}, \bibfnamefont {A.-L.}},\ }\bibfield  {title} {\enquote {\bibinfo {title} {Science communication. response to comment on ``quantifying long-term scientific impact''},}\ }\href {https://doi.org/10.1126/science.1248961} {\bibfield  {journal} {\bibinfo  {journal} {Science}\ }\textbf {\bibinfo {volume} {345}},\ \bibinfo {pages} {149--149} (\bibinfo {year} {2014})}\BibitemShut {NoStop}%
\bibitem [{\citenamefont {Wang}, \citenamefont {Mei},\ and\ \citenamefont {Hicks}(2014)}]{Wang2014-zu}%
  \BibitemOpen
  \bibfield  {author} {\bibinfo {author} {\bibnamefont {Wang}, \bibfnamefont {J.}}, \bibinfo {author} {\bibnamefont {Mei}, \bibfnamefont {Y.}}, and\ \bibinfo {author} {\bibnamefont {Hicks}, \bibfnamefont {D.}},\ }\bibfield  {title} {\enquote {\bibinfo {title} {Comment on ``quantifying long-term scientific impact''},}\ }\href {https://doi.org/10.1126/science.1248770} {\bibfield  {journal} {\bibinfo  {journal} {Science}\ }\textbf {\bibinfo {volume} {345}},\ \bibinfo {pages} {149--149} (\bibinfo {year} {2014})}\BibitemShut {NoStop}%
\bibitem [{\citenamefont {Wang}, \citenamefont {Yu},\ and\ \citenamefont {Yu}(2009)}]{Wang2009-cr}%
  \BibitemOpen
  \bibfield  {author} {\bibinfo {author} {\bibnamefont {Wang}, \bibfnamefont {M.}}, \bibinfo {author} {\bibnamefont {Yu}, \bibfnamefont {G.}}, and\ \bibinfo {author} {\bibnamefont {Yu}, \bibfnamefont {D.}},\ }\bibfield  {title} {\enquote {\bibinfo {title} {Effect of the age of papers on the preferential attachment in citation networks},}\ }\href {https://doi.org/10.1016/j.physa.2009.05.008} {\bibfield  {journal} {\bibinfo  {journal} {Physica A: Statistical Mechanics and its Applications}\ }\textbf {\bibinfo {volume} {388}},\ \bibinfo {pages} {4273--4276} (\bibinfo {year} {2009})}\BibitemShut {NoStop}%
\bibitem [{\citenamefont {Way}\ \emph {et~al.}(2019)\citenamefont {Way}, \citenamefont {Morgan}, \citenamefont {Larremore},\ and\ \citenamefont {Clauset}}]{Way2019-av}%
  \BibitemOpen
  \bibfield  {author} {\bibinfo {author} {\bibnamefont {Way}, \bibfnamefont {S.~F.}}, \bibinfo {author} {\bibnamefont {Morgan}, \bibfnamefont {A.~C.}}, \bibinfo {author} {\bibnamefont {Larremore}, \bibfnamefont {D.~B.}}, and\ \bibinfo {author} {\bibnamefont {Clauset}, \bibfnamefont {A.}},\ }\bibfield  {title} {\enquote {\bibinfo {title} {Productivity, prominence, and the effects of academic environment},}\ }\href {https://doi.org/10.1073/pnas.1817431116} {\bibfield  {journal} {\bibinfo  {journal} {Proc. Natl. Acad. Sci. U. S. A.}\ }\textbf {\bibinfo {volume} {166}},\ \bibinfo {pages} {10729--10733} (\bibinfo {year} {2019})}\BibitemShut {NoStop}%
\bibitem [{\citenamefont {Wilsdon}(2015)}]{Wilsdon2015-gk}%
  \BibitemOpen
  \bibfield  {author} {\bibinfo {author} {\bibnamefont {Wilsdon}, \bibfnamefont {J.}},\ }\bibfield  {title} {\enquote {\bibinfo {title} {We need a measured approach to metrics},}\ }\href {https://doi.org/10.1038/523129a} {\bibfield  {journal} {\bibinfo  {journal} {Nature}\ }\textbf {\bibinfo {volume} {523}},\ \bibinfo {pages} {129} (\bibinfo {year} {2015})}\BibitemShut {NoStop}%
\bibitem [{\citenamefont {Wilsdon}\ \emph {et~al.}(2015)\citenamefont {Wilsdon}, \citenamefont {Allen}, \citenamefont {Belfiore}, \citenamefont {Campbell}, \citenamefont {Curry}, \citenamefont {Hill}, \citenamefont {Jones}, \citenamefont {Kain}, \citenamefont {Kerridge}, \citenamefont {Thelwall}, \citenamefont {Tinkler}, \citenamefont {Viney}, \citenamefont {Wouters}, \citenamefont {Hill},\ and\ \citenamefont {Johnson}}]{Wilsdon2015-bd}%
  \BibitemOpen
  \bibfield  {author} {\bibinfo {author} {\bibnamefont {Wilsdon}, \bibfnamefont {J.}}, \bibinfo {author} {\bibnamefont {Allen}, \bibfnamefont {L.}}, \bibinfo {author} {\bibnamefont {Belfiore}, \bibfnamefont {E.}}, \bibinfo {author} {\bibnamefont {Campbell}, \bibfnamefont {P.}}, \bibinfo {author} {\bibnamefont {Curry}, \bibfnamefont {S.}}, \bibinfo {author} {\bibnamefont {Hill}, \bibfnamefont {S.}}, \bibinfo {author} {\bibnamefont {Jones}, \bibfnamefont {R.}}, \bibinfo {author} {\bibnamefont {Kain}, \bibfnamefont {R.}}, \bibinfo {author} {\bibnamefont {Kerridge}, \bibfnamefont {S.}}, \bibinfo {author} {\bibnamefont {Thelwall}, \bibfnamefont {M.}}, \bibinfo {author} {\bibnamefont {Tinkler}, \bibfnamefont {J.}}, \bibinfo {author} {\bibnamefont {Viney}, \bibfnamefont {I.}}, \bibinfo {author} {\bibnamefont {Wouters}, \bibfnamefont {P.}}, \bibinfo {author} {\bibnamefont {Hill}, \bibfnamefont {J.}}, and\ \bibinfo {author} {\bibnamefont {Johnson}, \bibfnamefont {B.}},\ }\href {https://doi.org/10.13140/RG.2.1.4929.1363} {\enquote {\bibinfo {title} {Metric tide: Report of the independent review of the role of metrics in research assessment and management},}\ }\bibinfo {type} {Tech. Rep.}\ (\bibinfo  {institution} {Higher Education Funding Council for England},\ \bibinfo {year} {2015})\BibitemShut {NoStop}%
\bibitem [{\citenamefont {Woods}\ \emph {et~al.}(2022)\citenamefont {Woods}, \citenamefont {Brumberg}, \citenamefont {Kaltenbrunner}, \citenamefont {Pinfield},\ and\ \citenamefont {Waltman}}]{Woods2022-xp}%
  \BibitemOpen
  \bibfield  {author} {\bibinfo {author} {\bibnamefont {Woods}, \bibfnamefont {H.~B.}}, \bibinfo {author} {\bibnamefont {Brumberg}, \bibfnamefont {J.}}, \bibinfo {author} {\bibnamefont {Kaltenbrunner}, \bibfnamefont {W.}}, \bibinfo {author} {\bibnamefont {Pinfield}, \bibfnamefont {S.}}, and\ \bibinfo {author} {\bibnamefont {Waltman}, \bibfnamefont {L.}},\ }\href {https://doi.org/10.31235/osf.io/qaksd} {\enquote {\bibinfo {title} {Innovations in peer review in scholarly publishing: a meta-summary},}\ } (\bibinfo {year} {2022})\BibitemShut {NoStop}%
\bibitem [{\citenamefont {Wouters}\ \emph {et~al.}(2019)\citenamefont {Wouters}, \citenamefont {Sugimoto}, \citenamefont {Larivi{\`e}re}, \citenamefont {McVeigh}, \citenamefont {Pulverer}, \citenamefont {de~Rijcke},\ and\ \citenamefont {Waltman}}]{Wouters2019-cr}%
  \BibitemOpen
  \bibfield  {author} {\bibinfo {author} {\bibnamefont {Wouters}, \bibfnamefont {P.}}, \bibinfo {author} {\bibnamefont {Sugimoto}, \bibfnamefont {C.~R.}}, \bibinfo {author} {\bibnamefont {Larivi{\`e}re}, \bibfnamefont {V.}}, \bibinfo {author} {\bibnamefont {McVeigh}, \bibfnamefont {M.~E.}}, \bibinfo {author} {\bibnamefont {Pulverer}, \bibfnamefont {B.}}, \bibinfo {author} {\bibnamefont {de~Rijcke}, \bibfnamefont {S.}}, and\ \bibinfo {author} {\bibnamefont {Waltman}, \bibfnamefont {L.}},\ }\bibfield  {title} {\enquote {\bibinfo {title} {Rethinking impact factors: better ways to judge a journal},}\ }\href {https://doi.org/10.1038/d41586-019-01643-3} {\bibfield  {journal} {\bibinfo  {journal} {Nature}\ }\textbf {\bibinfo {volume} {569}},\ \bibinfo {pages} {621--623} (\bibinfo {year} {2019})}\BibitemShut {NoStop}%
\bibitem [{\citenamefont {Wu}, \citenamefont {Wang},\ and\ \citenamefont {Evans}(2019)}]{Wu2019-yv}%
  \BibitemOpen
  \bibfield  {author} {\bibinfo {author} {\bibnamefont {Wu}, \bibfnamefont {L.}}, \bibinfo {author} {\bibnamefont {Wang}, \bibfnamefont {D.}}, and\ \bibinfo {author} {\bibnamefont {Evans}, \bibfnamefont {J.~A.}},\ }\bibfield  {title} {\enquote {\bibinfo {title} {Large teams develop and small teams disrupt science and technology},}\ }\href {https://doi.org/10.1038/s41586-019-0941-9} {\bibfield  {journal} {\bibinfo  {journal} {Nature}\ }\textbf {\bibinfo {volume} {566}},\ \bibinfo {pages} {378--382} (\bibinfo {year} {2019})}\BibitemShut {NoStop}%
\bibitem [{\citenamefont {Yin}\ and\ \citenamefont {Wang}(2017)}]{Yin2017-la}%
  \BibitemOpen
  \bibfield  {author} {\bibinfo {author} {\bibnamefont {Yin}, \bibfnamefont {Y.}}and\ \bibinfo {author} {\bibnamefont {Wang}, \bibfnamefont {D.}},\ }\bibfield  {title} {\enquote {\bibinfo {title} {The time dimension of science: Connecting the past to the future},}\ }\href {https://doi.org/10.1016/j.joi.2017.04.002} {\bibfield  {journal} {\bibinfo  {journal} {Journal of Informetrics}\ }\textbf {\bibinfo {volume} {11}},\ \bibinfo {pages} {608--621} (\bibinfo {year} {2017})},\ \Eprint {https://arxiv.org/abs/1704.04657} {arXiv:1704.04657 [physics.soc-ph]} \BibitemShut {NoStop}%
\bibitem [{\citenamefont {Zeng}\ \emph {et~al.}(2017)\citenamefont {Zeng}, \citenamefont {Shen}, \citenamefont {Zhou}, \citenamefont {Wu}, \citenamefont {Fan}, \citenamefont {Wang},\ and\ \citenamefont {Stanley}}]{Zeng2017-tg}%
  \BibitemOpen
  \bibfield  {author} {\bibinfo {author} {\bibnamefont {Zeng}, \bibfnamefont {A.}}, \bibinfo {author} {\bibnamefont {Shen}, \bibfnamefont {Z.}}, \bibinfo {author} {\bibnamefont {Zhou}, \bibfnamefont {J.}}, \bibinfo {author} {\bibnamefont {Wu}, \bibfnamefont {J.}}, \bibinfo {author} {\bibnamefont {Fan}, \bibfnamefont {Y.}}, \bibinfo {author} {\bibnamefont {Wang}, \bibfnamefont {Y.}}, and\ \bibinfo {author} {\bibnamefont {Stanley}, \bibfnamefont {H.~E.}},\ }\bibfield  {title} {\enquote {\bibinfo {title} {The science of science: From the perspective of complex systems},}\ }\href {https://doi.org/10.1016/J.PHYSREP.2017.10.001} {\bibfield  {journal} {\bibinfo  {journal} {Phys. Rep.}\ }\textbf {\bibinfo {volume} {714-715}},\ \bibinfo {pages} {1--73} (\bibinfo {year} {2017})}\BibitemShut {NoStop}%
\bibitem [{\citenamefont {Zhang}, \citenamefont {Rousseau},\ and\ \citenamefont {Sivertsen}(2017)}]{Zhang2017-yo}%
  \BibitemOpen
  \bibfield  {author} {\bibinfo {author} {\bibnamefont {Zhang}, \bibfnamefont {L.}}, \bibinfo {author} {\bibnamefont {Rousseau}, \bibfnamefont {R.}}, and\ \bibinfo {author} {\bibnamefont {Sivertsen}, \bibfnamefont {G.}},\ }\bibfield  {title} {\enquote {\bibinfo {title} {Science deserves to be judged by its contents, not by its wrapping: Revisiting seglen's work on journal impact and research evaluation},}\ }\href {https://doi.org/10.1371/journal.pone.0174205} {\bibfield  {journal} {\bibinfo  {journal} {PLoS One}\ }\textbf {\bibinfo {volume} {12}},\ \bibinfo {pages} {e0174205} (\bibinfo {year} {2017})}\BibitemShut {NoStop}%
\end{thebibliography}%

\end{document}